\documentclass[12pt]{article}
\usepackage[T1]{fontenc}
\usepackage[utf8]{inputenc}
\usepackage{amsmath}
\usepackage{amssymb}
\usepackage{graphicx}
\usepackage{enumerate}
\usepackage{float}
\usepackage{natbib}
\usepackage{url}
\usepackage{bm}
\usepackage{bbm}
\usepackage{subfigure}
\usepackage{multirow}
\usepackage{booktabs}
\usepackage{array}
\usepackage[margin=1in]{geometry}
\usepackage[hidelinks]{hyperref}
\graphicspath{{fig/}}
\allowdisplaybreaks[4]
\newtheorem{thm}{Theorem}
\newtheorem{lemma}{Lemma}

\newcommand*{\dif}{\mathop{}\!\mathrm{d}}

\def\E{\mathrm{E}}
\def\Pr{\mathrm{P}}
\def\var{\mathrm{var}}

\def\cov{\mathrm{cov}}
\def\diag{\mathrm{diag}}

\def\tr{\mathrm{tr}}

\def\A{{\bf A}}

\def\B{{\bf B}}
\def\z{{\bm z}}

\def\R{{\bf R}}
\def\I{{\bf I}}

\def\bmu{{\bm \mu}}

\def\X{{\bm X}}

\def\W{{\bm W}}
\def\b{{\bm b}}
\def\L{{\bf \Lambda}}

\def\D{{\bf D}}

\def\bms{{\bm\Sigma}}

\def\cd{\mathop{\rightarrow}\limits^{d}}
\def\mR{\mathbb{R}}

\def\bmv{\bm \varepsilon}

\def\Q{{\bm Q}}

\hypersetup{
  pdftitle={Adaptive L-statistics for high dimensional test problem},
  pdfauthor={Huifang Ma, Long Feng, and Zhaojun Wang}
}

\begin{document}

\title{\textbf{Adaptive L-statistics for high dimensional test problem}}
\author{Huifang Ma \qquad Long Feng \qquad Zhaojun Wang\\[0.5em]
\small School of Statistics and Data Science, LEBPS, KLMDASR, AAIS and LPMC,\\[-0.1em]
\small Nankai University}
\date{}
\maketitle

\begin{abstract}
In this study, we focus on applying $L$-statistics to the high-dimensional one-sample location test problem. Intuitively, an $L$-statistic with $k$ parameters tends to perform optimally when the sparsity level of the alternative hypothesis matches $k$. We begin by deriving the limiting distributions for both $L$-statistics with fixed parameters and those with diverging parameters. To ensure robustness across varying sparsity levels of alternative hypotheses, we first establish the asymptotic independence between $L$-statistics with fixed and diverging parameters. Building on this, we propose a Cauchy combination test that integrates $L$-statistics with different parameters. Both simulation results and real data applications highlight the advantages of our proposed methods.
\end{abstract}

\noindent%
{\it Keywords:} Asymptotic independence; Cauchy combination; High-dimensional tests; $L$-statistics; Mean test.

\section{Introduction}\label{Sec: Introduction}

In statistical analysis, the mean test is a foundational problem. Traditionally, Hotelling's $T^2$ test has been extensively employed when the number of dimensions in the observations is fixed. However, in recent times, high-dimensional data has become prevalent in numerous fields, including finance, industries, and gene expression data. A defining characteristic of high-dimensional data is that the number of dimensions in the observations is either comparable to or even exceeds the sample sizes. When the dimensions surpass the sample sizes, Hotelling's $T^2$ test becomes inapplicable due to the non-invertibility of the sample covariance matrix. To address this challenge, recent research has focused on exploring high-dimensional location parameter testing problems.

One approach to addressing this problem is through sum-type tests. These tests involve summing the $r$-th power ($r=2,4,6,\ldots$) of the differences between the observed and null hypothesis means across all variables, resulting in a test statistic that captures the overall deviation from the null hypothesis. \cite{bai1996effect} proposed a solution where the sample covariance matrix in Hotelling's $T^2$ test statistic is replaced by the identity matrix. \cite{chen2010two} expanded on \cite{bai1996effect}'s test, allowing the dimension to grow exponentially with the sample sizes. \cite{srivastava2009test} introduced a scalar-invariant test that substitutes the sample covariance matrix with its diagonal matrix. \cite{feng2015two} proposed a bias-corrected test statistic that accommodates dimensions growing at a quadratic rate relative to the sample sizes. Additionally, several studies have explored robust sum-type test statistics for high-dimensional data, such as \cite{wang2015high}, \cite{feng2016multivariate}, and \cite{feng2016spatial}. Furthermore, \cite{XLWP16} and \cite{he2021asymptotically} considered sum-type test statistics with higher-order $r>2$, which outperform tests with $r=2$ when the sparsity level is a little higher. { \cite{saha2017some} investigate some nonparametric one-sample tests based on functions of interpoint distances. Most of the aforementioned methods rely on the assumption that the covariance matrix exhibits weak dependence. To address more general dependence structures, \cite{zhang2020simple} proposed an $L_2$-norm based test that accommodates arbitrary covariance matrices by employing the Welch–Satterthwaite $\chi^2$-approximation to determine critical values. In contrast, \cite{lopes2011more} introduced a random projection approach that reduces the high-dimensional data to a lower-dimensional space, allowing the application of the classical Hotelling’s $T^2$ test. Their results demonstrate that the method is particularly effective under strong dependence structures. More recently, \cite{liu2024projection} proposed an alternative projection-based technique tailored for high-dimensional mean testing. For a comprehensive overview of existing high-dimensional mean test procedures, we refer readers to \cite{huang2022overview}.} However, in high-dimensional scenarios, sum-type tests may become less effective when the true signal is highly sparse, meaning that only a small number of variables contain significant information.

An alternative method is the use of max-type tests. Rather than aggregating squared differences, these tests focus on the maximum deviation observed across all variables. This approach excels at identifying sparse signals within high-dimensional datasets. \cite{TonyCai2014TwosampleTO} investigated the limited null distribution and conducted a power analysis for the high-dimensional two-sample location problem under the assumption of weak correlation. For more complex correlation structures, Gaussian approximation \citep{Chernozhukov2013InferenceOC,chang2024central} is commonly employed to simulate the null distribution of max-type test statistics. However, traditional max-type tests, which only consider the maximum test statistic for each variable, may lose some efficiency when a small number of variables have non-zero means. To address this, some studies have proposed max-$k$-type tests that sum the $k$ largest test statistics, such as \cite{zhang2015,chang2017,zhou2018unified,chang2023testing,xia2024adaptive,ding2024gaussian}. Nevertheless, these papers do not establish the limiting null distribution for max-$k$-type tests. {  \cite{zhong2013tests} proposed a Higher Criticism (HC) test tailored for sparse alternatives, which involves constructing a thresholded $L_1$ or $L_2$-based statistic and then taking the maximum over a family of $L_\gamma$-thresholded statistics.} Furthermore, max-type tests, including max-$k$ tests, may lack power when the signal is dense, meaning that many variables have non-zero means. 

To tackle the constraints of sum-type and max-type test procedures, numerous adaptive tests have emerged. These tests endeavor to merge the advantages of sum-type and max-type tests by dynamically modifying their operation according to data characteristics. 
{  \cite{fan2015power} propose a power enhancement procedure that uses a screening technique for high-dimensional tests. They combine the power enhancement component with an asymptotically pivotal statistic to strengthen the power under sparse
alternatives.} \cite{XLWP16} integrated multiple sum-of-powers test statistics with varying power indices ($r=2,4,6,\ldots,\infty$). Building on this, \cite{he2021asymptotically} eliminated redundant terms from \cite{XLWP16}'s test statistics and formulated a series of asymptotically independent U-statistics, subsequently proposing adaptive tests based on these U-statistics.  Meanwhile, \cite{zhou2018unified} developed a broad spectrum of tests rooted in a novel $L_r$ norm variant with differing $r$ values, amalgamating these tests to devise a data-adaptive test that exhibits robustness across various alternative scenarios. \cite{feng2024asymptotic} delved into the realm of adaptive tests within the context of a broader and more general correlation matrix.  As demonstrated in these studies, test statistics with large $r$ excel under sparse alternatives, whereas those with small $r$ perform well under dense alternatives. Despite the impressive performance of methods combining $L_r$ norm test statistics across a wide array of alternatives, they suffer from two primary drawbacks. Firstly, there is a significant computational load for large $r$ values. {  Secondly, due to computational constraints, the parameter $r$ in their test procedures is typically restricted to a limited set of values, such as $r = 2, 4, 6, \infty$. This limited selection significantly narrows the adaptability of their methods, resulting in strong performance only within a relatively small range of sparsity levels. As a consequence, these procedures may exhibit suboptimal power when the true sparsity level deviates from the predefined values, highlighting the need for more flexible and broadly robust testing strategies.} To circumvent this problem, we propose utilizing $L$-statistics to devise an adaptive test procedure tailored for high-dimensional testing problems.

$L$-statistics are composed of linear combinations of functions of order statistics, forming a well-recognized class of statistics in robust and non-parametric theory, as documented in various studies \citep{Chernoff1967,Stigler1969,Chanda1990,Puri1993}. Intuitively, in high dimensional location testing problem, when we know that $k$ variables have nonzero means, we can sum the squares of the $k$ largest order statistics derived from the classic $t$-test for each variable. This directly constitutes an $L$-statistic with a specific function. In this paper, we begin by analyzing the theoretical properties of these $L$-statistics for varying parameters $k$. For a fixed $k$, the corresponding null distribution of the $L$-statistic converges to a Gumbel-type distribution, whereas for large $k$ tending to infinity, the $L$-statistic follows an asymptotic normal distribution. In practical scenarios, the number of variables with nonzero means is often unknown. To address this, we propose combining $L$-statistics with different $k$ values using a Cauchy combination test procedure, leveraging the asymptotic independence between $L$-statistics with fixed and diverging $k$. Simulation studies and real-world data applications demonstrate that our newly proposed adaptive $L$-statistic test procedure outperforms existing adaptive test procedures.

This paper makes three primary contributions:
\begin{itemize}
	\item Firstly, we determine the limited null distribution of the $L$-statistic with a fixed parameter $k$, which aligns with the max-$k$-type test statistics. This $L$-statistic demonstrates strong performance under sparse alternatives, outperforming the classic approach of solely considering the maximum value of $t$-test statistics.
	\item Secondly, we establish the asymptotic multivariate normality of the $L$-statistic with numerous diverging parameters $k$. This $L$-statistic is highly efficient under dense alternatives. {  We provide a simple power analysis for this dense alternative, demonstrating that the optimal choice of the parameter $k$ is achieved when $k$ matches the true number of nonzero mean components in the alternative hypothesis. This result highlights the importance of aligning the test procedure with the underlying sparsity structure to maximize statistical power.}
	\item Lastly, we demonstrate the asymptotic independence between the $L$-statistic with diverging parameters $k$ and fixed parameters $k$. The corresponding Cauchy combination test procedure exceeds the performance of existing adaptive methods.
\end{itemize}

This paper is structured as follows: In Section \ref{Sec: One sample location problem}, we introduce the test statistics for addressing the high-dimensional one-sample location problem. Section \ref{Sec: Simulation} presents the results of simulation studies conducted to evaluate the performance of the proposed test statistics. 
Section \ref{Sec: Conclusion} summarizes our findings and conclusions. The real-data application, additional simulation results, further discussion, and detailed proofs are provided in the Supplementary Material.

\section{One sample location problem}\label{Sec: One sample location problem}

Assume $\X_1,\ldots,\X_n$ are
independent and identically distributed $p$-dimensional random vectors with mean $\bmu$ and covariance matrix $\bms$. The
classical one-sample mean testing problem considers
\begin{align}\label{hypothesis}
H_0: \bmu=\bm 0 ~~\textrm{versus}~~H_1: \bmu\ne \bm 0.
\end{align}
or equivalently
\begin{align*}
H_0: \mu_i=0, i=1,\ldots,p ~~\textrm{versus}~~H_1: \mu_i\not=0~~ \text{for~~some}~~ i.
\end{align*}
For each variable, the classic $t$-test statistic is
\begin{align}\label{t_test statistic}
t_i=\frac{\sqrt{n}\bar{X}_{.i}}{\hat{\sigma}_{ii}^{1/2}},
\end{align}
where $\bar{X}_{.i}$ and $\hat{\sigma}_{ii}$ are the sample mean and variance of the $i-$th coordinate of the population vector.

\cite{XLWP16} defined a sum-of-powers test statistic with power index $r$ as
\begin{align*}
L(r)=\sum_{i=1}^p \bar{X}_{.i}^r.
\end{align*}
The power of $L(r)$ with large $r$ gets larger as the sparsity of the alternative hypothesis. So they proposed the following adaptive test to combine the sum-of-power tests to improve the test power:
\begin{align*}
T_{adaQ}=\min_{r \in \Gamma} p_{r},
\end{align*}
where $p_{r}$ is the $p$-value of $L(r)$ and $\Gamma=\{1,2,\ldots,r_u,\infty\}$. \cite{zhou2018unified} introduced a $(s_0,r)$-norm based test statistic:
\begin{align}\label{sg}
T_{(s_0,r)}=\sum_{i=1}^{s_0} |t|^r_{(p+1-i)},
\end{align}
where $|t|_{(1)}\le\ldots\le |t|_{(p)}$ are the order statistic of $\{|t_1|,\ldots, |t_p|\}$.
They showed the consistency of the Gaussian approximation method under the assumption $s_0^2\log (p)=o(n^\delta),0<\delta<1/7$. However, they did not establish the limiting null distribution of their proposed test statistics for both small and large values of $s_0$. They also proposed an adaptive test by taking the minimum $p$-values of all individual tests, i.e.
\begin{align*}
T_W=\min_{r \in \Gamma} p_{(s_0,r)}.
\end{align*}
Both methods rely on combining test procedures with varying power parameters $r$. According to the articles, $r$ is typically set to $2, 4, 6,$ or $\infty$ for the adaptive procedures. Odd values of $r$ are not preferred because, in practice, we do not know the signs of the signals $\mu_i$, and if $\sum_{i=1}^p \mu_i = 0$, test procedures with odd $r$ would be ineffective. For large $r \geq 8$, the computational load becomes excessive, and the performance is comparable to $r = \infty$. Therefore, if we select $\Gamma = \{2, 4, 6, \infty\}$, the adaptive test's effectiveness would be limited to a narrow range of sparsity for the alternative hypothesis.

Instead of considering the power function of $t_i$, we introduce the following L-statistic based on the ordered $t_i^2$'s,
\begin{align}\label{L-statistic}
T_k=\sum_{i=1}^k t^2_{(p+1-i)},
\end{align}
where $t^2_{(1)}\le\ldots\le t^2_{(p)}$ are the order statistic of $\{t^2_1,\ldots,t^2_p\}$. In this paper, we first consider the $L$-statistic in \eqref{L-statistic} with different integer $k$, then also proposed an adaptive test procedure which combine different $L$-statistics with different $k$. It is important to note that our proposed test statistic $T_k$ is equivalent to $T_{(k,2)}$ as defined in \eqref{sg}. While the test statistic itself is not novel, establishing its theoretical properties presents a significant challenge. This paper's primary contribution lies in addressing this challenge and providing a rigorous theoretical foundation for the test statistic. 

\subsection{L-test with fixed parameter}

We will establish the asymptotic properties of $T_k$ with fixed $k$ when $\min(n,p)\to\infty$. To proceed, we first introduce some assumptions.

Let $\D$ denote the diagonal matrix of $\bms$ and let $\R=(r_{ij})_{p\times p}=\D^{-1/2}\bms \D^{-1/2} $denote the correlation matrix. Let $\lambda_{\min}(\cdot)$ and $\lambda_{\max}(\cdot)$ denote the minimum and maximum
eigenvalues of a matrix respectively. For a set of multivariate random vectors $\z=\{z_j:j\ge 1\}$ and integers $a<b$, let $\mathcal{Z}_a^b$ be the $\sigma$-algebra generated by $\{z_j:j\in[a,b]\}$. For each $s\ge 1$, define the $\alpha$-mixing coefficient $\alpha_Z(s)=\sup_{t\ge 1}\{|\Pr(A\cap B)-\Pr(A)\Pr(B)|:A\in\mathcal{Z}_1^t,B\in\mathcal{Z}_{t+s}^{\infty}\}$.
\begin{itemize}
	\item[(A1)] $\X_i=(X_{i1},\ldots,X_{ip})^\top,i=1,\ldots,n$ are
	independent and identically distributed with $\E(\X_1)=\bmu$ and $\var(\X_1)=\bms$, where $\bms=(\sigma_{ij})_{p\times p}$ is strictly positive definite. $X_{ij}$'s have one of the following three types of tails: (\romannumeral1) sub-Gaussian-type tails, i.e. there exist $\eta>0$ and $K>0$ such that $\E\{\exp(\eta X_{ij}^2/\sigma_{jj})\}\le K$ for $1\le j\le p$, where $p$ satisfies $\log p=o(n^{1/4})$; (\romannumeral2) sub-exponential-type tails, i.e. there exist $\eta>0$ and $K>0$ such that $\E\{\exp(\eta |X_{ij}|/\sigma_{jj}^{1/2})\}\le K$ for $1\le j\le p$, where $p$ satisfies $\log p=o(n^{1/4})$; (\romannumeral3) sub-polynomial-type tails, i.e. for some constants $\gamma_0,\epsilon>0$ and $K>0$, $\E|X_{ij}/\sigma_{jj}^{1/2}|^{2\gamma_0+2+\epsilon}\le K$ for $1\le j\le p$,  where $p$ satisfies $p\le c_0n^{\gamma_0}$ for some constant $c_0>0$.
	
	\item[(A2)] (\romannumeral1) There exists $c_1>0$ such that $c_1^{-1}\le \lambda_{\min}(\R)\le \lambda_{\max}(\R)\le c_1$; (\romannumeral2) there exists $r_1>0$ such that $\max_{1\le i<j\le p}|r_{ij}|\le r_1<1$.
	
	\item[(A3)] $\{(X_{ij},i=1,\ldots,n):1\le j \le p\}$ is $\alpha$-mixing with $\alpha_X(s)\le C\delta^s$, where $\delta\in(0,1)$ and $C>0$ is some constant.
\end{itemize}
{  Condition (A1) is analogous to Conditions 6 and 7 in \cite{TonyCai2014TwosampleTO}, which ensure that the error distribution does not deviate significantly from normality. This mild moment condition guarantees the validity of asymptotic approximations under non-Gaussian settings. Condition (A2) corresponds to Conditions 1–3 in \cite{TonyCai2014TwosampleTO}, which impose standard regularity assumptions on the eigenvalue spectrum of the covariance matrix. Such assumptions are commonly employed in high-dimensional analysis to prevent degenerate scenarios and ensure stable estimation.

Condition (A3) imposes a weak temporal or spatial dependence structure among the variables and has been adopted in various works including \cite{bickel2008regularized,XLWP16,he2021asymptotically,chen2019two}. Specifically, it requires the data to be $\alpha$-mixing with a sufficiently fast decay rate, which allows for certain degrees of dependence while still permitting valid asymptotic results. This condition is particularly relevant in applications such as genome-wide association studies (GWAS), where variables (e.g., genetic markers) exhibit local dependence due to their physical proximity on chromosomes. In such cases, correlations tend to decay with increasing genomic distance. It is worth noting that Condition (A3) is invariant under a permutation of variable indices—thus, as long as a reordering exists that satisfies the mixing condition, the theoretical results remain applicable.

}

We now state our first main result, which is about the limiting distribution of each $t_{(j)}^2$ and the joint limiting distribution of all $t_{(j)}^2$'s.
\begin{thm}\label{asymptotic distribution of ordered t_i}
	Suppose Assumptions (A1)-(A3) hold. Then as $\min(n,p)\to\infty$, we have
	
	\noindent(\romannumeral1) for all integer $1\le s\le p$ and $x\in\mathbb{R}$,
	\begin{align*}
	\Pr_{H_0}\left(t_{(p+1-s)}^2-b_p\le x \right)\to \Lambda(x)\sum_{i=0}^{s-1}\frac{\{\log \Lambda^{-1}(x)\}^i}{i!},
	\end{align*}
	\noindent(\romannumeral2) for all integer $2\le k\le p$ and $x_1\ge \ldots\ge x_k\in\mR$
	\begin{align*}
	&\Pr_{H_0}\left(\bigcap_{j=1}^k\left(t_{(p+1-j)}^2-b_p \leq x_j\right)\right)\\
	\to&\Lambda\left(x_k\right) \sum_{\sum_{i=2}^j k_i \leq j-1, j=2, \ldots, k} \prod_{i=2}^k \frac{\{\log \Lambda^{-1}(x_i)-\log \Lambda^{-1}(x_{i-1})\}^{k_i}}{k_i!},
	\end{align*}
	where $b_p=2\log p-\log (\log p)$ and $\Lambda(x)=\exp\{-\pi^{-1/2}\exp (-x/2)\}$.
\end{thm}

According to Theorem \ref{asymptotic distribution of ordered t_i}, through convolution, for fixed $k$ the asymptotic null distribution of $T_k$ exists. Specially, $\Pr_{H_0}(T_1-b_p\le x)\to \Lambda(x)$. When $k>1$, the asymptotic null distribution of $T_k$ is complex. {   The limiting null distribution of $T_k$ plays a crucial role in establishing the asymptotic independence between $T_k$ statistics with fixed and diverging $k$. This theoretical justification underpins the construction of our adaptive test procedure.}

{  As highlighted in several studies (e.g., \cite{liu2013convergence}, \cite{feng2024asymptotic}), the convergence of max-type test statistics to their asymptotic distributions can be remarkably slow, particularly in high-dimensional settings. This slow convergence may lead to substantial size distortions when using asymptotic critical values in finite samples.}
Hence, we adopt the wild bootstrap method to calculate the $p$-values. First, we centered the sample as $\tilde{\X}_i=\X_i-\hat{\bmu}$ where $\hat{\bmu}$ is the sample mean. Then, we generate the bootstrap sample as $\X_i^*=\tilde{\X}_i*\xi_i$ where $\xi_i$ is generated from Rademacher distribution. Next, we calculate {  the bootstrap $t$-test statistic $t_i^*=\sqrt{n}\bar{X}^*_{.i}/\sqrt{\hat{\sigma}_{ii}}$ and }the corresponding test statistic as $T_k^*$ with different $k$ based on the bootstrap samples $\{\X_1^*,\cdots,\X_n^*\}$. Repeat this procedure $B$ times, we obtain the corresponding bootstrap test statistics $T_{k,1}^*,\cdots,T_{k,B}^*$. For fixed $k$, we calculate the $p$-values as
\begin{align}\label{p_value of T_k}
p_{T_k}=\frac{1}{B}\sum_{i=1}^B I(T_k<T_{k,i}^*).
\end{align}
We reject the null hypothesis at the significant level $\alpha$ if $p_{T_k}\le \alpha$. {  We present the theoretical justification of the proposed bootstrap method in the Supplementary Material.
}

Next, we turn to analyze the power of the $T_k$-based testing procedure.
\begin{thm}\label{power of T_k}
	Suppose Assumptions (A1)-(A3) hold. If $k^{7/2}n^{-1/2}\log^{7/2}(np)=o(1)$, then as $\min(n,p)\to\infty$,
	\begin{align*}
	\inf_{\bmu\in\mathcal{A}_{k,\epsilon}}\Pr(p_{T_k}<\alpha)\to 1,
	\end{align*}
	for some $\epsilon>0$. Here,
	\begin{equation*}
	\mathcal{A}_{k,\epsilon}:=\left\lbrace \bmu=(\mu_1,\ldots,\mu_p)^\top:\sum_{i=1}^k\check{\mu}^2_{(p+1-i)}\ge (2k+\epsilon)\frac{\log p}{n} \right\rbrace,
	\end{equation*}
	where $\check{\mu}^2_{(i)}$ is the $i$-th order statistic of $\{\mu_i^2/\sigma_{ii}\}_{i=1}^p$.
\end{thm}

Theorem \ref{power of T_k} shows that the $T_k$-based test (fixed $k$) is effective in detecting sparse alternatives.
When k = 1, $\mathcal{A}_{k,\epsilon}$ reduces to $\max_{1\le i\le p}|\mu_i|/\sqrt{\sigma_{ii}}\ge \sqrt{(2+\epsilon)\log(p)/n}$
for some $\epsilon$. According to Theorem 3 of \cite{TonyCai2014TwosampleTO}, the separation rate $\sqrt{\log(p)/n}$ is minimax optimal.

{ 
Finally, we analyze the power function of the test statistic \( T_k \) for a fixed \( k \). Let \( G_k(x, \boldsymbol{\mu}) \) and \( g_k(x, \boldsymbol{\mu}) \) denote the cumulative distribution function (CDF) and probability density function (PDF) of \( T_k \), respectively. Under assumptions (A1)–(A3), and following a similar argument to the proof of Theorem \ref{asymptotic distribution of ordered t_i}, we can show that
\[
G_k(x, \boldsymbol{\mu}) = P\left( \sum_{i=1}^k Z_{(p+1-i)}^2 \le x \right),
\]
where \( Z_{(i)} \) is the \( i \)-th order statistic of the sample \( Z_1, \dots, Z_p \), and \( (Z_1, \dots, Z_p)^\top \sim N(\boldsymbol{\mu}, \mathbf{I}_p) \). Related results can be found in Theorem 5 of \cite{ding2024gaussian}.

Based on this representation, the power function of \( T_k \) is given by
\[
\beta_k(\boldsymbol{\mu}) = 1 - G_k(q_k(1 - \alpha), \boldsymbol{\mu}),
\]
where \( q_k(1 - \alpha) \) is the \( (1 - \alpha) \)-quantile of \( G_k(x, \boldsymbol{0}) \).

The optimal choice of \( k \) is then defined as
\[
k^* = \underset{1 \le k \le K}{\arg\max}~ \beta_k(\boldsymbol{\mu}),
\]
where \( K \) is the maximum number of nonzero components allowed under the sparse alternative. In our simulation studies, we observed that when \( K > 20 \), the limiting distribution of \( T_k \) approaches normality. This is further supported by the theoretical results in the next subsection. Thus, we fix \( K = 20 \) in our analysis.

Since the explicit forms of \( G_k \) and \( g_k \) are analytically intractable, we estimate them via simulation. For simplicity, we consider a sparse mean vector of the form:
\[
\mu_i = \delta \text{ for } i = 1, \dots, k_0, \quad \mu_i = 0 \text{ for } i > k_0,
\]
i.e., only the first \( k_0 \) elements of the mean vector are nonzero. In this setting, we denote the power function as \( \beta_k(\boldsymbol{\mu}) = H(k, k_0) \).

Figure \ref{kk} presents the curves of \(H(k,k_0)\) for the odd values \(k_0=1,3,\ldots,19\), with significance level \(\alpha=0.05\) and signal strength parameter \(\delta=4-(k_0-1)/8\). This choice of \( \delta \) ensures that the signal becomes weaker as the number of nonzero components increases, maintaining a roughly comparable signal-to-noise ratio across different sparsity levels.

From Figure \ref{kk}, we observe a clear pattern. When the alternative hypothesis is highly sparse (i.e., \( k_0 < 10 \)), the function \( H(k, k_0) \)—and hence the power—is maximized when \( k \) is chosen equal to the true sparsity level \( k_0 \). In contrast, when \( k_0 > 10 \), the curves of \( H(k, k_0) \) become relatively flat for \( k \in (k_0, 20) \), suggesting that the power is not very sensitive to the exact choice of \( k \) in this range. This corresponds to the regime where \( k \) diverges with the dimension and the distribution of \( T_k \) approaches normality, as discussed in the next subsection.

Based on these findings, we fix \( k = 5 \) in our simulation studies as a reasonable compromise between detection power under sparse and moderately dense alternatives. However, in practice, the optimal choice of \( k \) may depend on the underlying sparsity level and signal strength, and alternative values of \( k \) can be considered accordingly.

\begin{figure}[htb]
	\centering
	\includegraphics[width=\linewidth]{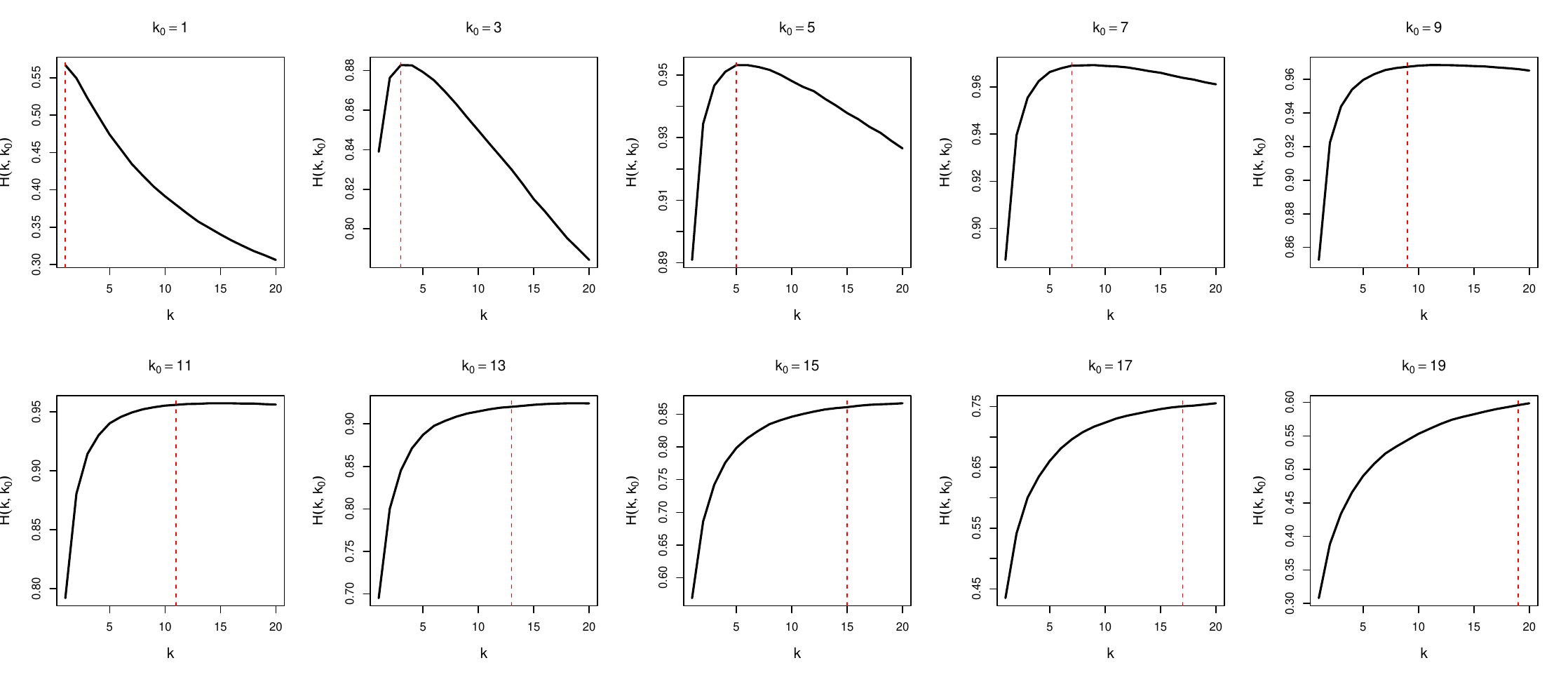} 
 \caption{Power curves of \( H(k, k_0) \) under different values of \( k_0 \). The red dashed line indicates the true value of \( k_0 \).
}\label{kk}
\end{figure}
}

\subsection{L-test with diverging parameter}

To detect dense alternatives, we consider a test based on $T_k$ with diverging $k=\lceil \gamma p\rceil$ and { fixed} $\gamma\in(0,1)$. We will establish the asymptotic normality of $T_{\lceil \gamma p\rceil}$. Here $\lceil x\rceil=\inf\{y\in\mathbb{Z}:y\ge x\}$. To proceed, define
\begin{align}\label{check T_k}
\check{T}_{\lceil \gamma p\rceil}=\sum_{i=1}^{\lceil \gamma p\rceil}\check{t}^2_{(p+1-i)}~~\text{with}~~\check{t}_i=\frac{\sqrt{n}\bar{X}_{.i}}{\sigma_{ii}^{1/2}}.
\end{align}
The difference between $T_{\lceil \gamma p\rceil}$ and $\check{T}_{\lceil \gamma p\rceil}$ is negligible under some mild conditions. So, it suffices to show the asymptotic normality of $\check{T}_{\lceil \gamma p\rceil}$. 
{  Let $\hat{v}_\gamma=\check{t}^2_{\lceil(1-\gamma)p\rceil+1}$ be the $(1-\gamma)$th sample quantile of $\{\check{t}_i^2\}_{1\le i\le p}$, and $v_\gamma$ be the $(1-\gamma)$th quantile of $\chi^2_1$. $\mathbb{I}(\cdot)$ represents the indicator function. Motivated by the Bahadur type expansion \citep{Wang2023NonparametricEO}, in the Supplementary Material, we show that under $H_0$
\begin{align*}
    \check{T}_{\lceil \gamma p\rceil}=\sum_{i=1}^p\check{t}_i^2\mathbb{I}(\check{t}_i^2\ge \hat{v}_\gamma)=\sum_{i=1}^p\left\{(\check{t}_i^2-v_\gamma)\mathbb{I}(\check{t}_i^2\ge v_\gamma)+\gamma v_\gamma\right\}+o_p(p^{1/2}),
\end{align*}
which,via the Berstein's block method \citep{Ibragimov1971IndependentRV}, leads to the asymptotic normality as the following theorem.
\begin{thm}\label{asymptotic normality}
	Under Assumptions (A1)-(A3), for all { fixed} $\gamma_l\in(0,1),l=1,\ldots,s$ with fixed $s$, if $p=o(n^2)$ and $\gamma_l\log(p)n\sum_{i=1}^{\lceil\gamma_1 p\rceil}\check{\mu}^2_{(p+1-i)}=o(1)$, then as $\min(n,p)\to\infty$,
	\begin{align*}
	\left(\frac{T_{\lceil \gamma_1 p\rceil}-\mu_{\gamma_1}-n\sum_{i=1}^{\lceil\gamma_1 p\rceil}\check{\mu}^2_{(p+1-i)}}{\sqrt{p\sigma_{\gamma_1\gamma_1}}},\ldots,\frac{T_{\lceil \gamma_s p\rceil}-\mu_{\gamma_s}-n\sum_{i=1}^{\lceil\gamma_s p\rceil}\check{\mu}^2_{(p+1-i)}}{\sqrt{p\sigma_{\gamma_s\gamma_s}}} \right)
	\end{align*}
	converges in distribution to $\mathcal{N}_s(\bm 0,\bm \Xi)$, where 
    \begin{align*}
        \mu_\gamma:=&\sum_{i=1}^p\E_{H_0}\left\{(\check{t}_i^2-v_\gamma)\mathbb{I}(\check{t}_i^2\ge v_\gamma)+\gamma v_\gamma\right\},\nonumber\\
        \sigma_{\gamma_t\gamma_q}:=&\cov_{H_0}\left\{\frac{1}{\sqrt{p}}\sum_{i=1}^p(\check{t}_i^2-v_{\gamma_t})\mathbb{I}(\check{t}_i^2\ge v_{\gamma_t}),\frac{1}{\sqrt{p}}\sum_{i=1}^p(\check{t}_i^2-v_{\gamma_q})\mathbb{I}(\check{t}_i^2\ge v_{\gamma_q})\right\}\nonumber\\
        =&\frac{1}{p}\sum_{i=1}^p\cov_{H_0}\{(\check{t}^2_i-v_{\gamma_t})\mathbb{I}(\check{t}^2_i\ge v_{\gamma_t}), (\check{t}^2_i-v_{\gamma_q})\mathbb{I}(\check{t}^2_i\ge v_{\gamma_q}) \}\nonumber\\
        &+\frac{2}{p}\sum_{1\le i<j\le p}\cov_{H_0}\{(\check{t}^2_i-v_{\gamma_t})\mathbb{I}(\check{t}^2_i\ge v_{\gamma_t}), (\check{t}^2_j-v_{\gamma_q})\mathbb{I}(\check{t}^2_j\ge v_{\gamma_q}) \}=O(1),
\end{align*}
and $\bm\Xi=(\Xi_{tq})_{s\times s}$ satisfies for $1\le t,q\le s$, $\Xi_{tq}=\sigma_{\gamma_t\gamma_q}/\sqrt{\sigma_{\gamma_t\gamma_t}\sigma_{\gamma_q\gamma_q}}$.
\end{thm}
}
Note that $\mu_\gamma$ and $\sigma_{\gamma\gamma}$ are unknown. We therefore calculate the $p$-value based on the wild bootstrap test statistics $T_{\lceil\gamma p\rceil,1}^*,\ldots,T_{\lceil\gamma p\rceil,B}^*$ as
\begin{align}\label{p-value of T_gammap}
p_{T_{\lceil \gamma p\rceil}}=\frac{1}{B}\sum_{i=1}^B I(T_{\lceil \gamma p\rceil,i}^*<T_{\lceil \gamma p\rceil}).
\end{align}
We reject the null hypothesis at the significant level $\alpha$ if $p_{T_{\lceil\gamma p\rceil}}\le \alpha$. According to Theorem \ref{asymptotic normality}, the $T_{\lceil\gamma p\rceil}$-based test is powerful against dense alternatives. 

{  
Next, we also give a simple power analysis for different $\gamma$.
According to Theorem \ref{asymptotic normality}, the power function of $T_{\lceil \gamma p\rceil}$ is
\begin{align*}
\beta_{\lceil \gamma p\rceil}(\delta)=\Phi\left(-z_{\alpha}+\frac{n\sum_{i=1}^{\lceil\gamma p\rceil}\check{\mu}^2_{(p+1-i)}}{\sqrt{p\sigma_{\gamma\gamma}}}\right),
\end{align*}
where $z_\alpha$ is the upper $\alpha$-quantile of $N(0,1)$.
So the optimal $\gamma$ should be
\begin{align*}
\gamma^*=\underset{{\gamma \in (0,1)} }{{\arg\max}}\frac{n\sum_{i=1}^{\lceil\gamma p\rceil}\check{\mu}^2_{(p+1-i)}}{\sqrt{p\sigma_{\gamma\gamma}}}.
\end{align*}
For simplicity, we only consider $\bms$ is diagonal and $\check{\mu}_{(p+1-i)}^2=\delta$ for $i=1,\cdots, \lceil \gamma_0 p\rceil$. In this case,
\begin{align*}
&\frac{n\sum_{i=1}^{\lceil\gamma p\rceil}\check{\mu}^2_{(p+1-i)}}{\sqrt{p\sigma_{\gamma\gamma}}}=\frac{n\delta \min\{\lceil \gamma_0 p\rceil, \lceil \gamma p \rceil\}}{\sqrt{p\sigma_{\gamma\gamma}}}
\end{align*}
and we could proof that $\gamma^*=\gamma_0$. 
Define  $h(\gamma,\gamma_0)=\frac{\min\{\lceil \gamma_0 \rceil, \lceil \gamma  \rceil\}}{\sqrt{\sigma_{\gamma\gamma}}}$. Figure~\ref{hr} illustrates the curves of \( h(\gamma, \gamma_0) \) for various values of \( \gamma_0 \), specifically \( \gamma_0 = 0.1, 0.3, 0.5, 0.7, \) and \( 0.9 \). It is observed that each curve of \( h(\gamma, \gamma_0) \) reaches its maximum at the point \( \gamma = \gamma_0 \). This finding aligns with the theoretical predictions.

\begin{figure}[htb]
	\centering
	\includegraphics[width=\linewidth]{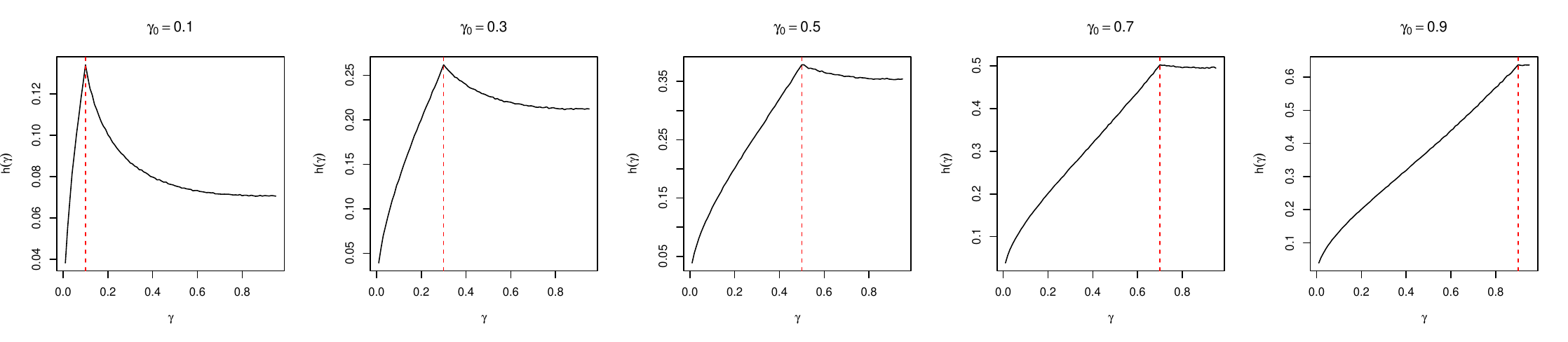}
 \caption{Curves of \( h(\gamma, \gamma_0) \) under various \( \gamma_0 \). The red dashed line marks the true \( \gamma_0 \).
}\label{hr}
\end{figure}

If the sparsity level of the alternative, denoted by $\gamma_0$, were known in advance, the optimal choice of the $L$-statistic would be $T_{\lceil \gamma_0 p \rceil}$. This selection naturally aligns with the statistician's intuition: to maximize detection power, one should aggregate test statistics corresponding to the proportion of true signals. By concentrating on the most informative subset of components—neither too few to miss relevant signals, nor too many to dilute the signal with noise—the statistic achieves the best trade-off between signals and variance. Thus, knowing $\gamma_0$ essentially allows for a targeted construction of the test that is tailored to the true alternative sparsity level.

In practice, however, $\gamma_0$ is typically unknown, which motivates the development of adaptive procedures that can automatically adjust to an unknown range of sparsity levels while retaining near-optimal performance.

}

\subsection{Adaptive test}

In practice, we seldom know whether the regime is sparse or dense. In order to adapt to various alternative behaviors, we propose an adaptive test combining different $L$-statistics in \eqref{L-statistic} with different $k$. The key message is that $T_k$ with fixed $k$ and $T_{\lceil \gamma p\rceil}$ for $\gamma\in(0,1)$ are asymptotically independent if $H_0$ holds.
\begin{thm}\label{asymptotic independence}
	Suppose Assumptions (A1)-(A3) hold. If {  $p=o(n^2)$}, then as $\min(n,p)\to\infty$, we have
	
	\noindent(\romannumeral1) for all integer $1\le s\le p$ and { fixed} $\gamma\in(0,1)$,
	\begin{align*}
	\Pr_{H_0}\left(t_{(p+1-s)}^2-b_p\le x, \frac{T_{\lceil \gamma p\rceil}-\mu_{\gamma}}{\sqrt{p\sigma_{\gamma\gamma}}}\le y \right)\to \Lambda(x)\sum_{i=0}^{s-1}\frac{\{\log \Lambda^{-1}(x)\}^i}{i!}\Phi(y),
	\end{align*}
	
	\noindent(\romannumeral2) for all integer $2\le k\le p$ and { fixed}  $\gamma\in(0,1)$,
	\begin{align*}
	&\Pr_{H_0}\left(\bigcap_{j=1}^k\left(t_{(p+1-j)}^2-b_p \leq x_j\right),\frac{T_{\lceil \gamma p\rceil}-\mu_{\gamma}}{\sqrt{p\sigma_{\gamma\gamma}}}\le y\right)\\
	&\to \Lambda\left(x_k\right) \sum_{\sum_{i=2}^j k_i \leq j-1, j=2, \ldots, k} \prod_{i=2}^k \frac{\{\log \Lambda^{-1}(x_i)-\log \Lambda^{-1}(x_{i-1})\}^{k_i}}{k_i!}\Phi(y).
	\end{align*}
\end{thm}

According to Theorem \ref{asymptotic independence}, {  if $p=o(n^2)$}, then $T_k$ is asymptotically independent with $(T_{\lceil\gamma_1 p\rceil},\ldots,T_{\lceil\gamma_s p\rceil})$. In our simulations, we observed that when $\lceil \gamma p\rceil$ is greater than or equal to 20, the distribution of $T_{\lceil \gamma p\rceil}$ tends towards normality. Therefore, in practice, we select $\lceil 2^{-i}p\rceil$ where $i$ ranges from 1 to $K$ with $K=\lfloor \log (p/20)/\log 2\rfloor$. Here, $\lfloor x\rfloor$ denotes the integer part of $x$. We proposed the following Cauchy combination test
\begin{align}\label{Cauchy combine test}
T_C&=\frac{1}{K+1}\tan\left\{\left(\frac{1}{2}-p_{T_5}\right)\pi\right\}+\frac{1}{K+1}\sum_{i=1}^K\tan\left\{\left(\frac{1}{2}-p_{T_{\lceil 2^{-i}p\rceil}}\right)\pi\right\}.
\end{align}
Based on the asymptotic independence between $T_5$ and $T_{\lceil2^{-i} p\rceil}$'s and the asymptotic normality of $T_{\lceil2^{-i} p\rceil}$'s, the corresponding $p$-value is $p_C=1-G(T_C)$ with $G(\cdot)$ being the C.D.F. of the the standard Cauchy distribution. If $p_C\le\alpha$, then we reject $H_0$.

{  The asymptotic independence results of Theorem \ref{asymptotic independence} serve two important purposes. First, to the best of our knowledge, the theoretical validity of the Cauchy combination test procedure under arbitrary dependence structures requires the baseline test statistics to follow a joint normal distribution, as discussed in \cite{liu2020cauchy} and \cite{li2023}. Currently, there are no theoretical results establishing the validity of the Cauchy combination test when combining statistics with mixed limiting distributions—such as a combination of Gaussian and Gumbel-type statistics. Therefore, for conservative and theoretically justified application, we restrict the use of the Cauchy combination procedure to the case where all components (e.g., $T_k$ with diverging $k$) satisfy the required joint normality condition. However, our established asymptotic independence between $T_k$ with fixed and diverging $k$ allows us to safely combine them using the Cauchy combination approach. Furthermore, alternative combination strategies, such as Fisher's method \citep{littell1971asymptotic}, could also be employed to construct new hybrid test procedures in this context, potentially accommodating a wider variety of limiting distributions. { Note that some alternative combination strategies (e.g., Fisher-type combinations) typically require additional calibration—such as a permutation or bootstrap procedure—to obtain valid critical values when the component tests are dependent. In contrast, the Cauchy combination approach remains analytically tractable under dependence and therefore does not require an extra resampling step for critical-value determination in our setting.
} }

Next, we analyze the power of the adaptive testing procedure.
\begin{thm}\label{asymptotic independence under alternative hypothesis}
	Suppose Assumptions (A1)-(A3) hold. If { $\gamma\log(p)n\sum_{i=1}^{\lceil\gamma p\rceil}\check{\mu}^2_{(p+1-i)}=o(1)$ and $p=o(n^2)$}. Then as $\min(n,p)\to\infty$, we have
	
	\noindent(\romannumeral1) for all integer $1\le s\le p$ and { fixed}  $\gamma\in(0,1)$,
	\begin{align*}
	&\Pr\left(t_{(p+1-s)}^2-b_p\le x, \frac{T_{\lceil \gamma p\rceil}-\mu_{\gamma}-n\sum_{i=1}^{\lceil\gamma p\rceil}\check{\mu}^2_{(p+1-i)}}{\sqrt{p\sigma_{\gamma\gamma}}}\le y \right)\\
	&\to \Pr\left(t_{(p+1-s)}^2-b_p\le x\right)\Pr\left(\frac{T_{\lceil \gamma p\rceil}-\mu_{\gamma}-n\sum_{i=1}^{\lceil\gamma p\rceil}\check{\mu}^2_{(p+1-i)}}{\sqrt{p\sigma_{\gamma\gamma}}}\le y \right),
	\end{align*}
	
	\noindent(\romannumeral2) for all integer $2\le k\le p$ and { fixed}  $\gamma\in(0,1)$,
	\begin{align*}
	&\Pr\left(\bigcap_{j=1}^k\left(t_{(p+1-j)}^2-b_p \leq x_j\right),\frac{T_{\lceil \gamma p\rceil}-\mu_{\gamma}-n\sum_{i=1}^{\lceil\gamma p\rceil}\check{\mu}^2_{(p+1-i)}}{\sqrt{p\sigma_{\gamma\gamma}}}\le y\right)\\
	&\to\Pr\left(\bigcap_{j=1}^k\left(t_{(p+1-j)}^2-b_p \leq x_j\right)\right)\Pr\left(\frac{T_{\lceil \gamma p\rceil}-\mu_{\gamma}-n\sum_{i=1}^{\lceil\gamma p\rceil}\check{\mu}^2_{(p+1-i)}}{\sqrt{p\sigma_{\gamma\gamma}}}\le y\right).
	\end{align*}
\end{thm}

According to Theorem \ref{asymptotic independence under alternative hypothesis}, under the following alternative hypothesis
\begin{align}\label{alternative hypothesis}
H_1: \gamma\log(p)n\sum_{i=1}^{\lceil\gamma p\rceil}\check{\mu}^2_{(p+1-i)}=o(1),
\end{align}
the asymptotic independence between the $T_k$-based test and $T_{\lceil\gamma p\rceil}$-based test still hold. \cite{li2023} showed that the power of Cauchy combination-based test would be powerful than that of the test based on $\min\{p_{T_k},p_{T_{\lceil\gamma p\rceil}}\}$ (referred to as the minimal p-value combination), say $\beta_{T_k\wedge T_{\lceil\gamma p\rceil}, \alpha}=P(\min\{p_{T_k},p_{T_{\lceil\gamma p\rceil}}\}\le 1-\sqrt{1-\alpha})$. Obviously,
\begin{align}\label{power_H1}
\beta_{T_k\wedge T_{\lceil\gamma p\rceil},\alpha}\ge&P(\min\{p_{T_k},p_{T_{\lceil\gamma p\rceil}}\}\le\alpha/2)\nonumber\\
=&\beta_{T_k,\alpha/2}+\beta_{T_{\lceil\gamma p\rceil},\alpha/2}-P(p_{T_k}\le\alpha/2,p_{T_{\lceil\gamma p\rceil}}\le\alpha/2)\nonumber\nonumber\\
\ge&\max\{\beta_{T_k,\alpha/2},\beta_{T_{\lceil\gamma p\rceil},\alpha/2}\}.
\end{align}
On the other hand, under $H_{1}$ in \eqref{alternative hypothesis}, we have
\begin{align}\label{power_H1np}
\beta_{T_k\wedge T_{\lceil\gamma p\rceil},\alpha}\ge \beta_{T_k,\alpha/2}+\beta_{T_{\lceil\gamma p\rceil},\alpha/2}-\beta_{T_k,\alpha/2}\beta_{T_{\lceil\gamma p\rceil},\alpha/2}+o(1),
\end{align}
due to the asymptotic independence entailed by Theorem \ref{asymptotic independence under alternative hypothesis}. For a small $\alpha$, the difference between $\beta_{T_k,\alpha}$ and $\beta_{T_k,\alpha/2}$ should be small, and the same fact applies to $\beta_{T{\lceil\gamma p\rceil},\alpha}$. Consequently, by \eqref{power_H1}-\eqref{power_H1np}, the power of the adaptive test would be no smaller than or even significantly larger than that of either $T_k$-based test or $T_{\lceil\gamma p\rceil}$-based test.

{  Finally, the various conditions and conclusions corresponding to different regimes of $p$ and $n$ in Theorems \ref{asymptotic distribution of ordered t_i}–\ref{asymptotic independence under alternative hypothesis} are consolidated in Table \ref{tabp-n} for ease of comparison and clarity. Overall, when $k$ is fixed, under light–tailed distributional assumptions, the dimension $p$ can grow at an exponential rate in $n$. In contrast, for heavy–tailed distributions, $p$ can grow only up to a polynomial order in $n$, which is consistent with related findings in the literature. When $k$ diverges with $n$, the accumulation of bias introduced by the variance estimation limits $p$ to at most the quadratic order of $n$. This growth constraint on $p$ in turn restricts the order for which the subsequent asymptotic independence results hold. These distinctions highlight the trade–off between the tail behavior of the underlying distribution, the growth of $k$, and the admissible rates of $p$, providing useful guidance for applying the proposed methods in practice.
}
\begin{table}[htbp]
\centering
\small
\renewcommand{\arraystretch}{0.9}
\caption{Conditions and conclusions on $p$ and $n$ in all theorems}
\begin{tabular}{@{}>{\raggedright\arraybackslash}p{0.25\textwidth} >{\raggedright\arraybackslash}p{0.40\textwidth} >{\raggedright\arraybackslash}p{0.27\textwidth}@{}} \hline\hline
Theorems & Conditions & Conclusions\\ \hline
Theorem \ref{asymptotic distribution of ordered t_i} and Theorem \ref{power of T_k}
& sub-Gaussian-type, $\log p=o(n^{1/4})$; sub-exponential-type tail, $\log p=o(n^{1/4})$; sub-polynominal-type tail, $p\le cn^{\gamma_0}$
& Joint distribution of $t_{(j)}^2$'s \\
Theorem \ref{asymptotic normality} & $p=o(n^2)$ & CLT for $T_{\lceil\gamma p\rceil}$\\
Theorem \ref{asymptotic independence} & $p=o(n^2)$ & Asymptotic independence under $H_0$ \\
Theorem \ref{asymptotic independence under alternative hypothesis} & $p=o(n^2)$ & Asymptotic independence under $H_1$ \\ \hline\hline
\end{tabular}
\label{tabp-n}
\end{table}

\section{Simulation}\label{Sec: Simulation}

We generate the sample as $\X_i=\bmu+\bms^{1/2}\bmv_i$ where the covariance matrix $\bms=(0.5^{|i-j|})_{1\le i,j\le p}$ and $\bmv_i=(\varepsilon_{i1},\ldots,\varepsilon_{ip})^\top$. We consider three distributions for $\varepsilon_{ij}$: (\romannumeral1) normal distribution $N(0,1)$; (\romannumeral2) standardized $t$-distribution $t(3)/\sqrt{3}$; (\romannumeral3) standardized mixture normal distribution $\{0.9N(0,1)+0.1N(0,9)\}/\sqrt{1.8}$. Two sample sizes and three dimensions are considered, i.e. $n=100,200; p=100,200,400$. Here we only show the sizes of $T_5,T_{\lceil0.25p\rceil},T_{\lceil0.5p\rceil}$ for $n=100$, $T_5,T_{\lceil0.125p\rceil},T_{\lceil0.25p\rceil},T_{\lceil0.5p\rceil}$ for $n=200$ and Cauchy combination test $T_C$. Table \ref{tab1} reports the empirical sizes of the five tests under different scenarios. We found that all the proposed tests can control the empirical sizes very well in all cases. 
\begin{table}[htbp]
	\centering
  \renewcommand{\arraystretch}{0.5}
	\caption{Sizes of L-statistics and Cauchy Combination test}
	\begin{tabular}{@{}l lll lll lll@{}} \hline\hline
		&\multicolumn{3}{c}{Normal} &\multicolumn{3}{c}{$t(3)$}&\multicolumn{3}{c}{Mixture Normal}\\ \cmidrule(r){2-4} \cmidrule(r){5-7} \cmidrule(r){8-10}
		$p$&100&200&400 &100&200&400 &100&200&400 \\ \hline
		& \multicolumn{9}{c}{$n=100$}\\ \hline
		$T_5$&0.049	&0.060	&0.064	&0.060	&0.056	&0.039	&0.047	&0.042	&0.045\\
		$T_{\lceil 0.25p\rceil}$&0.061	&0.052	&0.061	&0.068	&0.062	&0.051	&0.070	&0.049	&0.051\\
		$T_{\lceil 0.5p\rceil}$&0.057	&0.054	&0.054	&0.058	&0.065	&0.054	&0.074	&0.051	&0.050\\
		$T_C$&0.059	&0.051	&0.062	&0.064	&0.055	&0.049	&0.059	&0.050	&0.057\\ \hline
		& \multicolumn{9}{c}{$n=200$}\\ \hline
		$T_5$&0.052	&0.058	&0.044	&0.060	&0.048	&0.049	&0.052	&0.041	&0.042\\
		$T_{\lceil 0.125p\rceil}$&0.062&0.062	&0.051	&0.053	&0.058	&0.051	&0.059	&0.052	&0.059\\
		$T_{\lceil 0.25p\rceil}$&0.058	&0.066	&0.049	&0.057	&0.062	&0.056	&0.065	&0.058	&0.061\\
		$T_{\lceil 0.5p\rceil}$&0.061	&0.070	&0.045	&0.057	&0.065	&0.056	&0.060	&0.053	&0.058\\
		$T_C$&0.058	&0.064	&0.061	&0.055	&0.058	&0.060	&0.059	&0.059	&0.061\\ \hline\hline
	\end{tabular}
	\label{tab1}
\end{table}

\begin{figure}[htbp]
	\centering
	\subfigure[$n=100,p=100$]{
		\includegraphics[width=0.9\textwidth]{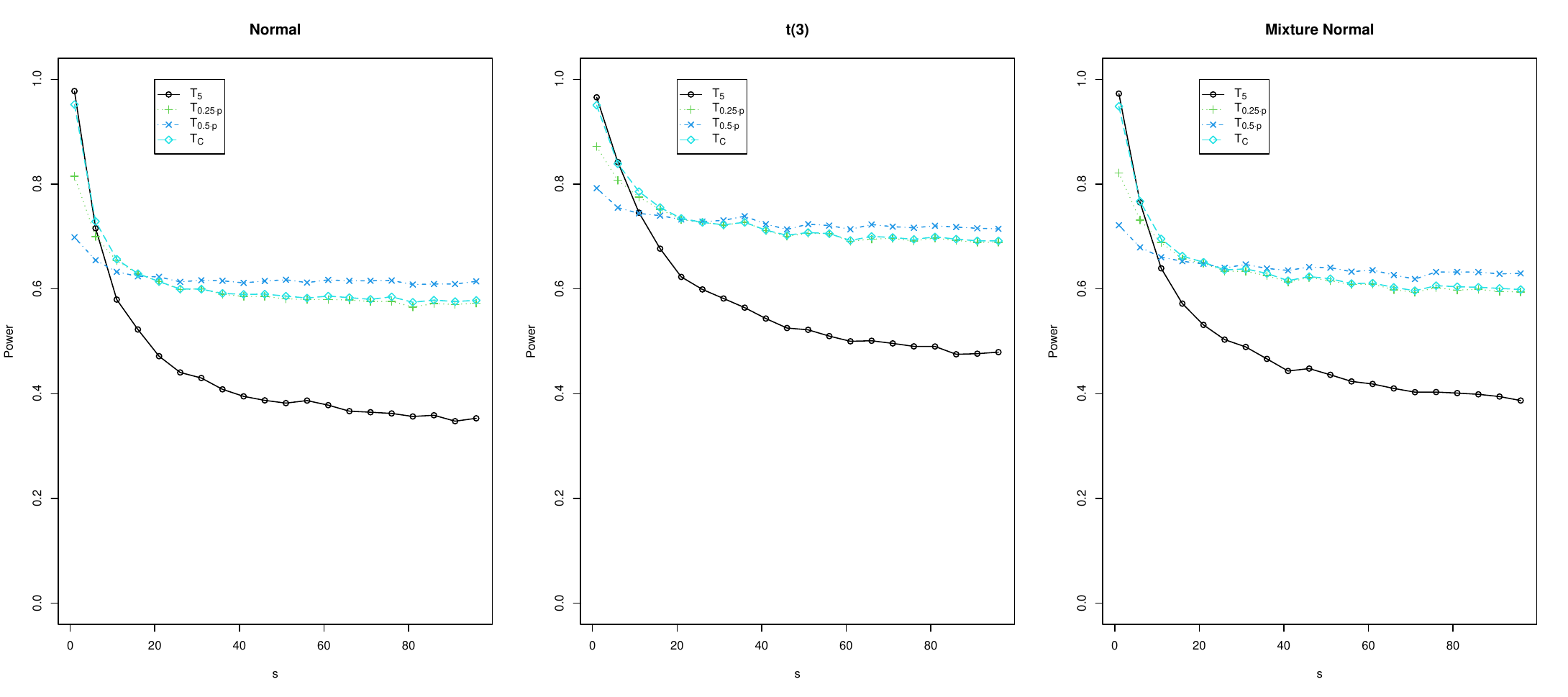}
		\label{fig1a}
	}\\
	\subfigure[$n=100,p=200$]{
		\includegraphics[width=0.9\textwidth]{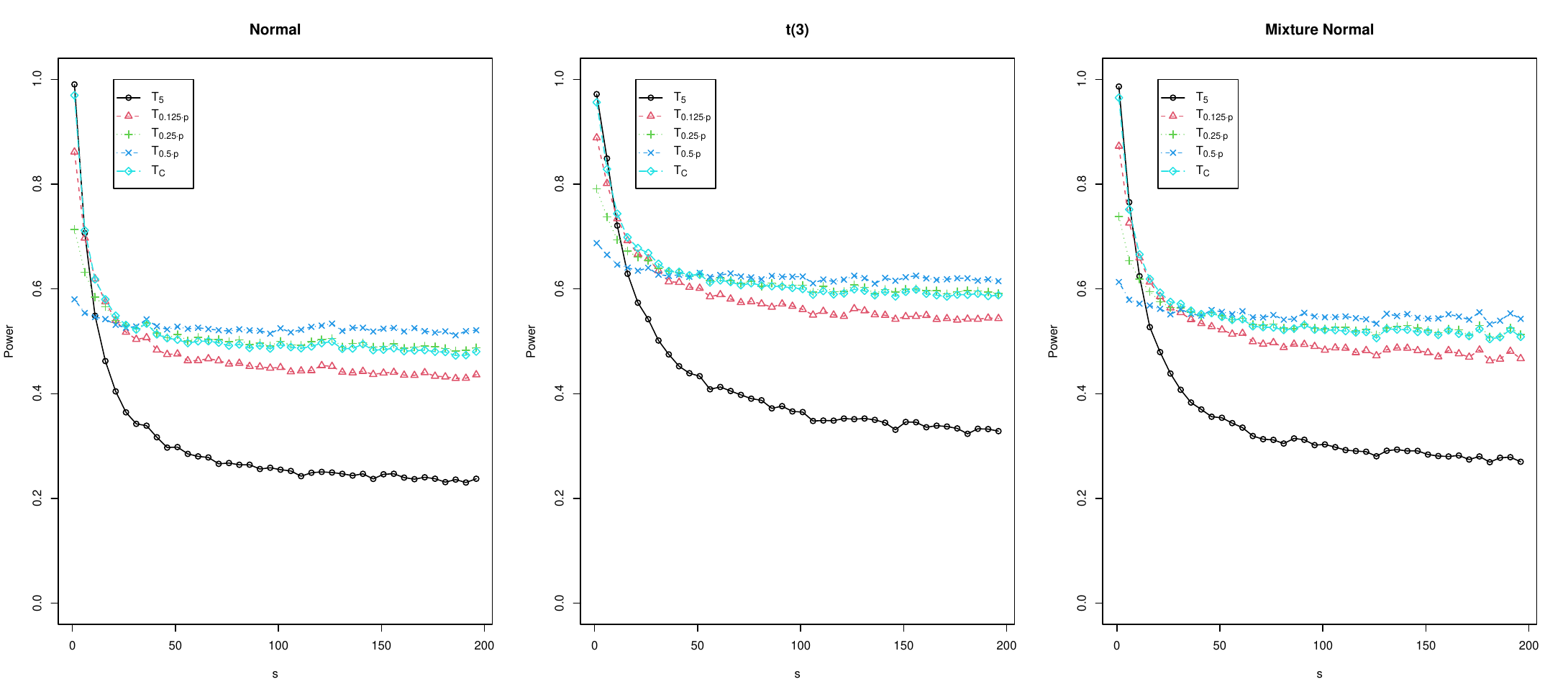}
		\label{fig1b}
	}
	\caption{Power of individual L-tests { and adaptive L-test} with different numbers of nonzero means.}
	\label{fig1}
\end{figure}

For fair comparison, we consider size-corrected power comparison of each test. First, we consider the power comparison of the individual $L$-statistics tests with Cauchy combination test with different sparsity. For power comparison, we consider $\bmu=\kappa_s(1,\ldots,1,0,\ldots,0)^\top$, where the first $s$ components are equal to $\kappa_s$ and the others are all equal to zero. Here we set $\kappa_s=3\sqrt{\log p/(ns)}$. Figure \ref{fig1} reports the power of the five tests with different number of nonzero means. We found that $T_5$ outperforms $T_{\lceil 0.25p\rceil}$ and $T_{\lceil 0.5p\rceil}$ under sparse alternatives, i.e. $s<10$. While for the dense alternatives, $T_{\lceil 0.25p\rceil}$ and $T_{\lceil 0.5p\rceil}$ perform better than $T_5$. This is consistent with the intuition. The adaptive test $T_C$ performs very robust under different sparsity.

\begin{figure}[H]
	\centering
		\includegraphics[width=0.9\textwidth]{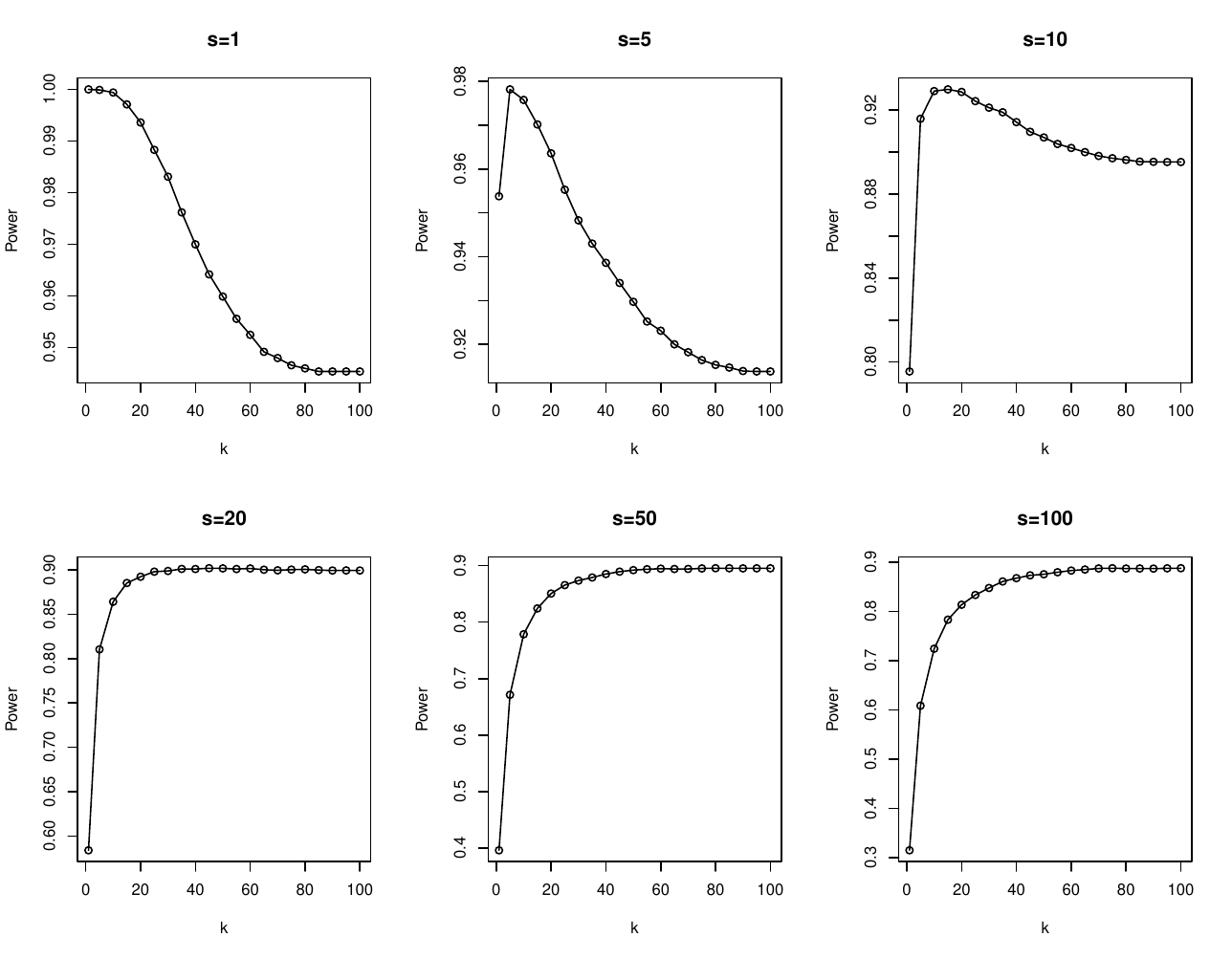}
	\caption{Power of individual L-tests with different parameter $k$.}
	\label{figk}
\end{figure}

{  In addition, we conduct a more detailed investigation by fixing the sparsity level $s$ and examining the performance of individual $T_k$ tests with varying choices of the parameter $k$, as shown in Figure~\ref{figk}. Specifically, we consider the case where $\kappa_s = 4 \sqrt{\log p / (ns)}$ with $p = 100$, while all other simulation settings remain the same as described earlier. This setting allows us to evaluate the sensitivity of the $T_k$ test to the choice of $k$ under a fixed signal strength.

Our findings reveal several important patterns. When the true number of nonzero means is small (e.g., $s < 10$), the $T_k$ test achieves its best performance precisely when $k$ is correctly specified or very close to the true sparsity level, i.e., $k = s$. In such sparse scenarios, choosing $k$ significantly smaller than $s$ may still yield decent power, but selecting $k$ larger than $s$ leads to noticeable power loss. Conversely, when the number of nonzero means becomes large (i.e., a dense alternative), the $T_k$ test continues to perform well when $k$ is moderately large, and its performance stabilizes across a wide range of large $k$ values.

These observations highlight a key practical challenge: while the $T_k$ test can be powerful when $k$ is chosen appropriately, its performance is sensitive to mis-specification of $k$, particularly when $k$ is set too small. This further justifies the need for an adaptive approach.

Our proposed Cauchy combination test addresses this issue by aggregating $T_k$ statistics over a carefully selected set of $k$ values, such as $T_5$ and $T_{\lceil 2^{-i}p \rceil}$ for several choices of $i$. This strategy effectively covers a wide range of sparsity levels and ensures strong power performance without requiring prior knowledge of the true sparsity. Empirically, this adaptive method consistently delivers near-optimal performance across both sparse and dense regimes, making it a robust and practical solution for high-dimensional mean testing problems.

}

\begin{figure}[H]
	\centering
	\subfigure[$n=100,p=100$]{
		\includegraphics[width=0.9\textwidth]{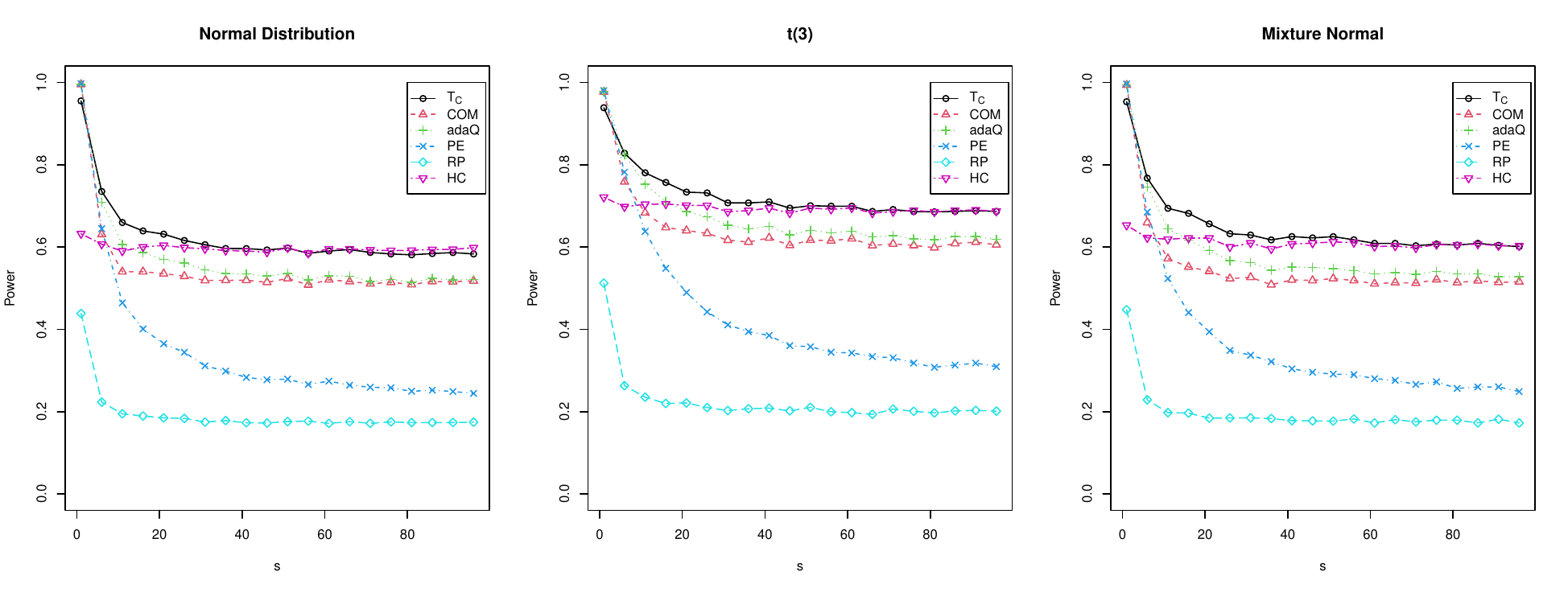}
		\label{fig4a}
	}\\[-0.5em]
	\subfigure[$n=100,p=200$]{
		\includegraphics[width=0.9\textwidth]{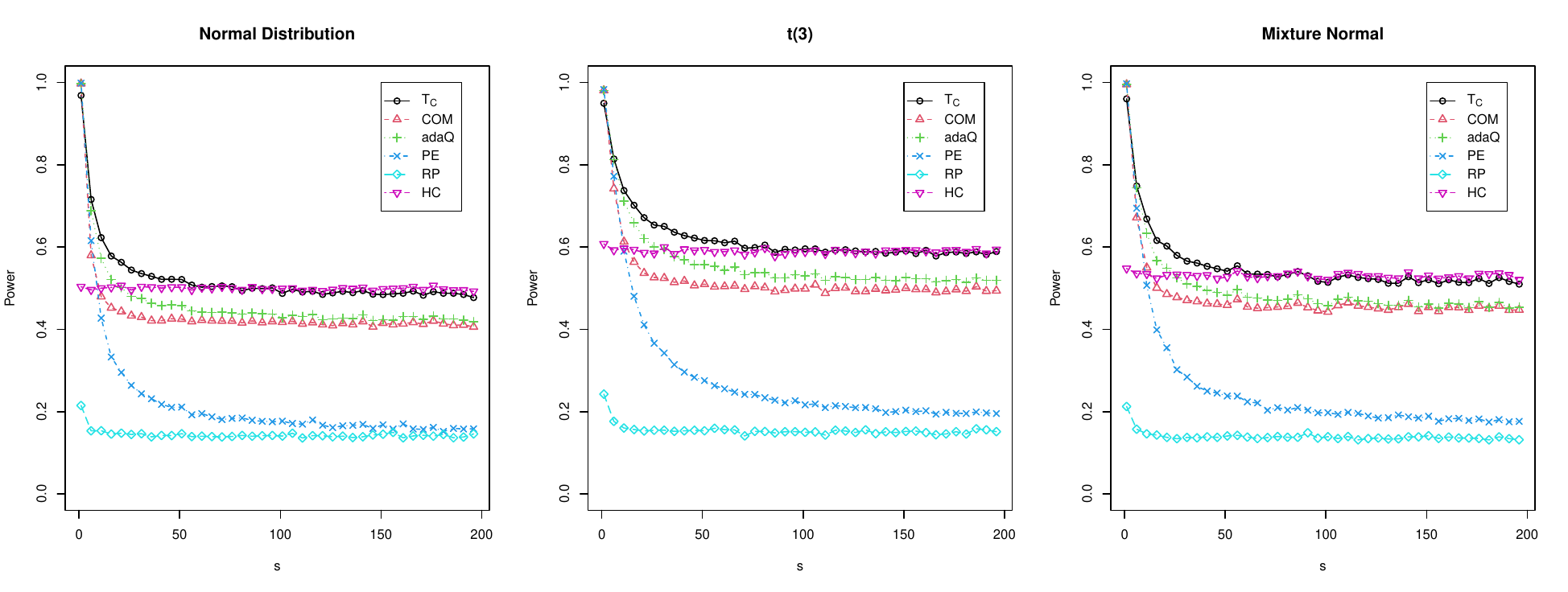}
		\label{fig4b}
	}
	\caption{Power of tests with different numbers of nonzero means.}
	\label{fig4}
\end{figure}

Next, we compare our adaptive test $T_C$ with the following tests:
\begin{itemize}
\item The COM test proposed by \cite{feng2024asymptotic} which only combine \cite{srivastava2009test}'s sum-type test $T_{p}$ and max-type test statics $T_1$;
\item The the adaptive test proposed by \cite{XLWP16} (abbreviated as adaQ) which consider four $L_q$-norms of the sample mean, i.e. $q=2,4,6,\infty$; { 
\item The power enhancement test proposed by \cite{fan2015power} (abbreviated as PE);
\item The random projection test proposed by \cite{lopes2011more} (abbreviated as RP);
\item The higher criticism test proposed by \cite{zhong2013tests} (abbreviated as HC) with parameter $\gamma=2$.}
\end{itemize}
For a fair comparison, we also employ the Cauchy Combination test procedure to construct an adaptive test, rather than relying on the minimum $p$-values. In this context, we are specifically focusing on the scenarios where $(n,p)$ is either $(100,100)$ or $(100,200)$. It's worth noting that in our additional simulation studies, we found that the performance of the three tests remains consistent across different sample sizes and dimensions. Figure \ref{fig4} reports the power curves of those tests with different sparsity.   

{  We find that our proposed test statistic $T_C$ consistently achieves the best overall performance across different scenarios. The COM test, which only combines $L_2$- and $L_\infty$-based statistics, lacks the flexibility to fully adapt to various sparsity structures, and thus does not attain optimal power. In contrast, our $L$-statistic offers a more nuanced approach and outperforms the $L_q$-norm-based procedures by leveraging a wider range of sparsity adaptivity.

The PE test performs comparably to $T_C$ when the alternative is extremely sparse, but its power diminishes significantly under dense alternatives. This is not surprising, as the PE test is primarily designed to detect alternatives where ${\bf \Sigma}^{-1/2} \bm{\mu}$ is dense, rather than $\bm{\mu}$ itself. Similar observations have been reported in recent studies, such as \cite{chen2024asymptotic} and \cite{yan2025high}, where the structure of the signal relative to the covariance matrix plays a critical role in the test's efficacy.

Additionally, while the HC (Higher Criticism) test performs well under dense alternatives and shows similar power to $T_C$ in those settings, our test is clearly preferred in the sparse regime due to its adaptivity and robustness. On the other hand, random projection (RP)-based methods fail to deliver satisfactory performance when the correlation among variables is weak, highlighting their sensitivity to the dependence structure.

For completeness, we further investigate the performance of all competing methods under strong correlation settings in the supplementary material. The results reaffirm the robustness of our proposed test $T_C$, which maintains its superiority across a broad spectrum of correlation structures and sparsity patterns.

}

{

We conduct a real-data analysis on weekly returns of S\&P 500 constituents from February 2009 to September 2018 to illustrate the empirical utility of the proposed procedure. The scientific question is whether the cross-sectional mean vector of excess returns is zero, i.e., whether there exists any security with a nonzero expected return beyond the risk-free rate. Using a Ljung--Box screening step to mitigate serial dependence, our analysis provides clear evidence against $H_0$, and all competing tests reject the null across the considered subsample sizes. Moreover, the rejection-rate comparison suggests that the underlying alternative is neither extremely sparse nor extremely dense: an intermediate $L$-statistic such as $T_{\lceil p/8\rceil}$ performs best among $T_k$ test, while our adaptive procedure $T_C$ achieves the strongest (or nearly strongest) overall performance and is competitive with other adaptive methods (e.g., COM and adaQ). We further identify a small set of securities with significant excess returns under BH-FDR control. Due to space limitations, we present only a brief summary here; the complete dataset description, implementation details, and all supporting figures/tables are deferred to the Supplementary Material.

}

\section{Conclusion}\label{Sec: Conclusion}

In this paper, we introduce a new adaptive testing procedure for the high-dimensional one-sample location test problem, which is based on $L$-statistics. We determine the limited null distribution of $L$-statistics for various parameters and demonstrate the asymptotic independence between $L$-statistics with fixed and diverging parameters. Both simulation studies and real-world data applications indicate that our proposed adaptive method outperforms existing methods. {  Additional details regarding the conditions and future research directions are provided in the supplementary material.}

\medskip
\noindent\textbf{Code availability.}
The R code and reproducibility materials for the proposed methods, simulations,
and empirical analysis are publicly available at
\url{https://github.com/flnankai/ALtest}.

\clearpage
\setcounter{section}{0}
\setcounter{subsection}{0}
\setcounter{equation}{0}
\setcounter{figure}{0}
\setcounter{table}{0}
\setcounter{lemma}{0}
\setcounter{thm}{5}
\renewcommand{\thesection}{S\arabic{section}}
\renewcommand{\thesubsection}{S\arabic{section}.\arabic{subsection}}
\renewcommand{\theequation}{S\arabic{equation}}
\renewcommand{\thefigure}{S\arabic{figure}}
\renewcommand{\thetable}{S\arabic{table}}
\renewcommand{\theHsection}{supp.\arabic{section}}
\renewcommand{\theHsubsection}{supp.\arabic{section}.\arabic{subsection}}
\renewcommand{\theHequation}{supp.\arabic{equation}}
\renewcommand{\theHfigure}{supp.\arabic{figure}}
\renewcommand{\theHtable}{supp.\arabic{table}}
\providecommand{\theHsubfigure}{}
\renewcommand{\theHsubfigure}{supp.\arabic{figure}.\arabic{subfigure}}

\begin{center}
{\Large\bfseries Supplementary Material for\\[0.25em]
``Adaptive L-statistics for high dimensional test problem''}
\end{center}
\medskip

Section~S1 presents additional simulation results.  
Section~S2 provides a real-data application.  
Section~S3 offers further discussion of the main paper.  
Section~S4 contains useful lemmas, and Section~S5 presents the detailed proofs of Theorems~1--5.  
Section~S6 reports the theoretical results of the bootstrap method.

\section{Additional Simulation Results}

\begin{figure}[htbp]
	\centering
	\subfigure[$n=100,p=100$]{
		\includegraphics[width=0.9\textwidth]{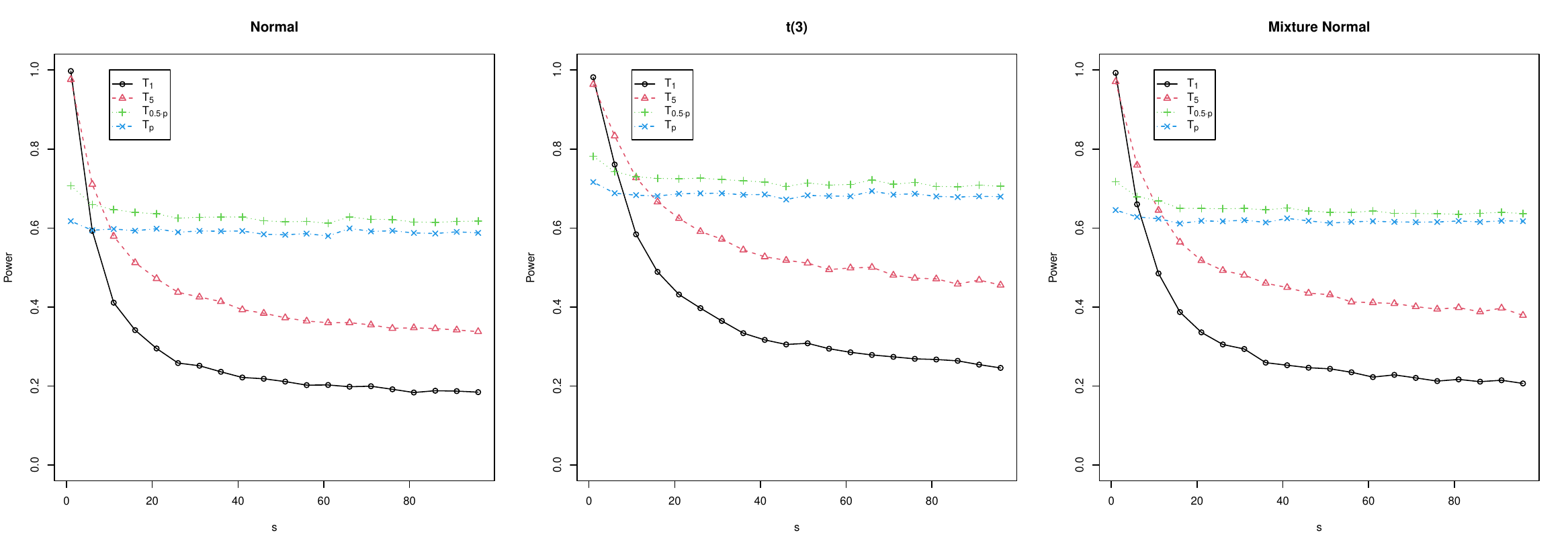}
		\label{fig3a}
	}\\
	\subfigure[$n=100,p=200$]{
		\includegraphics[width=0.9\textwidth]{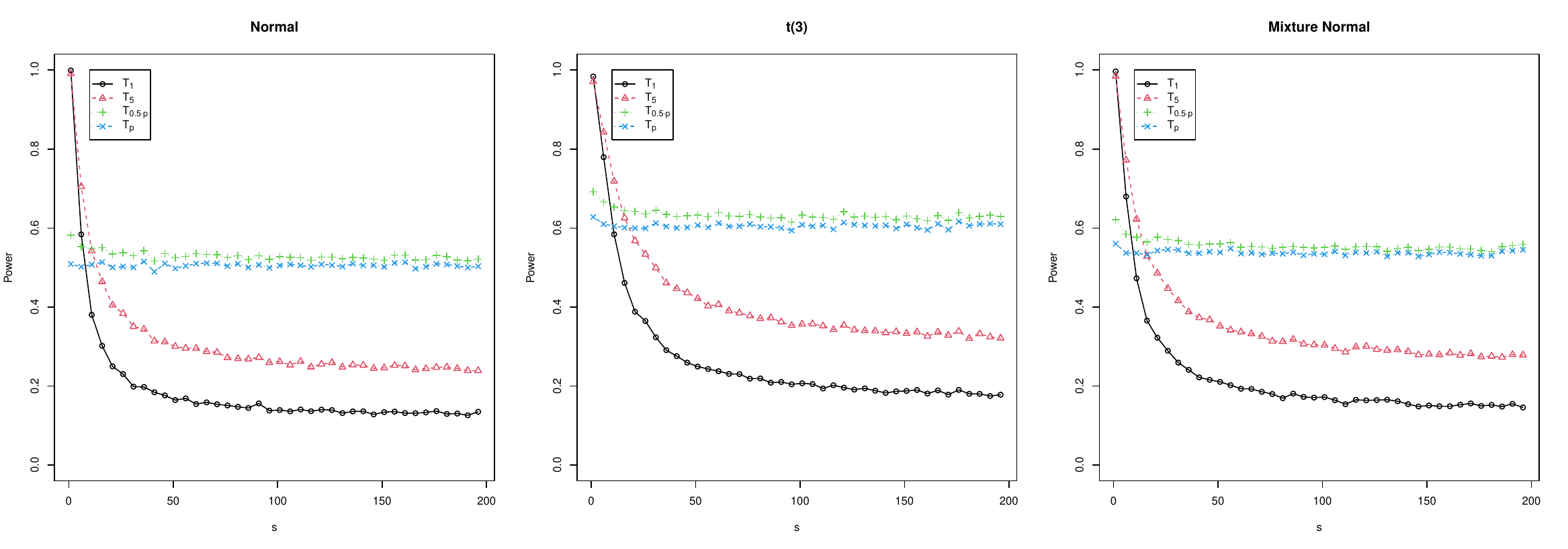}
		\label{fig3b}
	}
	\caption{Power of $T_1,T_5,T_p,T_{\lceil 0.5p\rceil}$ tests with different numbers of nonzero means.}
	\label{fig3}
\end{figure}

In our study, we juxtapose our proposed $L$-statistics with two prevalent methods, namely MAX ($T_1$) and SUM ($T_p$). The settings are all the same as Section 3 in the main text. The power performance of the $T_1$, $T_5$, $T_{p}$ and $T_{\lceil 0.5p\rceil}$ tests, corrected for size, is depicted in Figure \ref{fig3}. Our findings indicate that the $T_5$ tests consistently outperform the MAX test, except in instances where only one variable has a nonzero mean. In all instances of sparsity, the $T_{\lceil 0.5p\rceil}$ test is uniformly more potent than the SUM test. Therefore, it is not surprising that our proposed Cauchy combination test $T_C$ performs better than the COM test. This also underscores the superiority of our proposed $L$-statistic over the existing MAX and SUM tests.

\begin{figure}[H]
	\centering
		\includegraphics[width=0.9\textwidth]{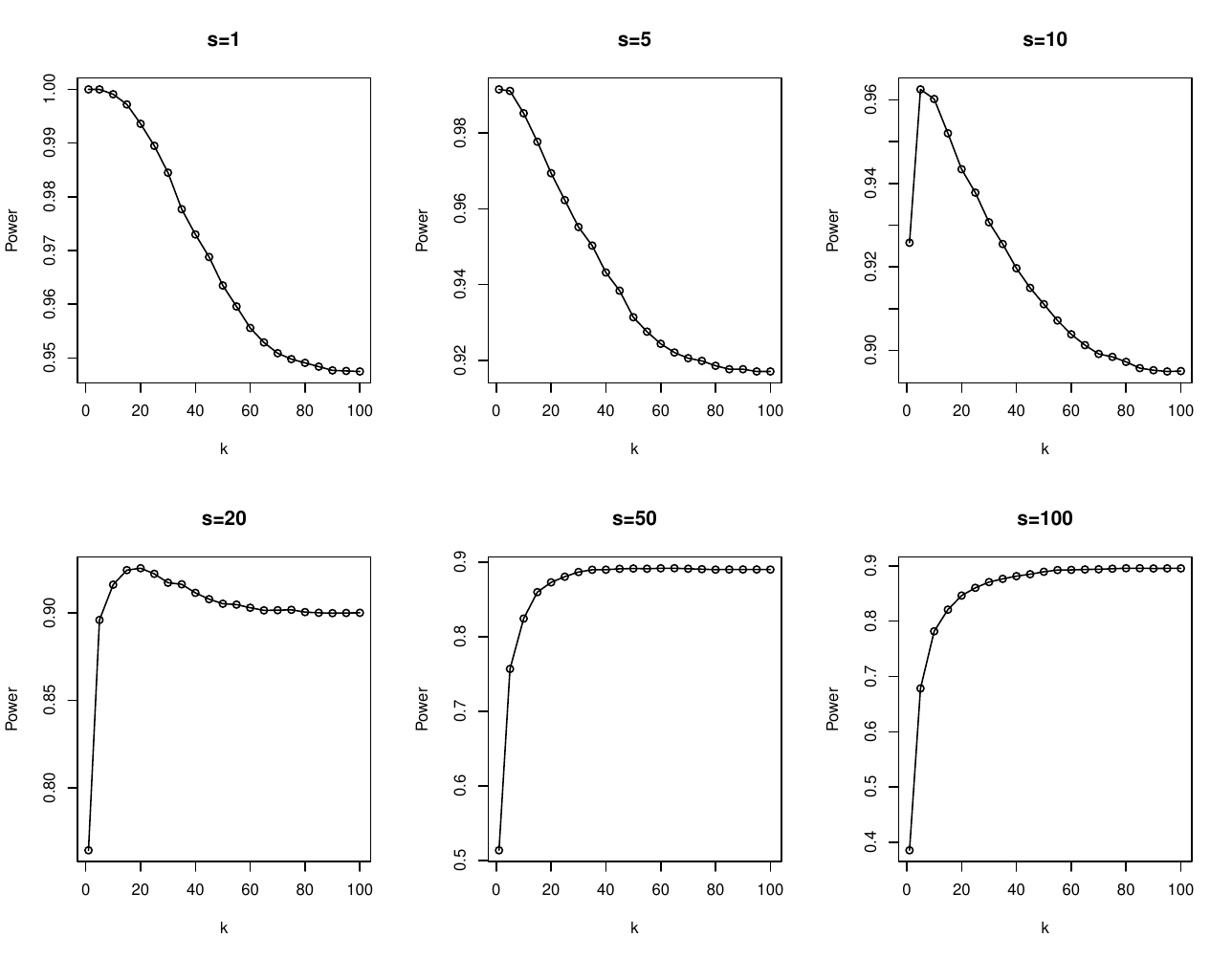}
	\caption{Power of individual L-tests with different parameter $k$  under ${\bf \Sigma}=(0.5^{|i-j|})_{1\le i,j\le p}$ and $\mu=\kappa_s(1,\ldots,s,0,\ldots,0)^\top$.}
	\label{figks}
\end{figure}

\begin{figure}[H]		
	\centering
	\subfigure[$n=100,p=100$]{
		\includegraphics[width=0.9\textwidth]{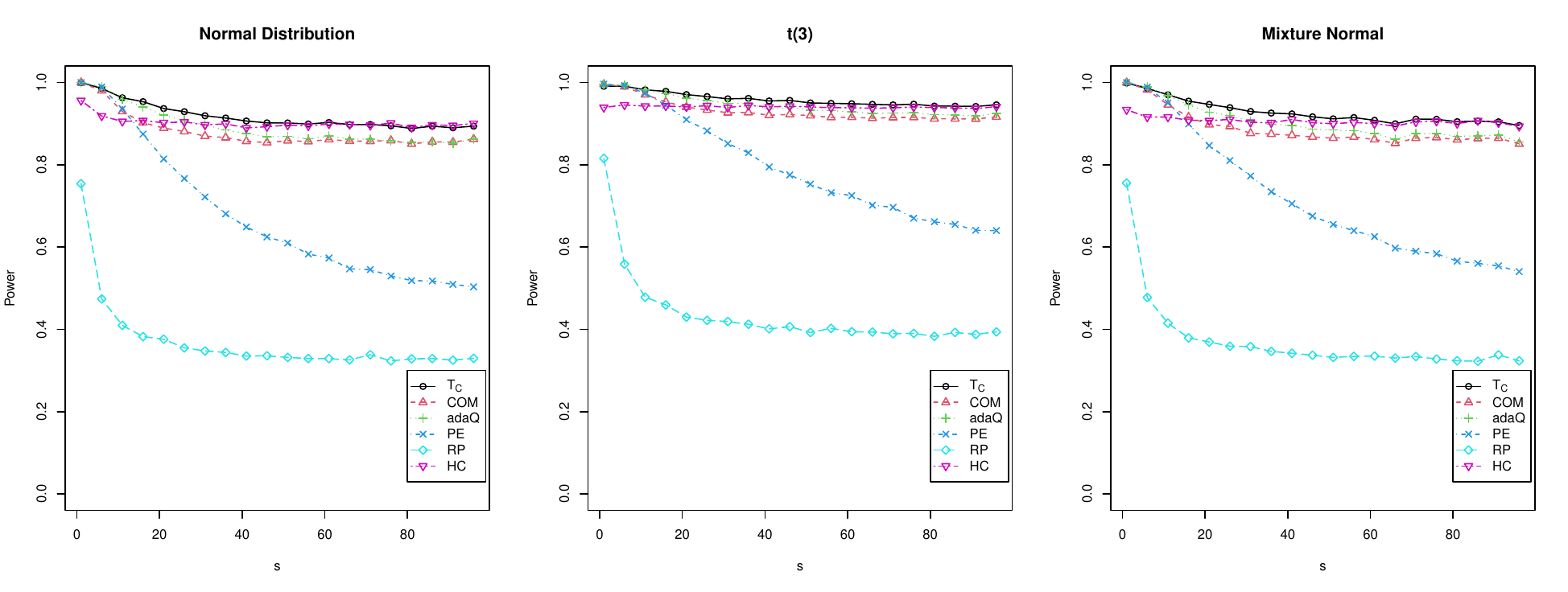}
	}\\
	\subfigure[$n=100,p=200$]{
		\includegraphics[width=0.9\textwidth]{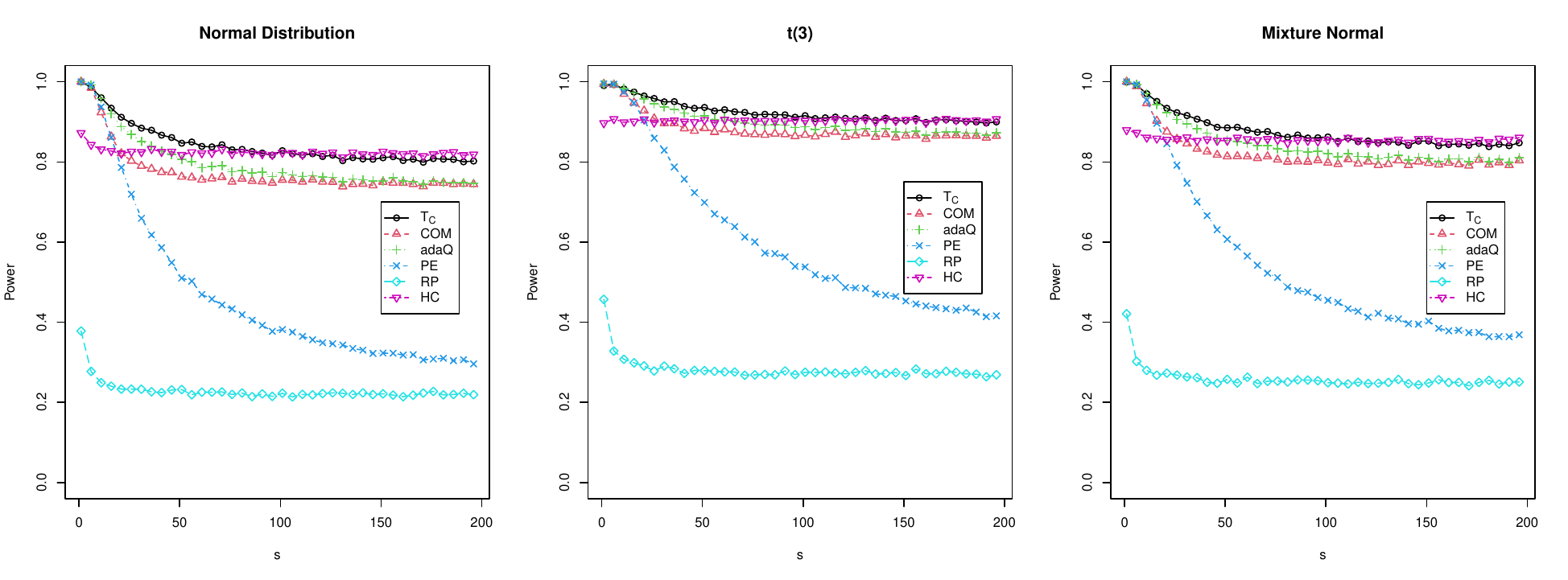}
	}
	\caption{Power of tests with different numbers of nonzero means under ${\bf \Sigma}=(0.5^{|i-j|})_{1\le i,j\le p}$ and $\mu=\kappa_s(1,\ldots,s,0,\ldots,0)^\top$.}\label{figs0}
\end{figure}

{  
Next, we examine an alternative setting where the nonzero mean vector is specified as $\mu = \kappa_s(1,\ldots,s,0,\ldots,0)^\top$, with $\kappa_s = 4\sqrt{6\log p/(ns(s+1)(2s+1))}$ and $p = 100$. All other configurations are consistent with those in Figure 4 of the main text. Figure \ref{figks} presents the power curves of the individual $L$-tests with varying values of the parameter $k$. The results closely resemble those in Figure 4 of the main text, confirming that the optimal choice of $k$ aligns with the true number of nonzero mean components. Furthermore, when the number of nonzero means is large, setting $k = p/2$ yields comparable performance to the optimal choice. For completeness, we also compare the power of all competing tests in Figure \ref{figs0} under this alternative setting. The conclusions are consistent with those in the main text: our proposed test statistic $T_C$ continues to demonstrate the best overall performance across different sparsity levels.
} 

{ 

Next, we explore various covariance structures to further demonstrate the effectiveness of our proposed method. Specifically, we consider three additional covariance configurations as follows:
\begin{itemize}
\item[(I)] Strong correlation: ${\bf \Sigma} = (\sigma_{ij})_{1 \le i, j \le p}$, $\sigma_{ii} = 1$ and $\sigma_{ij} = 0.5$ for $i \ne j$.
\item[(II)] $\bms=\D^{1/2}\R\D^{1/2}$ with
$\D=\diag(\sigma_1^2,\cdots,\sigma_p^2)$ and $\R=\I_p+\bm b\bm b
^{T}-\check{\B}$,
where $\sigma_i^2$ are generated independently from $Uniform(1,2)$,
$\bm b=(b_1,\cdots,b_p)^{T}$ and
$\check{\B}=\diag(b_1^2,\cdots,b_p^2)$.
The first $[p^{0.5}]$ entries of $\b$ are
independently sampled from $Uniform(0.7,0.9)$, and the remaining entries  are set to be zero, where $[\cdot]$ denotes taking integer part.

\item[(III)]
$\bms=\bm \gamma \bm
\gamma^{T}+(\I_p-\rho_{\epsilon}\W)^{-1}(\I_p-\rho_{\epsilon}\W^{T})^{-1},$ where
$\bm \gamma=(\gamma_1,\cdots,\gamma_{[p^{\delta_{\gamma}}]},0,0,\cdots,0)^{T}$.
Here  $\gamma_i$ with $i=1,\cdots, [p^{\delta_{\gamma}}]$ are generated independently from $Uniform(0.7,0.9)$. Let
$\rho_{\epsilon}=0.5$ and $\delta_{\gamma}=0.5$. Let $\W=(w_{i_1i_2})_{1\le i_1,i_2\le p}$ have a so-called rook form, i.e.,
all elements of $\W$ are zero except that
$w_{i_1+1,i_1}=w_{i_2-1,i_2}=0.5$ for $i_1=1,\cdots,p-2$ and
$i_2=3,\cdots,p$, and $w_{1,2}=w_{p,p-1}=1$.
\end{itemize}
Note that Scenario I exhibits a strong correlation structure, whereas Scenarios II and III violate the $\alpha$-mixing condition specified in Condition (A3).

All other simulation settings are kept consistent with those described in Section 3. The empirical sizes of the individual $L$-tests and the proposed Cauchy combination test are reported in Table \ref{tabs1}-\ref{tabs3}. As shown in the results, both the individual $L$-tests and the Cauchy combination test maintain the nominal significance level accurately, even under the strong correlation structure. This demonstrates the robustness and effectiveness of the proposed wild bootstrap procedure in controlling type I error rates. The ability to preserve size accuracy under such dependence further supports the applicability of these methods in practical high-dimensional settings where strong correlations are common. }

\begin{table}[H]
	\centering
	\caption{Sizes of L-statistics and Cauchy Combination test under Scenario I}
	\begin{tabular}{@{}l lll lll lll@{}} \hline\hline
		&\multicolumn{3}{c}{Normal} &\multicolumn{3}{c}{$t(3)$}&\multicolumn{3}{c}{Mixture Normal}\\ \cmidrule(r){2-4} \cmidrule(r){5-7} \cmidrule(r){8-10}
		$p$&100&200&400 &100&200&400 &100&200&400 \\ \hline
		& \multicolumn{9}{c}{$n=100$}\\ \hline
$T_5$&0.04&0.051&0.041&0.047&0.046&0.052&0.05&0.036&0.052\\
$T_{[0.125p]}$&0.046&0.05&0.039&0.051&0.042&0.051&0.057&0.034&0.049\\
$T_{[0.25p]}$&0.046&0.05&0.039&0.058&0.044&0.053&0.058&0.037&0.050\\
$T_{[0.5p]}$&0.043&0.051&0.039&0.06&0.045&0.054&0.057&0.039&0.051\\
$T_{C}$&0.047&0.058&0.042&0.058&0.045&0.054&0.058&0.039&0.053\\ \hline
		& \multicolumn{9}{c}{$n=200$}\\ \hline
$T_5$&0.052&0.055&0.053&0.049&0.042&0.054&0.051&0.055&0.040\\
$T_{[0.125p]}$&0.057&0.052&0.052&0.052&0.038&0.05&0.047&0.051&0.044\\
$T_{[0.25p]}$&0.056&0.05&0.056&0.05&0.037&0.05&0.046&0.05&0.044\\
$T_{[0.5p]}$&0.057&0.048&0.058&0.049&0.037&0.052&0.046&0.053&0.044\\
$T_{C}$&0.06&0.053&0.06&0.056&0.042&0.055&0.051&0.054&0.047\\ \hline\hline
	\end{tabular}
	\label{tabs1}
\end{table}

\begin{figure}[H]		
	\centering
	\subfigure[$n=100,p=100$]{
		\includegraphics[width=0.9\textwidth]{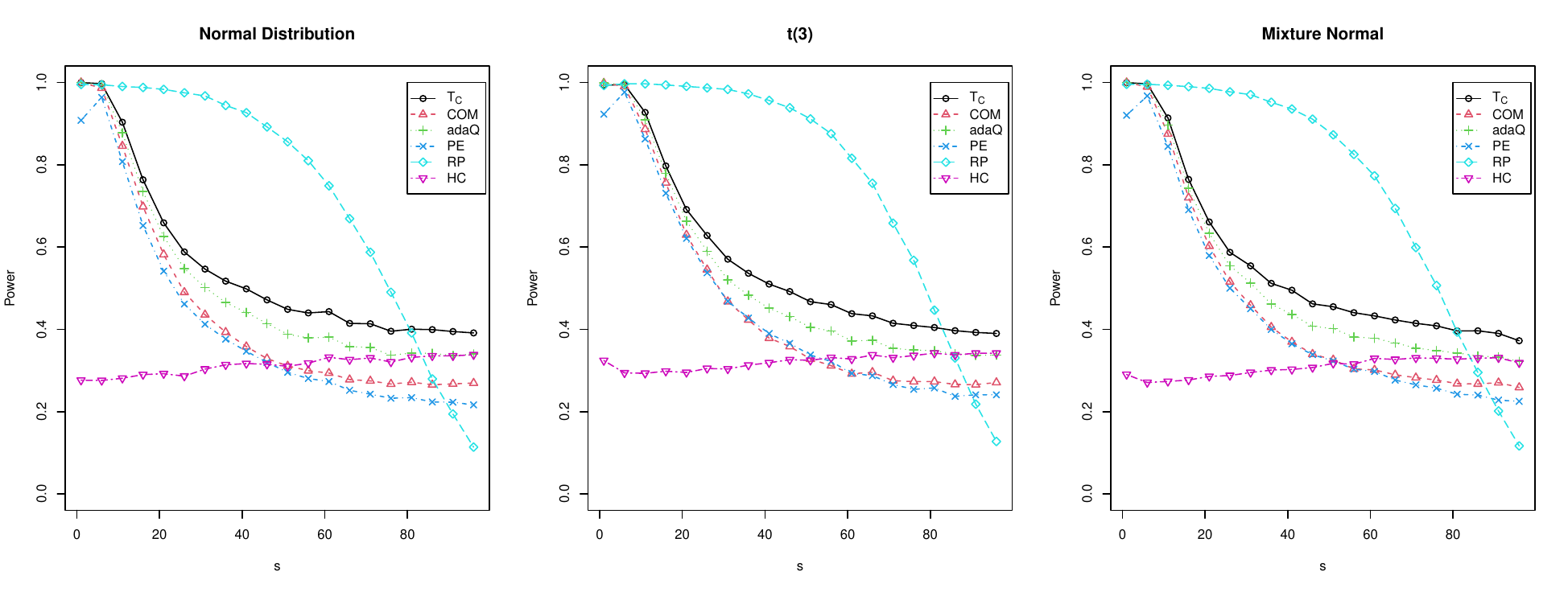}
	}\\
	\subfigure[$n=100,p=200$]{
		\includegraphics[width=0.9\textwidth]{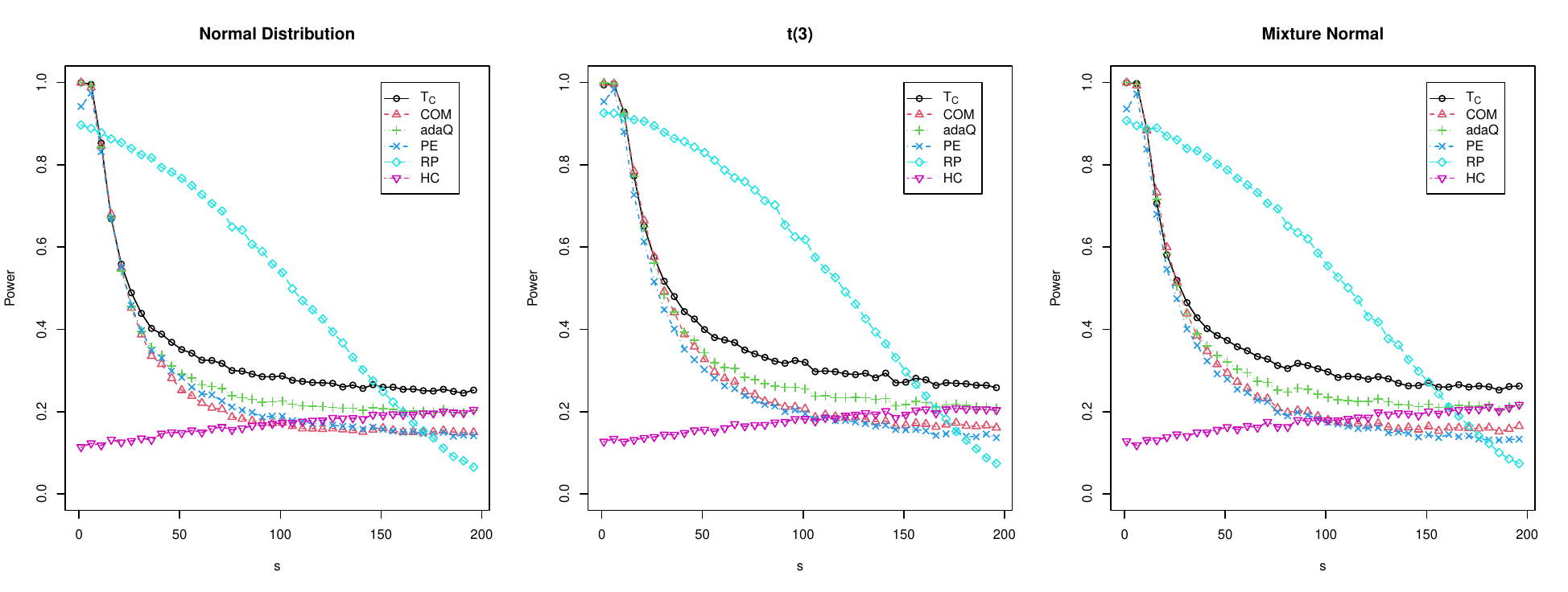}

	}
	\caption{Power of tests with different numbers of nonzero means under Scenario I.}\label{figs2}
\end{figure}

\begin{figure}[H]		
	\centering
	\subfigure[$n=100,p=100$]{
		\includegraphics[width=0.9\textwidth]{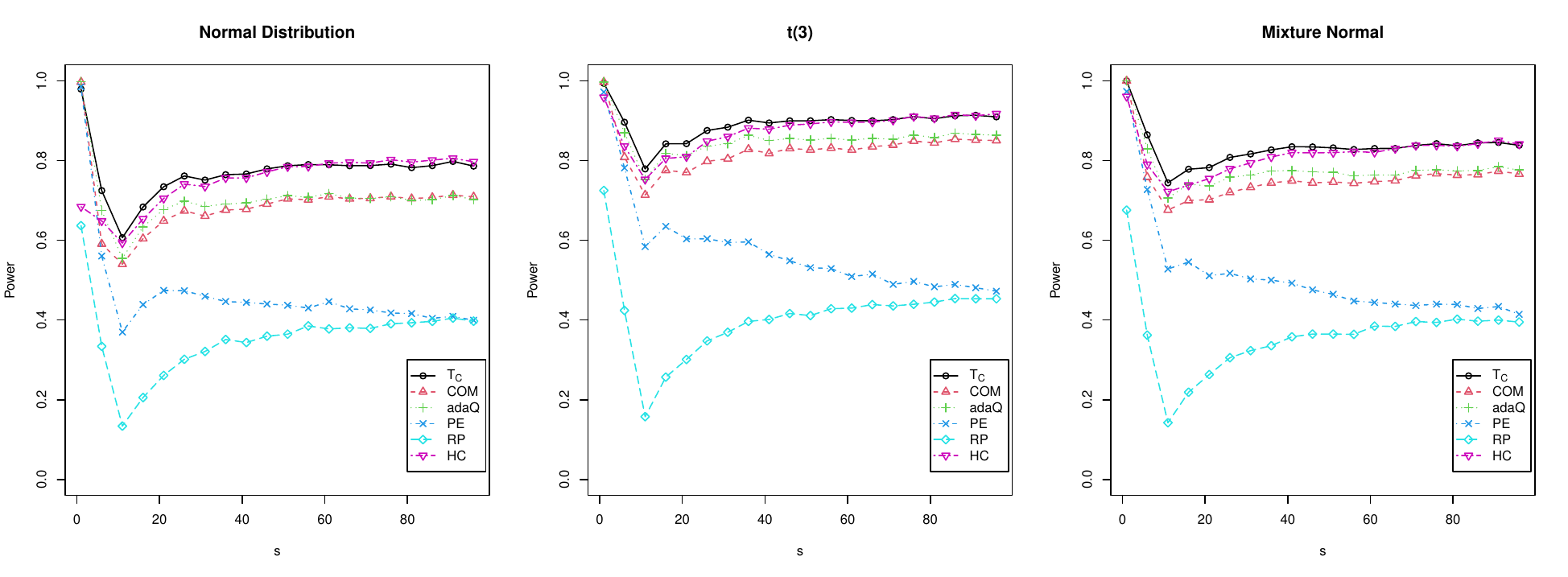}
	}\\
	\subfigure[$n=100,p=200$]{
		\includegraphics[width=0.9\textwidth]{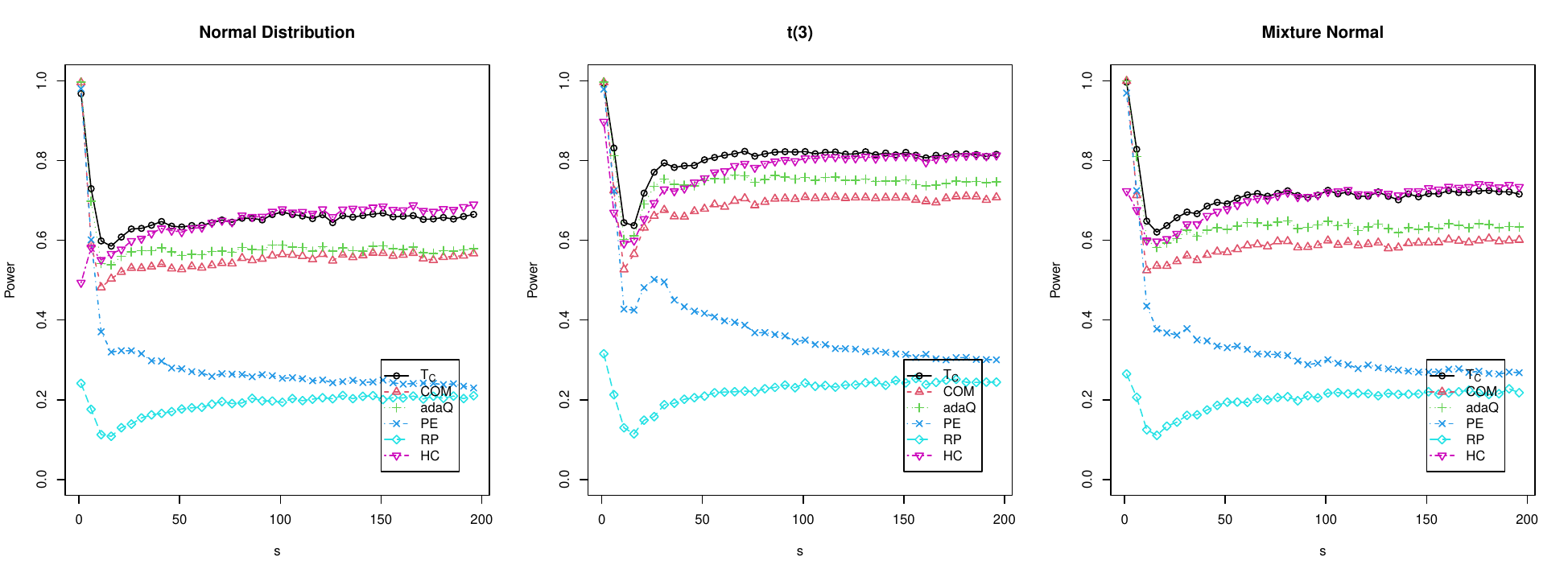}

	}
	\caption{Power of tests with different numbers of nonzero means under Scenario II.}\label{figs3}
\end{figure}

\begin{figure}[H]		
	\centering
	\subfigure[$n=100,p=100$]{
		\includegraphics[width=0.9\textwidth]{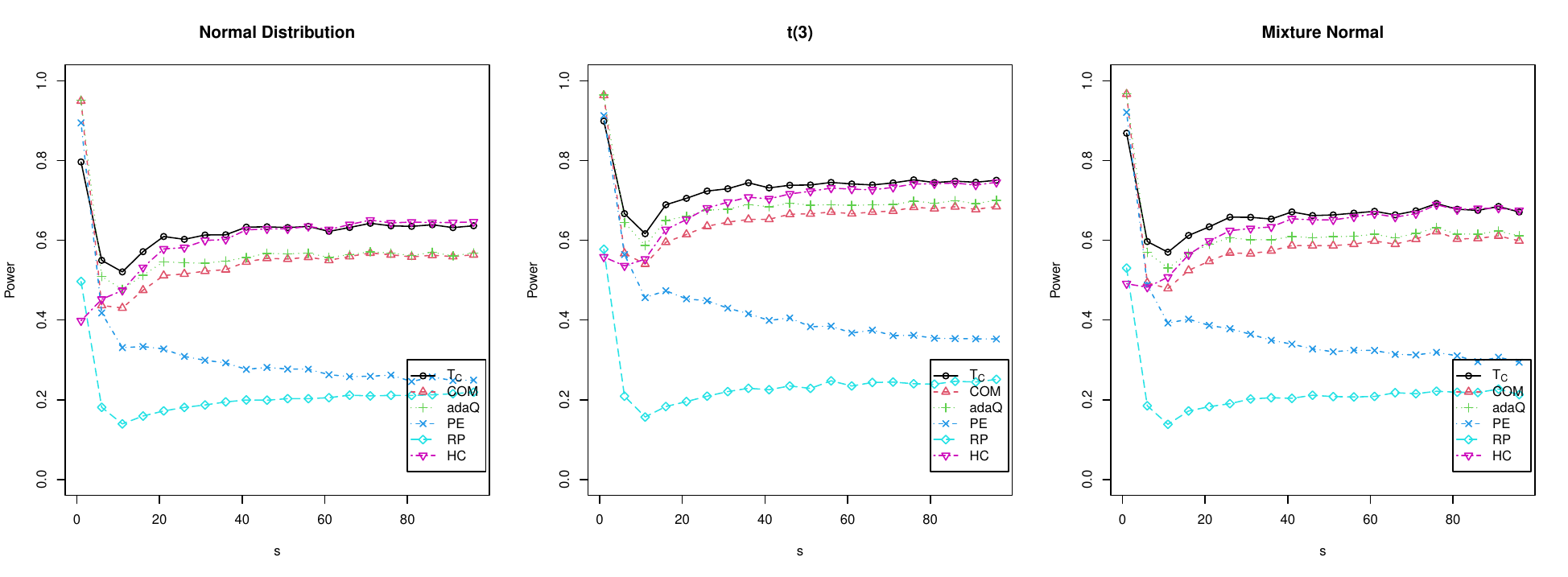}
	}\\
	\subfigure[$n=100,p=200$]{
		\includegraphics[width=0.9\textwidth]{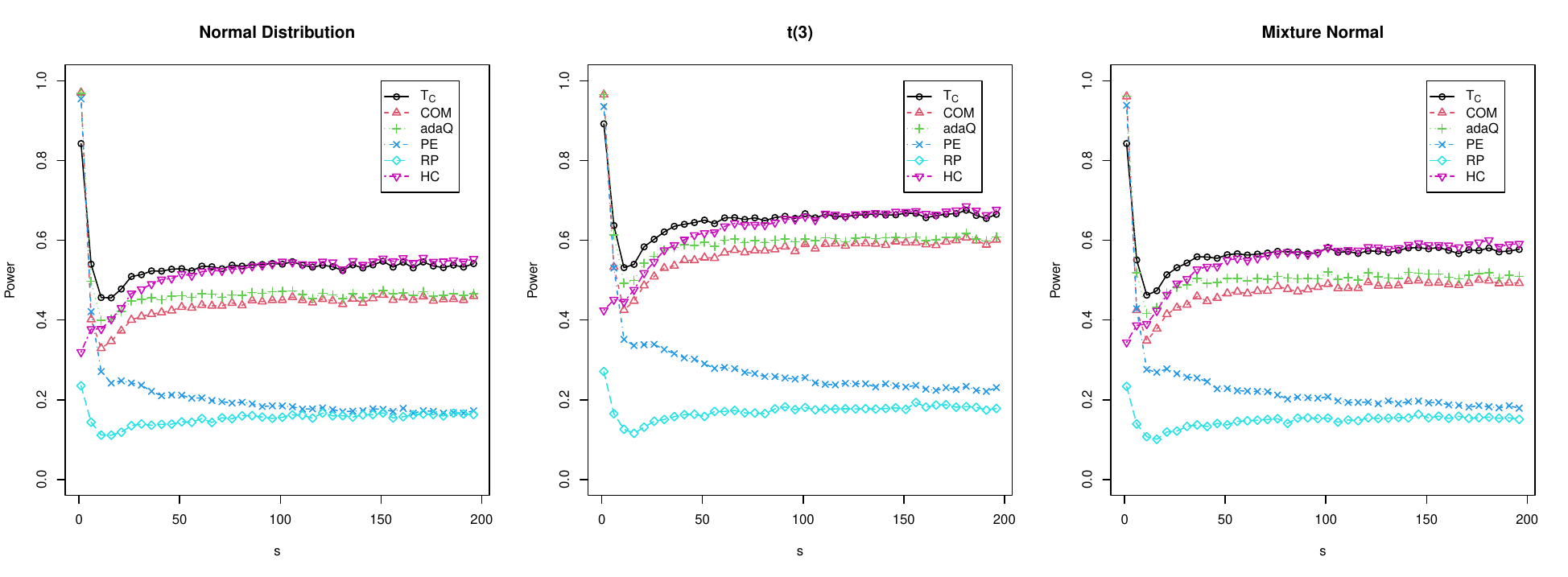}

	}
	\caption{Power of tests with different numbers of nonzero means under Scenario III.}\label{figs4}
\end{figure}

{ 
 We also conduct a comprehensive comparison between our proposed methods and several existing adaptive testing procedures. For power comparison, we set $\kappa_s = 5\sqrt{\log p / (ns)}$ for Scenario I and 
$\kappa_s = 4\sqrt{\log p / (ns)}$ for Scenario II-III.  Figure \ref{figs2} presents the power curves of each method with varying levels of signal sparsity under Scenario I. From the results, we observe that our proposed adaptive $L$-test consistently outperforms the other adaptive procedures, including COM, adaQ, HC, and PE, across a wide range of scenarios. Notably, the Random Projection (RP) method exhibits strong performance when the number of nonzero mean components is moderate—that is, neither too sparse nor too dense. This observation is aligned with previous findings, such as those in \cite{lopes2011more}, which highlight the efficiency of RP-based methods in settings with strong correlations among variables. 

Figures \ref{figs3}–\ref{figs4} present the power curves of each test under Scenario II and Scenario III, respectively. The simulation results remain consistent with those reported in Figure 5 of the main text. In particular, our proposed test statistic $T_C$ continues to demonstrate superior performance, especially under sparse covariance structures. This highlights the robustness and adaptability of $T_C$ across a variety of dependence settings, including those that deviate from standard assumptions such as $\alpha$-mixing.

The superior performance of our adaptive $L$-test, especially under such correlation structures, further illustrates its robustness and versatility in high-dimensional inference tasks. Overall, the results demonstrate that our approach provides a powerful and reliable alternative to existing methods across different sparsity levels and dependence structures. }

{

\begin{table}[H]
	\centering
	\caption{Sizes of L-statistics and Cauchy Combination test under Scenario II}
	\begin{tabular}{@{}l lll lll lll@{}} \hline\hline
		&\multicolumn{3}{c}{Normal} &\multicolumn{3}{c}{$t(3)$}&\multicolumn{3}{c}{Mixture Normal}\\ \cmidrule(r){2-4} \cmidrule(r){5-7} \cmidrule(r){8-10}
		$p$&100&200&400 &100&200&400 &100&200&400 \\ \hline
		& \multicolumn{9}{c}{$n=100$}\\ \hline
$T_5$&0.051&0.065&0.056&0.035&0.035&0.041&0.042&0.048&0.053\\
$T_{[0.125p]}$&0.05&0.056&0.061&0.038&0.046&0.042&0.039&0.038&0.057\\
$T_{[0.25p]}$&0.053&0.053&0.053&0.037&0.05&0.049&0.04&0.042&0.056\\
$T_{[0.5p]}$&0.051&0.057&0.048&0.034&0.049&0.047&0.044&0.039&0.056\\
$T_{C}$&0.06&0.067&0.07&0.044&0.049&0.05&0.043&0.045&0.064\\ \hline
		& \multicolumn{9}{c}{$n=200$}\\ \hline
$T_5$&0.063&0.052&0.047&0.051&0.042&0.044&0.052&0.043&0.057\\
$T_{[0.125p]}$&0.065&0.045&0.049&0.049&0.047&0.041&0.051&0.052&0.048\\
$T_{[0.25p]}$&0.052&0.04&0.051&0.053&0.047&0.042&0.051&0.049&0.057\\
$T_{[0.5p]}$&0.046&0.044&0.051&0.048&0.041&0.043&0.053&0.051&0.047\\
$T_{C}$ &0.064&0.049&0.061&0.056&0.052&0.048&0.059&0.054&0.064\\ \hline\hline
	\end{tabular}
	\label{tabs2}
\end{table}

\begin{table}[H]
	\centering
	\caption{Sizes of L-statistics and Cauchy Combination test under Scenario III}
	\begin{tabular}{@{}l lll lll lll@{}} \hline\hline
		&\multicolumn{3}{c}{Normal} &\multicolumn{3}{c}{$t(3)$}&\multicolumn{3}{c}{Mixture Normal}\\ \cmidrule(r){2-4} \cmidrule(r){5-7} \cmidrule(r){8-10}
		$p$&100&200&400 &100&200&400 &100&200&400 \\ \hline
		& \multicolumn{9}{c}{$n=100$}\\ \hline
$T_5$&0.048&0.054&0.049&0.057&0.052&0.043&0.039&0.045&0.045\\
$T_{[0.125p]}$&0.048&0.059&0.053&0.056&0.05&0.035&0.04&0.034&0.06\\
$T_{[0.25p]}$&0.048&0.053&0.054&0.057&0.052&0.038&0.044&0.035&0.064\\
$T_{[0.5p]}$&0.044&0.051&0.064&0.059&0.051&0.039&0.046&0.038&0.068\\
$T_{C}$&0.052&0.07&0.061&0.068&0.058&0.046&0.05&0.046&0.069\\ \hline
		& \multicolumn{9}{c}{$n=200$}\\ \hline
$T_5$&0.052&0.066&0.04&0.049&0.042&0.051&0.053&0.046&0.049\\
$T_{[0.125p]}$&0.046&0.064&0.051&0.047&0.055&0.05&0.045&0.056&0.049\\
$T_{[0.25p]}$&0.053&0.054&0.05&0.047&0.052&0.05&0.042&0.05&0.057\\
$T_{[0.5p]}$&0.051&0.06&0.054&0.041&0.05&0.061&0.032&0.049&0.063\\
$T_{C}$&0.052&0.071&0.058&0.05&0.057&0.055&0.044&0.059&0.067\\ \hline \hline
	\end{tabular}
	\label{tabs3}
\end{table}

}

{

\begin{figure}[htb]
	\centering
	\includegraphics[width=\linewidth]{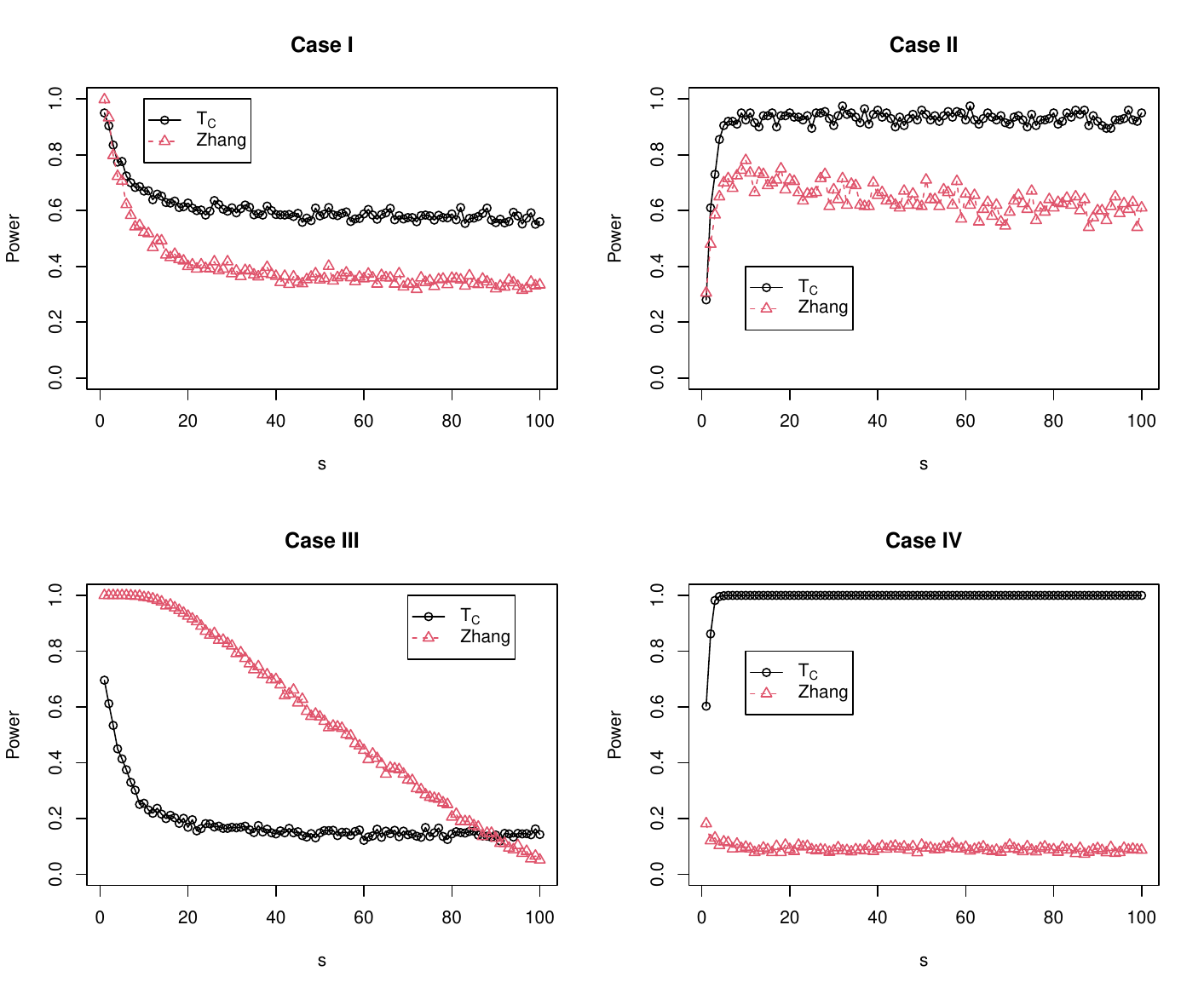}
	\caption{Power curves of $T_C$ and Zhang's test.}\label{zhang}
\end{figure}

We further compare the proposed test with the procedure of \citet{zhang2015} through a set of simulation studies. Here, we restrict our comparison to the test statistic $\tilde{T}_{fe,n}(M)$ introduced in Section~2.4 of \citet{zhang2015}, which incorporates an adaptive mechanism through the choice of the truncation parameter $k$. In our implementation, we set $M=40$, following common practice, to reflect this adaptive testing feature.
 We consider the high-dimensional setting with sample size $n=100$ and dimension $p=100$, and examine four representative configurations that differ in the covariance structure and the form of the mean vector. In Cases (I) and (II), the covariance matrix follows an autoregressive structure $\boldsymbol{\Sigma}=(0.5^{|i-j|})_{1\le i,j\le p}$, while in Cases (III) and (IV) it follows an equicorrelated structure $\boldsymbol{\Sigma}=0.5\mathbf{I}_p+0.5\boldsymbol{1}\boldsymbol{1}^\top$. For Cases (I) and (III), the mean vector is specified as $\boldsymbol{\mu}=3\sqrt{\log p/(ns)}(1,\ldots,1,0,\ldots,0)^\top$, where the first $s$ components are nonzero, corresponding to a sparse mean alternative. In contrast, Cases (II) and (IV) consider alternatives of the form $\boldsymbol{\mu}=(3/2)\sqrt{\log p/(ns)}\,\boldsymbol{\Sigma}(1,\ldots,1,0,\ldots,0)^\top$, which induce sparsity in $\boldsymbol{\Sigma}^{-1}\boldsymbol{\mu}$ rather than in $\boldsymbol{\mu}$ itself. Here we only consider multivariate normal distribution $N(\boldsymbol{\mu},\mathbf{\Sigma})$.

Figure~\ref{zhang} reports the empirical power curves of the proposed test statistic $T_C$ and Zhang’s test. We observe that the proposed test exhibits substantially higher power in Cases (I), (II), and (IV), whereas Zhang’s test outperforms $T_C$ in Case (III). This behavior is consistent with the intrinsic design principles of the two procedures. Specifically, Zhang’s test is tailored to detect sparse alternatives in $\boldsymbol{\Sigma}^{-1}\boldsymbol{\mu}$, while the proposed test targets sparsity directly in the mean vector $\boldsymbol{\mu}$. As a result, the relative performance of the two methods depends critically on the interaction between the covariance structure and the form of the mean signal. These results highlight that neither test uniformly dominates the other, and that their effectiveness is inherently case-dependent, underscoring the importance of selecting testing procedures that are well aligned with the underlying signal structure.

}

{
We also consider an alternative choice of the candidate set for the tuning parameter $k$.
Specifically, we construct a Cauchy combination test based on
\[
k\in\left\{5,10,\left\lceil 2^{-K}p\right\rceil,\ldots,\left\lceil p/2\right\rceil\right\},
\qquad
K=\Big\lfloor \log(p/20)/\log 2 \Big\rfloor,
\]
which we denote by {\it adaK2}, and compare it with our original version {\it adaK1} based on
\[
k\in\left\{5,\left\lceil 2^{-K}p\right\rceil,\ldots,\left\lceil p/2\right\rceil\right\}.
\]
We set $(n,p)\in\{(100,100),(100,200),(100,400),(200,100),(200,200),(200,400)\}$. Table~\ref{tab:2k} reports the empirical sizes
of {\it adaK2} as the same settings as Table 2 in the main paper. Overall, the new test controls the empirical size well, with rejection
frequencies fluctuating around the nominal level $0.05$ across the distributions considered.

\begin{table}[htbp]
	\centering
	\caption{Empirical sizes of the Cauchy combination test {\it adaK2}.}
	\begin{tabular}{@{}l lll lll lll@{}} \hline\hline
		&\multicolumn{3}{c}{Normal} &\multicolumn{3}{c}{$t(3)$}&\multicolumn{3}{c}{Mixture Normal}\\ \cmidrule(r){2-4} \cmidrule(r){5-7} \cmidrule(r){8-10}
		$p$&100&200&400 &100&200&400 &100&200&400 \\ \hline
		& \multicolumn{9}{c}{$n=100$}\\ \hline
		$T_C$&0.052	&0.047	&0.056	&0.058	&0.049	&0.053	&0.051	&0.046	&0.054\\ \hline
		& \multicolumn{9}{c}{$n=200$}\\ \hline
		$T_C$&0.051	&0.055	&0.048	&0.047	&0.052	&0.050	&0.053	&0.049	&0.051\\ \hline\hline
	\end{tabular}
	\label{tab:2k}
\end{table}

For power comparison, we only consider $n=100,p=100$ and keep all other
settings the same as in Figure~5 of the main paper.
Figure~\ref{fig:2k} compares the power curves of {\it adaK1} and {\it adaK2}. The two procedures
exhibit very similar power across the scenarios considered. Moreover, when the alternative becomes
denser (i.e., the number of nonzero signals increases), {\it adaK1} can be slightly more powerful
than {\it adaK2}, which is consistent with the intuition that adding an additional fixed-$k$ component
(e.g., $k=10$) brings limited benefit for dense alternatives and may mildly dilute the overall power
in the combination.

\begin{figure}[htb]
	\centering
	\includegraphics[width=\linewidth]{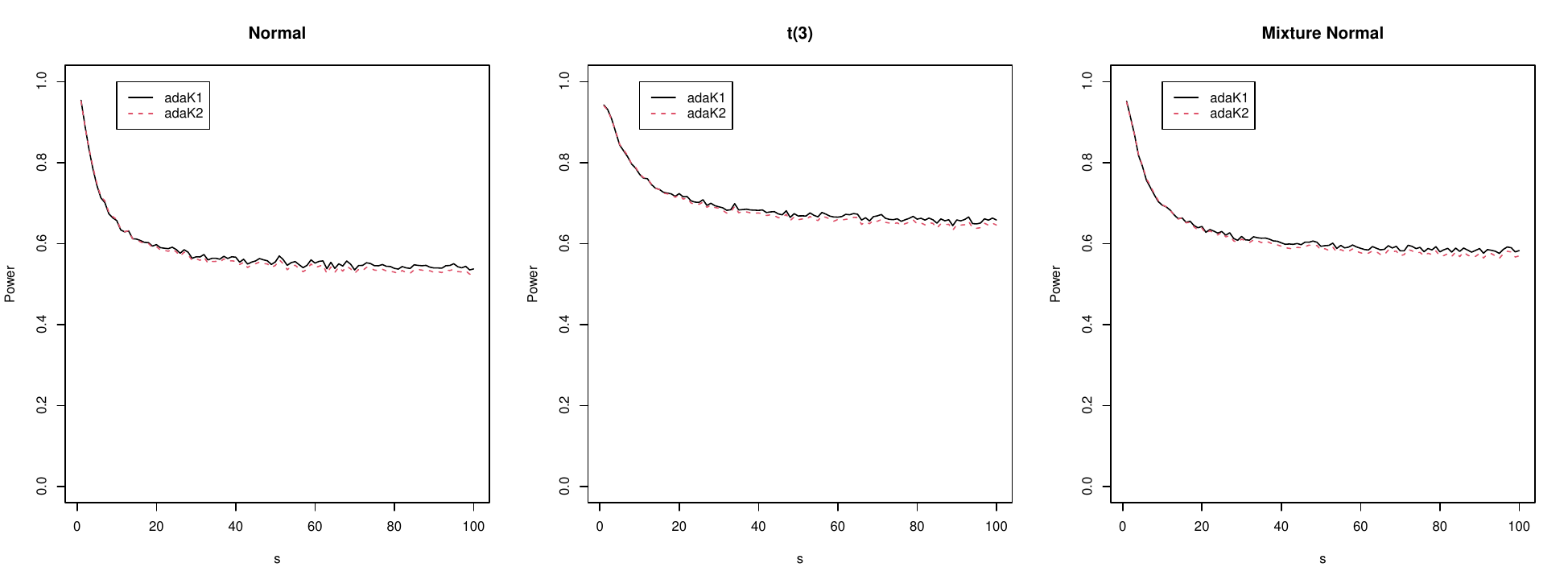}
	\caption{Power curves of two Cauchy combination tests: {\it adaK1} versus {\it adaK2}.}
	\label{fig:2k}
\end{figure}

}

\section{Real data application}\label{Sec: Real data application}

In this study, we apply statistical methods to analyze a financial dataset comprising weekly returns of stocks in the S\&P 500 index from February 2009 to September 2018. {  This dataset is publicly available from financial data sources such as Yahoo Finance, and can also be obtained from the authors upon request.} Our objective is to test whether the return of stock securities is the same as the risk-free return.

Let $R_{ij}$ denote the return of the $j$-th stock security at time $i$, and let $r_{i}$ be the risk-free return rate. We define $X_{ij} = R_{ij} - r_i$ as the excess return of the $j$-th stock at time $i$. Let $\X_i = (X_{i1}, \cdots, X_{ip})^\top$ be the vector of excess returns for all $p$ stocks at time $i$, and let $\bmu = E(\X_i)$ be the vector of expected excess returns. We consider the following hypothesis testing problem:
\begin{align}\label{tdata}
H_0: E(\bmu) = \bm 0~~\text{vs.}~~H_1: E(\bmu) \ne \bm 0,
\end{align}
which tests whether the excess return of any asset is zero on average.

{  From an economic standpoint, this test (\ref{tdata}) evaluates the existence of abnormal returns beyond the risk-free rate, a quantity central to asset pricing theory. Under the Capital Asset Pricing Model (CAPM) and related equilibrium frameworks, $\bmu=\bm 0$ is consistent with the absence of mispricing, implying that all assets are fairly priced given their risk exposures. Rejecting $H_0$ would indicate that at least one security earns a statistically significant abnormal return, potentially signaling market inefficiency, omitted risk factors, or arbitrage opportunities.}

\begin{figure}
	\centering
	\includegraphics[width=0.9\textwidth]{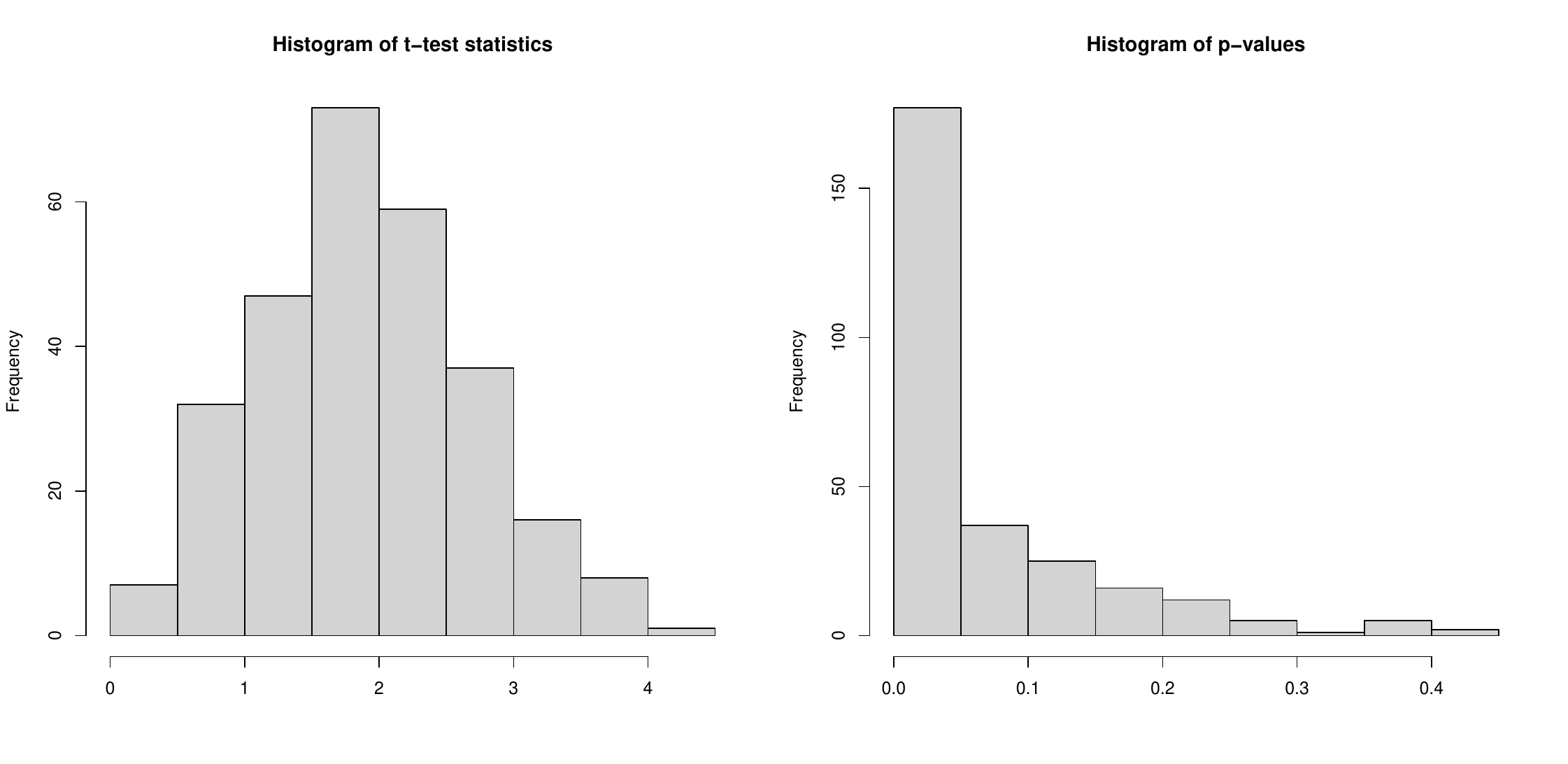}
	\caption{Histogram of t-test statistics and corresponding $p$-values of each security.}
	\label{figdata}
\end{figure}

The dataset consists of $T = 501$ weekly returns for 424 stocks that were consistently included in the S\&P 500 index during the study period. To ensure independence of observations, we applied the level-0.05 Ljung-Box test \citep{ljung1978} to each stock to check for zero autocorrelations. We retained $p = 280$ stocks for which the Ljung-Box test did not reject the null hypothesis of no autocorrelation {  under the significant level $0.05$}. Figure \ref{figdata} presents the histogram of $t$-test statistics and the corresponding $p$-values for each stock. We observe that many $p$-values are smaller than 0.05, suggesting evidence against the null hypothesis. We then applied the test procedures from a simulation study to test the problem \eqref{tdata}. All tests rejected the null hypothesis, indicating that the excess returns of all assets are not all equal to zero. It is worth pointing out that rejecting the null hypothesis, i.e. claiming the return rates are not only composed by the risk-free rates on average is consistent with the views of many economists \citep{fama1993,fama2015}. {  Such a finding would imply the presence of non-zero risk premia, aligning with multi-factor asset pricing theories \citep{Ross1976} where returns are explained by exposures to systematic risk factors such as size, value, and momentum. From a practical standpoint, it suggests that certain securities or portfolios may offer abnormal returns beyond compensation for risk-free investing, potentially due to market inefficiencies, behavioral biases, or unpriced sources of risk. This has direct implications for portfolio construction, active management strategies, and the evaluation of market efficiency.}

To evaluate the performance of our test and other competing tests, we randomly sampled $n$ observations from the 501 weekly returns and recorded the rejection rates of each test based on 1000 replications. Table \ref{tabdata} reports the results of each test. We found that the L-statistic $T_{\lceil p/8\rceil}$ performed best among all tests, suggesting that the alternative hypothesis is not very dense or very sparse. When we applied the Benjamini-Hochberg procedure \citep{bh} with a False Discovery Rate (FDR) of $\alpha = 0.01$, we found that 17 securities had significant excess returns. Compared to other adaptive test procedures, such as COM and adaQ, our proposed test $T_C$ performed best ({  The performance of $T_c$ and adaQ is close.}), demonstrating the advantage of our newly proposed adaptive procedures. 

\begin{table}
	\centering
	\caption{The rejection rates of testing excess returns of the S\&P stocks with $p=280$.}
	\begin{tabular}{@{}l ccccccccc@{}} \hline\hline
		&\multicolumn{1}{c}{$T_1(MAX)$} &\multicolumn{1}{c}{$T_5$} &\multicolumn{1}{c}{$T_{\lceil p/8\rceil}$} &\multicolumn{1}{c}{$T_{\lceil p/4\rceil}$} &\multicolumn{1}{c}{$T_{\lceil p/2\rceil}$} &\multicolumn{1}{c}{$T_p(SUM)$} &\multicolumn{1}{c}{$T_C$} &\multicolumn{1}{c}{COM} &adaQ\\ \hline
		$n=100$ &0.286 &0.351 &0.429 &0.418 &0.409 &0.399 &0.421  &0.363 &0.417\\
		$n=150$ &0.404 &0.498 &0.576 &0.573 &0.568 &0.550 &0.574  &0.511 &0.570\\
		$n=200$ &0.492 &0.589 &0.669 &0.666 &0.656 &0.649 &0.658  &0.614 &0.653 \\
		$n=250$ &0.585 &0.721 &0.821 &0.810 &0.798 &0.800 &0.808  &0.744 &0.802\\
		$n=300$ &0.686 &0.825 &0.908 &0.908 &0.903 &0.896 &0.909  &0.864 &0.895\\ \hline\hline
	\end{tabular}
	\label{tabdata}
\end{table}

\section{Additional Discussion}
{  Condition (A3) imposes a certain degree of dependence structure through the $\alpha$-mixing assumption, which may be somewhat restrictive for modeling complex or strongly correlated covariance structures encountered in real-world applications. Nevertheless, simulation studies presented in the supplementary material demonstrate that our proposed methods remain highly effective and robust across a wide range of covariance structures, including those that do not strictly satisfy Condition (A3).
These empirical findings suggest that the practical applicability of our approach may extend well beyond the theoretical assumptions currently required. However, rigorously establishing the asymptotic validity of the proposed procedures under more general dependence conditions—such as long-range dependence, block dependence, or non-stationary structures—remains an important open problem. Extending the theoretical framework to accommodate these more general settings is a promising direction for future research and would enhance the robustness and generalizability of the proposed methodology.}

The following are proposed areas for further research on this method: Firstly, our method relies on an independent component model that excludes heavy-tailed distributions like the multivariate t-distribution. As such, it poses an interesting challenge to devise robust testing procedures grounded in $L$-statistics. The spatial-sign based method, referenced in \cite{wang2015high,feng2016spatial,feng2016multivariate}, may serve as a suitable option. This research domain merits further investigation. Secondly, there are several other high-dimensional testing problems worth considering. These encompass the high dimensional covariance matrix test \citep{chen2010tests,cai2013two,zhu2022hypothesis,li2025tests}, the high-dimensional regression coefficient test within the generalized linear model, as explored in \cite{goeman2006testing,zhong2011tests,feng2013rank,guo2016tests,zhang2022new,zhu2022hypothesis}, the high-dimensional alpha test in the linear factor pricing model, as discussed in \cite{feng2022high,pesaran2023testing}, the high dimensional white noise test \citep{chang2017testing,li2019testing,feng2022white} and the high-dimensional change point test, mentioned in \citep{wang2023computationally}, among others. Extending the aforementioned methods to these problems presents an engaging challenge.

\section{Useful Lemmas}
Throughout the Supplementary Material, let $C$ be a generic constant which is different from line to line. For a set $S\subseteq\{1,\ldots,p\}$ and a matrix $\A$, let $\A_{S,S}$ be the submatrix of $\A$ that contains the rows and columns in $S$. For a subset of a Euclidean space $\mathcal{X}$ and $\epsilon>0$, a subset $\mathcal{X}_\epsilon$ of $\mathcal{X}$ is called an $\epsilon$-net of $\mathcal{X}$ if every point $x\in\mathcal{X}$ can be approximated to within $\epsilon$ by some point $y\in\mathcal{X}_\epsilon$, i.e. $|x-y|\le\epsilon$. We first restate some useful lemmas.

\begin{lemma}
	\label{Tonycai}
	Let $(Z_1,...,Z_p)^\top$ be a zero mean multivariate normal random vector with covariance matrix $\bms=(\sigma_{ij})_{p\times p}$ and diagonal $\sigma_{ii}=1$ for $1\leq i\leq p$. If $\max_{1\le i<j\le p}|\sigma_{ij}|\leq r<1$ and $\max_{1\le j\le p}\sum_{i=1}^p\sigma_{ij}^2\leq c$ for some $r$ and $c$, then for any $x\in\mathbb{R}$ as $p\to \infty$,
	\begin{equation*}
	\Pr\left(\max_{1\le i\le p}Z_i^2-2\log p+\log(\log p)\le x\right)\to\exp\left\lbrace-\pi^{-1/2}\exp(-x/2) \right\rbrace.
	\end{equation*}
\end{lemma}
\begin{proof}
	See Lemma 6 in \cite{TonyCai2014TwosampleTO}.
\end{proof}

\begin{lemma}
	\label{alpha mixing inequality}
	Suppose that $X$ and $Y$ are $\mathcal{F}_1^t$-measurable and $\mathcal{F}_{t+m}^\infty$-measurable random variables, respectively. If $\E(|X|^p)<\infty$, $\E(|Y|^q)<\infty$ for some $p,q,l>1$ with $p^{-1}+q^{-1}+l^{-1}=1$. Then, for some constant $C$,
	\begin{align*}
	|\E(XY)-\E(X)\E(Y)|\le C\alpha^{1/t}(m)\Vert X\Vert_p\Vert Y\Vert_q.
	\end{align*}
\end{lemma}
\begin{proof}
	See \cite{RI87alphamixinginequality}.
\end{proof}

\begin{lemma}
	\label{independence of main term}
	Let $\{(U,U_N,\widetilde{U}_N)\in\mathbb{R}^3;N\geq 1\}$ and $\{(V,V_N,\widetilde{V}_N)\in\mathbb{R}^3;N\geq 1\}$ be two sequences of random variables with $U_N\stackrel{d}{\to} U$ and $V_N\stackrel{d}{\to} V$ as $N\to\infty$. Assume $U$ and $V$ are continuous random variables. We assume that
	\begin{equation*}
	\widetilde{U}_N=U_N+o_p(1)~~\text{and}~~ \widetilde{V}_N=V_N+o_p(1).
	\end{equation*}
	If $U_N$ and $V_N$ are asymptotically independent, then $\widetilde{U}_N$ and $\widetilde{V}_N$ are also asymptotically independent.
\end{lemma}
\begin{proof}
	See Lemma 7.10 in \cite{feng2024asymptotic}.
\end{proof}

\begin{lemma}\label{Bonferroni inequality}
	For events $A_1,\ldots,A_p$, let $X=\sum_{i=1}^p\mathbb{I}(A_i)$, $S_j=\sum_{1\le i_1<\ldots<i_j\le p}P(A_{i_1}\cap\ldots\cap A_{i_j})$, then for any integer $k\in\{1,\ldots,p\}$, for any integer $1\le d\le [(p-k)/2]$,
	\begin{align*}
	\sum_{j=k}^{2d+k+1}(-1)^{k+j}\binom{j-1}{k-1}S_j \le \Pr(X\ge k) \le \sum_{j=k}^{2d+k}(-1)^{k+j}\binom{j-1}{k-1}S_j.
	\end{align*}
\end{lemma}
\begin{proof}
	See Chapter \uppercase\expandafter{\romannumeral4} in \cite{Feller1968Probability}.
\end{proof}

\begin{lemma}\label{lemma 6.1 in Z15}
	For any integer $k\in\{1,\ldots,p\}$, let $\mathcal{S}:=\{v\in\mathbb{S}^{p-1}:\Vert v\Vert_0\le k\}$ with $\mathbb{S}^{p-1}:=\{b\in\mathbb{R}^p:\Vert b\Vert=1\}$. For any $\epsilon>0$, there exists an $\epsilon$-net of $\mathcal{S}$, denoted by $\mathcal{S}_\epsilon$, such that $\Vert\mathcal{S}_\epsilon\Vert_0\le \{(2+\epsilon)ep/(\epsilon k)\}^k$ and
	\begin{align*}
	\bigcap_{v\in\mathcal{S}_\epsilon}\{w\in\mathbb{R}^p:w^\top v\le(1-\epsilon)\sqrt{x}\}\subseteq\bigcap_{v\in\mathcal{S}}\{w\in\mathbb{R}^p:w^\top v\le\sqrt{x}\}{ \subseteq}\bigcap_{v\in\mathcal{S}_\epsilon}\{w\in\mathbb{R}^p:w^\top v\le\sqrt{x}\}.
	\end{align*}
\end{lemma}
\begin{proof}
	See Lemma 6.1 in \cite{zhang2015}.
\end{proof}

\section{Proof of Theorems 1-5}
\subsection{Proof of (\romannumeral1) in Theorem 1}
\begin{proof}
	Define the point process of exceedances of level $x+b_p$ formed by $t_j^2,1\le j\le p$ as $N_p(B)=\sum_{j=1}^p \mathbb{I}(t_j^2>x+b_p,j/p\in B)$ for any Borel set $B$ on $(0,1]$. Then,
	\begin{align*}
	\{t_{(p+1-s)}^2\le x+b_p\}=\{N_p((0,1])\le s-1\}.
	\end{align*}
	Let $N$ be a Poisson
	point process on $(0,1]$ with intensity $\log \Lambda^{-1}(x)$, then
	\begin{align*}
	\Pr\left(N((0,1])\le s-1\right)=\Lambda(x)\sum_{i=0}^{s-1}\frac{\{\log \Lambda^{-1}(x)\}^i}{i!}.
	\end{align*}
	Hence, it suffices to show that $N_p\cd N$. According to Theorem A.1 in \cite{Leadbetter1983ExtremesAR}, it is sufficient to show that
	\begin{align}\label{A.1}
	\E\{N_p((a,b])\} \to \E\{N((a,b])\} =(b-a)\log \Lambda^{-1}(x),0<a<b\le 1,
	\end{align}
	and for $0<a_1<b_1\le a_2<b_2\le\ldots\le a_k<b_k\le 1$,
	\begin{align}\label{A.2}
	\Pr\left(\bigcap_{i=1}^k\{N_p((a_i,b_i])=0\} \right)\to& \Pr\left(\bigcap_{i=1}^k\{N((a_i,b_i])=0\} \right)\nonumber\\
	=&\exp\left[-\sum_{i=1}^k(b_i-a_i)\log \Lambda^{-1}(x)\right].
	\end{align}

	\noindent\textit{Step 1.} We prove \eqref{A.1}.  Recalling that $t_i=\sqrt{n}\bar{X}_{.i}/\hat{\sigma}_{ii}^{1/2}$ and $\check{t}_i=\sqrt{n}\bar{X}_{.i}/\sigma_{ii}^{1/2}$, define
	\begin{align*}
	\tilde{t}_i=n^{-1/2}\sum_{j=1}^nX_{ji}/\sigma_{ii}^{1/2}\mathbb{I}(X_{ji}/\sigma_{ii}^{1/2}<\tau_n),
	\end{align*}
	where $\tau_n=2\eta^{-1/2}\sqrt{\log(n+p)}$ (for sub-Gaussian-type tails), $\tau_n=4\eta^{-1}\log(n+p)$ (for sub-exponential-type tails) or $\tau_n=\sqrt{n\log^{-8} (p)}$ (for sub-polynomial-type tails) under assumption (A1). By Central limit theorem, $t_i^2-\check{t}_i^2=O_p(n^{-1/2})$. By Markov inequality, $\Pr(|\check{t}_i^2-\tilde{t}_i^2|\ge n^{-1/2})\le \Pr(|X_{ji}/\sigma_{ii}^{1/2}|\ge\tau_n)\to 0$. Thus,
	\begin{align}\label{A.3}
	\E\{N_p((a,b])\}&=\sum_{i\in (ap,bp]}\Pr(t_i^2>x+b_p)\nonumber\\
	&=\sum_{i\in (ap,bp]}\Pr(\tilde{t}_i^2>x+b_p+\tilde{t}_i^2-t_i^2)\nonumber\\
	&=([bp]-[ap])\{1-F(x+b_p)\}+o(1),
	\end{align}
	where $F(\cdot)$ is the c.d.f. of $\chi^2_1$. Here, the reason for the last equality is that
	\begin{align*}
	\Pr\left(|\tilde{t}_i|>\sqrt{x+b_p+cn^{-1/2}}\right)=&\Pr\left(|Z|>\sqrt{x+b_p+cn^{-1/2}}-\lambda_p\log^{-1/2}(p)\right)\\
	&+O[\exp\{-n^{1/2}\lambda_p\log^{-1/2}(p)\tau_n^{-1}\}]
	\end{align*}
	due to Theorem 1 in \cite{Zaitsev1987OnTG}, and that set $\lambda_p\to 0$ sufficiently slowly so that
	\begin{align*}
	\exp\{-n^{1/2}\lambda_p\log^{-1/2}(p)\tau_n^{-1}\}=o(p^{-1}).
	\end{align*}
	Let $Z_1,\cdots,Z_p$ be independently identically distributed standard normal, then
	\begin{align*}
	[1-\{1-F(x+b_p)\}]^p=\{F(x+b_p)\}^p=\Pr\left(\max_{1\le i\le p}Z_i^2\le x+b_p\right) \to\Lambda(x),
	\end{align*}
	due to Lemma \ref{Tonycai}. Then we must have $1-F(x+b_p)\to 0$. (For, otherwise, $1-F(x+b_{p_k})$ would be bounded away from zero for some subsequence $\{p_k\}$, leading to the conclusion $\Pr\left(\max_{1\le i\le p}Z_i^2\le x+b_p\right) \to 0$.) By taking logarithms, we get
	\begin{align*}
	-p\{1-F(x+b_p)\}\{1+o(1)\}\to \log \Lambda(x),
	\end{align*}
	which, together with \eqref{A.3}, implies \eqref{A.1}.
	
	\noindent\textit{Step 2.} We prove \eqref{A.2}. Let $E_k=\bigcup_{i=1}^k\{\lfloor a_ip\rfloor+1,\ldots,\lfloor b_ip\rfloor\}$, then
	\begin{align}\label{A.4}
	\Pr\left(\bigcap_{i=1}^k\{N_p((a_i,b_i])=0\} \right)=\Pr\left(\max_{i\in E_k}t_i^2\le x+b_p\right).
	\end{align}
	We claim that
	\begin{align}\label{A.5}
	\Pr\left(\max_{i\in E_k}\tilde{t}_i^2\le x+b_p\right)\to \exp\left\{-\sum_{i=1}^k(b_i-a_i)\log \Lambda^{-1}(x)\right\},
	\end{align}
	which, by setting $x=\log(\log p)/2$, implies that
	\begin{align*}
	\max_{i\in E_k}\tilde{t}_i^2=O_p(\log p).
	\end{align*}
	By the Markov inequality, we have for sub-Gaussian-type tails,
	\begin{align*}
	\Pr\left(\max_{i\in E_k}|\check{t}_i-\tilde{t}_i|\ge 1/\log p\right)&\le \Pr\left(\max_{i\in E_k} \max_{1\le j\le n} |X_{ji}/\sigma_{ii}^{1/2}|\ge\tau_n \right)\\
	&\le \sum_{i=1}^p\sum_{j=1}^n \Pr\left(|X_{ji}/\sigma_{ii}^{1/2}|\ge\tau_n \right)\\
	&\le Kpn\exp(-\eta \tau_n^2)\to 0,
	\end{align*}
	which still hold for sub-exponential-type tails and sub-polynomial-type tails. Notice that
	\begin{align*}
	\left|\max_{i\in E_k}t_i^2-\max_{i\in E_k}\check{t}_i^2\right|\le \max_{1\le i\le p}\left|\frac{\sigma_{ii}}{\hat{\sigma}_{ii}}-1\right|\cdot \max_{i\in E_k}\check{t}_i^2.
	\end{align*}
	Following form the proof of Lemma A.3 in \cite{Fan2011HighDC}, we have for some constant $C>0$,
	\begin{align}\label{A.6}
	\Pr\left(\max_{1\le i \le p}|\hat{\sigma}_{ii}-\sigma_{ii}|>C\sqrt{\frac{\log p}{n}}\right)\to 0~~\text{and}~~\Pr\left(\frac{4}{9}\le \frac{\hat{\sigma}_{ii}}{\sigma_{ii}} \le \frac{9}{4},i=1,\ldots,p\right)\to 1,
	\end{align}
	which imply that $\max_{1\le i le p}|\sigma_{ii}/\hat{\sigma}_{ii}-1|=O_p\{\sqrt{\log(p)/n}\}$. Combining all the results above, we get
	\begin{align}\label{A.7}
	\max_{i\in E_k}\tilde{t}_i^2-\max_{i\in E_k}t_i^2=O_p\left(\frac{\log^{3/2} p}{n^{1/2}}\right)=\max_{i\in E_k}\check{t}_i^2-\max_{i\in E_k}t_i^2,
	\end{align}
	which, together with \eqref{A.4} and \eqref{A.5}, implies \eqref{A.2}.
	
	Next, to complete the proof, we prove that claim \eqref{A.5} indeed holds. By the Bonferroni inequality in Lemma \ref{Bonferroni inequality}, for any fixed integer $b\le p_k/2$, where $p_k:=\sum_{i=1}^k(\lfloor b_ip\rfloor-\lfloor a_ip\rfloor)$, we have
	\begin{align*}
	\sum_{s=1}^{2b+2}(-1)^{s+1}\sum_{(i_1,\ldots,i_s)\in \mathcal{B}_s}&\Pr\left(|\tilde{t}_{i_1}|\ge x_p,\ldots, |\tilde{t}_{i_s}|\ge x_p\right)\le \Pr\left(\max_{i\in E_k}|\tilde{t}_i|\ge x_p \right)\\
	\le&\sum_{s=1}^{2b+1}(-1)^{s+1}\sum_{(i_1,\ldots,i_s)\in\mathcal{B}_s}\Pr\left(|\tilde{t}_{i_1}|\ge x_p,\ldots, |\tilde{t}_{i_s}|\ge x_p\right),
	\end{align*}
	where $x_p=\sqrt{x+b_p}$ and $\mathcal{B}_s=\{(i_1,\ldots,i_s)\in E_k|i_1<\ldots<i_s\}$. By Theorem 1 in \cite{Zaitsev1987OnTG} again, we have
	\begin{align*}
	\Pr\left(|\tilde{t}_{i_1}|\ge x_p,\ldots, |\tilde{t}_{i_s}|\ge x_p\right)=&\Pr\left(|\z|_{\min}\ge x_p-\kappa_p\log^{-1/2}(p)\right)\\
	&\qquad+c_1s^{5/2}\exp\left\{-\frac{n^{1/2}\kappa_p}{c_2S^3\tau_n\log^{1/2}(p)} \right\},
	\end{align*}
	where $c_1,c_2>0$ are two constants, $\kappa_p\to 0$ (specified later), $|z|_{\min}=\min_{1\leq l\leq s}|z_l|$, and $z=(z_1,\ldots,z_s)^\top$ is a $s$-dimensional normal vector, which has the same covariance as $(\tilde{t}_{i_1},\ldots,\tilde{t}_{i_s})^\top$. Let $\kappa_p\to 0$ sufficiently slowly so that
	\begin{align*}
	c_1s^{5/2}\exp\left\{-\frac{n^{1/2}\kappa_p}{c_2S^3\tau_n\log^{1/2}(p)} \right\}=O(p^{-\zeta}),
	\end{align*}
	for any large $\zeta>0$. Thus,
	\begin{align}\label{A.8}
	\Pr\left(\max_{i\in E_k}|\tilde{t}_i|\ge x_p \right)
	\le\sum_{s=1}^{2b+1}(-1)^{s+1}\sum_{(i_1,\cdots,i_s)\in\mathcal{B}_s}\Pr\left(|\z|_{\min}\ge x_p-\kappa_p\log^{1/2}(p)\right)+o(1).
	\end{align}
	Similarly, we have
	\begin{align}\label{A.9}
	\Pr\left(\max_{i\in E_k}|\tilde{t}_i|\ge x_p \right)
	\ge\sum_{s=1}^{2b+2}(-1)^{s+1}\sum_{(i_1,\cdots,i_s)\in\mathcal{B}_s}\Pr\left(|\z|_{\min}\ge x_p-\kappa_p\log^{1/2}(p)\right)+o(1).
	\end{align}
	
	Define
	\begin{align*}
	V_s:=\sum_{(i_1,\ldots,i_s)\in\mathcal{B}_s}P_{i_1,\ldots,i_s}=\sum_{(i_1,\ldots,i_s)\in\mathcal{B}_s} \Pr(|z_{i_1}|\geq \widetilde{x}_p,\ldots,|z_{i_s}|\geq \widetilde{x}_p),
	\end{align*}
	where $\widetilde{x}_p=x_p-\kappa_p\log^{1/2}(p)$. Write correlation matrix $\R=\Q^\top \L \Q$ with $\Q=(q_{ij})_{p\times p}$ being an orthogonal matrix and $\L=\diag(\lambda_1,\ldots,\lambda_p)$, $\lambda_i$'s being the eigenvalues of $\R$. Since $\sum_{1\leq j\leq p}r_{ij}^2$ is the $i$th diagonal element of $\R^2=\Q^\top\L^2\Q$, $\sum_{1\leq j\leq p}r_{ij}^2=\sum_{l=1}^pq_{li}^2\lambda_l^2\leq c_3^2$ for some constant $c_3$ due to assumption (A2). Thus, $\R$ satisfies the condition in Lemma \ref{Tonycai}.
	
	Define 
	\begin{align*}
	\mathcal{I}=\{(i_1,\ldots,i_s)\in\mathcal{B}_s:\max_{1\leq k<l\leq s}|\cov(z_{i_k},z_{i_l})|\ge p^{-a}\}, 
	\end{align*}
	where $a>0$ is a sufficiently small number to be specified later. For $2\leq d\leq s-1$, define 
	\begin{align*}
	\mathcal{I}_d=\{(i_1,\ldots,i_s)\in\mathcal{B}_s:&\text{card}(S)=d, ~~\text{with}~~S~~\text{being the largest subset of}~~\{i_1,\ldots,i_s\}\\
	&~~\text{such that}~~\forall i_k,i_t\in S,i_k\ne i_t,|\cov(z_{i_k},z_{i_t})|<p^{-a}\}.
	\end{align*}
	For $d=1$, define
	\begin{align*}
	\mathcal{I}_1=\{(i_1,\ldots,i_s)\in\mathcal{B}_s:|\cov(z_{i_k},z_{i_l})|\ge p^{-a},\forall 1\leq k<l\leq s\}.
	\end{align*}
	So we have $\mathcal{I}=\cap_{d=1}^{s-1}\mathcal{I}_d$. Here, $\text{card}(S)$ is the cardinality of $S$. We can show that $\text{card}(\mathcal{I}_d)\le Cp_k^d\cdot p^{2as}$. In fact, in $\mathcal{B}_s$, the total number of the subsets of $\{i_1,\ldots,i_s\}$ with cardinality $d$ is $\binom{p_k}{d}$. For a fixed subset $S$ with cardinality $d$, the number of $i$ such that $|\cov(z_i,z_j)|\ge p^{-a}$ for some $j\in S$ is no more than $Cdp^{2a}$. This implies that $\text{card}(\mathcal{I}_d)\le Cp_k^d\cdot p^{2as}$. Define $\mathcal{I}^c=\mathcal{B}_s\backslash \mathcal{I}$. Then the total number of elements in the sum $\sum_{(i_1,\cdots,i_s)\in\mathcal{I}^c}P_{i_1,\ldots,i_s}$ is $\binom{p_k}{s}-O(\sum_{d=1}^{s-1}p_k^d\cdot p^{2as})=\binom{p_k}{s}-O(p_k^{s-1}\cdot p^{2as})=\{1+o(1)\}\binom{p_k}{s}$. By equation (20) in the proof of Lemma 6 in \cite{TonyCai2014TwosampleTO}, we have
	\begin{align*}
	P_{i_1,\ldots,i_s}=\{1+o(1)\}\pi^{-s/2}p^{-s}\exp\left( -\frac{sx}{2}\right)
	\end{align*}
	uniformly in $(i_1,\ldots,i_t)\in\mathcal{I}^c$. Therefore,
	\begin{align}\label{A.10}
	\sum_{(i_1,\ldots,i_s)\in\mathcal{I}^c}P_{i_1,\ldots,i_s}=\{1+o(1)\}\frac{(p_k/p)^s}{s!}\pi^{-s/2}\exp\left( -\frac{sx}{2}\right).
	\end{align}
	By equation (21) in the proof of Lemma 6 in \cite{TonyCai2014TwosampleTO}, we have for $1\leq d\leq s-1$,
	\begin{align}\label{A.11}
	\sum_{(i_1,\ldots,i_s)\in\mathcal{I}_d}P_{i_1,\ldots,i_s}\to 0. 
	\end{align}
	Putting \eqref{A.8}-\eqref{A.11} together, we obtain that
	\begin{align*}
	\{1+o(1)\}S_{2b+2}\le P\left(\max_{i\in E_k}|\tilde{t}_i|\ge x_p \right) \le \{1+o(1)\}S_{2b+1},
	\end{align*}
	where
	\begin{align*}
	S_{b}=\sum_{s=1}^b(-1)^{s+1}\frac{(p_k/p)^s}{s!}\pi^{-s/2}\exp\left( -\frac{sx}{2}\right).
	\end{align*}
	Note that
	\begin{align*}
	\lim_{b\to\infty}S_{b}=1-\exp\left\{-\frac{p_k/p}{\sqrt{\pi}}\exp\left(-\frac{x}{2}\right)\right\}=1-\exp\left\{-\sum_{i=1}^k(b_i-a_i)\log \Lambda^{-1}(x)\right\}.
	\end{align*}
	By letting $p\to\infty$ first and then $d\to\infty$, we prove \eqref{A.5}.
\end{proof}

\subsection{Proof of (\romannumeral2) in Theorem 1}
\begin{proof}
	For Borel sets $B\in(0,1]\times R$, define the point process $\widetilde{N}_p$ of exceedances of levels $x_1+b_p,\ldots,x_k+b_p$ as
	\begin{align*}
	\widetilde{N}_p(B)\equiv \sum_{j=1}^k\widetilde{N}_p^{(j)}(B)=\sum_{j=1}^k\sum_{i=1}^p\mathbb{I}(t_i^2>x_j+b_p,(i/p,j)\in B).
	\end{align*}
	Then,
	\begin{align*}
	\bigcap_{j=1}^k\left\{ t_{(p+1-j)}^2-b_p \leq x_j \right\}= \bigcap_{j=1}^k\left\{ \widetilde{N}^{(j)}_p((0,1]\times (j-1/2,j+1/2] )\le j-1\right\}.
	\end{align*}
	Construct a point process $\widetilde{N}$ on $(0,1]\times R$ such that $\widetilde{N}(\cdot)=\sum_{j=1}^k\widetilde{N}^{(j)}(\cdot)$, where for each $1\le j\le k-1$, $\widetilde{N}^{(j)}$ is a Poisson process independent thinning of the Poisson
	process $\widetilde{N}^{(j+1)}$ with deleting probability $1-\frac{\log \Lambda^{-1}(x_j)}{\log \Lambda^{-1}(x_{j+1})}$, the initial Poisson process $\widetilde{N}^{(k)}$ has intensity $\log \Lambda^{-1}(x_k)$. According to Section 5 in \cite{Leadbetter1983ExtremesAR}, it is known that $\widetilde{N}^{(j)},j=1,\ldots,k$ are Poisson processes with intensity $\log \Lambda^{-1}(x_j)$ respectively, and are independent on disjoint intervals on the plane, and
	\begin{align*}
	&\Pr\left( \bigcap_{j=1}^k\left\{ \widetilde{N}^{(j)}((0,1]\times (j-1/2,j+1/2] )\le j-1\right\} \right)\\
	=& \Lambda\left(x_k\right) \sum_{\sum_{i=2}^j k_i \leq j-1, j=2, \ldots, s} \prod_{i=2}^s \frac{\{\log \Lambda^{-1}(x_i)-\log \Lambda^{-1}(x_{i-1})\}^{k_i}}{k_i!}.
	\end{align*}
	Hence, it suffices to show that $\widetilde{N}_p\cd\widetilde{N}$. According to Theorem A.1 in \cite{Leadbetter1983ExtremesAR}, it is sufficient to show that for $0<a<b\le 1$, $0<c<d\le k$,
	\begin{align}\label{A.12}
	\E\{\widetilde{N}_p((a,b]\times (c,d])\} \to \E\{\widetilde{N}((a,b]\times (c,d])\} =(b-a)\sum_{c<j\le d}\log \Lambda^{-1}(x_j),
	\end{align}
	and for $0<a_1<b_1\le a_2<b_2\le\ldots\le a_s<b_s\le 1$, $0<c_1<d_1\le c_2<d_2\le\ldots\le c_s<d_s\le k$,
	\begin{align}\label{A.13}
	\Pr\left(\bigcap_{l=1}^s\{\widetilde{N}_p((a_l,b_l]\times (c_l,d_l])=0\} \right)&\to \Pr\left(\bigcap_{l=1}^s\{\widetilde{N}((a_l,b_l]\times (c_l,d_l])=0\} \right)\nonumber\\
	&=\exp\left[-\sum_{l=1}^s(b_l-a_l)\log \Lambda^{-1}(x_{d_l})\right].
	\end{align}
	
	\noindent For \eqref{A.12}, according to the proof of \eqref{A.1},
	\begin{align*}
	\E\{\widetilde{N}_p((a,b]\times (c,d])\}&=\sum_{j\in(c,d]}\sum_{i\in (ap,bp]}\Pr(t_i^2>x_j+b_p) \to (b-a)\sum_{j\in(c,d]}\log \Lambda^{-1}(x_j).
	\end{align*}
	
	\noindent For \eqref{A.13}, by \eqref{A.7}, we have
	\begin{align}\label{A.14}
	\Pr\left(\bigcap_{l=1}^s\{\widetilde{N}_p((a_l,b_l]\times (c_l,d_l])=0\} \right)&=\Pr\left(\bigcap_{l=1}^s\left\{\max_{i\in (a_lp,b_lp]}t_i^2\le x_{d_l}+b_p \right\} \right)\nonumber\\
	&=\Pr\left(\bigcap_{l=1}^s\left\{\max_{i\in (a_lp,b_lp]}\tilde{t}_i^2\le x_{d_l}+b_p \right\} \right)+o(1).
	\end{align}
	Let $I_l$ denote the first $b_lp-a_lp-m_p$ elements of $(a_lp,b_lp]$, and $I_l^*$ the remaining $m_p$, where $m_p$ will be specified later. Then,
	\begin{align*}
	0\le & \Pr\left(\bigcap_{l=1}^s\left\{\max_{i\in I_l}\tilde{t}_i^2\le x_{d_l}+b_p \right\} \right)-\Pr\left(\bigcap_{l=1}^s\left\{\max_{i\in (a_lp,b_lp]}\tilde{t}_i^2\le x_{d_l}+b_p \right\} \right)\\
	\le &\sum_{l=1}^sP\left(\max_{i\in I_l^*}\tilde{t}_i^2> x_{d_l}+b_p\right)\\
	\le &s\max_{1\le i\le s}\Pr\left(\max_{i\in I_l^*}\tilde{t}_i^2> x_{d_l}+b_p\right)=:s\rho_p.
	\end{align*}
	By the strong mixing property (see Lemma 3.2.2 in \cite{Leadbetter1983ExtremesAR}), we have
	\begin{align*}
	\left|\Pr\left(\bigcap_{l=1}^s\left\{\max_{i\in I_l}\tilde{t}_i^2\le x_{d_l}+b_p \right\} \right)-\prod_{l=1}^s \Pr\left(\max_{i\in I_l}\tilde{t}_i^2\le x_{d_l}+b_p\right) \right|\le (s-1)\alpha_X(m_p).
	\end{align*}
	Further,
	\begin{align*}
	0\le& \prod_{l=1}^s \Pr\left(\max_{i\in I_l}\tilde{t}_i^2\le x_{d_l}+b_p\right) - \prod_{l=1}^s \Pr\left(\max_{i\in (a_lp,b_lp]}\tilde{t}_i^2\le x_{d_l}+b_p\right)\\
	\le &\prod_{l=1}^s\left\{\Pr\left(\max_{i\in (a_lp,b_lp]}\tilde{t}_i^2\le x_{d_l}+b_p\right)+\rho_p\right\} - \prod_{l=1}^s \Pr\left(\max_{i\in (a_lp,b_lp]}\tilde{t}_i^2\le x_{d_l}+b_p\right)\\
	\le &(1+\rho_p)^s-1,
	\end{align*}
	since $\prod_{l=1}^s(y_l+\rho_p)-\prod_{l=1}^sy_l$ is increasing in each $y_l$ when $\rho_p>0$. Similarly as \eqref{A.3}, we have
	\begin{align*}
	\rho_p=\max_{1\le i\le s}\Pr\left(\max_{i\in I_l^*}\tilde{t}_i^2> x_{d_l}+b_p\right)\le \max_{1\le i\le s}m_p\{1-F(x_{d_l}+b_p)\}+o(1)=o(1).
	\end{align*}
	Hence, combining all the results, we obtain
	\begin{align}\label{A.15}
	&\left| \Pr\left(\bigcap_{l=1}^s\left\{\max_{i\in (a_lp,b_lp]}\tilde{t}_i^2\le x_{d_l}+b_p \right\} \right)-\prod_{l=1}^s \Pr\left(\max_{i\in (a_lp,b_lp]}\tilde{t}_i^2\le x_{d_l}+b_p\right)\right|\nonumber\\
	\le& s\rho_p+(s-1)\alpha_X(m_p)+(1+\rho_p)^s-1\to 0,
	\end{align}
	by taking $m_p\to\infty$ slowly.
	According to the proof of \eqref{A.5}, we have
	\begin{align*}
	\Pr\left(\max_{i\in (a_lp,b_lp]}\tilde{t}_i^2\le x_{d_l}+b_p\right)\to \exp\left\{-(b_l-a_l)\log \Lambda^{-1}(x_{d_l})\right\},
	\end{align*}
	which, together with \eqref{A.14} and \eqref{A.15}, implies \eqref{A.13}.
\end{proof}

\subsection{Proof of Theorem 2}
\begin{proof}
	\emph{Step 1.} Establish the consistency of  $\check{T}_k=\max_{1\le i_1<\ldots<i_k\le p}\sum_{l=1}^kn\bar{X}_{.i_l}^2/\sigma_{i_li_l}$. Define $\check{\mu}_i:=\mu_i/\sqrt{\sigma_{ii}}$ and $\check{z}_i:=(\bar{X}_{.i}-\mu_i)/\sqrt{\sigma_{ii}}$. Note that $\bar{X}_{.i}^2/\sigma_{ii}=(\check{z}_i+\check{\mu}_i)^2$. Suppose that $\check{\mu}_{i_1^*}\ge\ldots\ge \check{\mu}_{i_k^*}$. Then,
	\begin{align*}
	\check{T}_k=&\max_{1\le i_1<\ldots<i_k\le p}\sum_{l=1}^kn(\check{z}_{i_l}^2+\check{\mu}_{i_l}^2+2\check{z}_{i_l}\check{\mu}_{i_l})\\
	\ge&n\sum_{l=1}^k(\check{z}_{i_l^*}^2+\check{\mu}_{i_l^*}^2+2\check{z}_{i_l^*}\check{\mu}_{i_l^*})\\
	\ge&n\sum_{l=1}^k\check{z}_{i_l^*}^2+n\sum_{l=1}^k\check{\mu}_{i_l^*}^2-2\left(n\sum_{l=1}^k\check{z}_{i_l^*}^2\right)^{1/2}\left(n\sum_{l=1}^k\check{\mu}_{i_l^*}^2\right)^{1/2}.
	\end{align*}
	For any $S:=\{i_1,\ldots,i_k\}\subseteq\{1,\cdots,p\}$, due to $\E(\check{z}_i)=0$ and $\E(\check{z}_i^2)=1/n$, we have $\E\{\sum_{l=1}^k(n\check{z}_{i_l}^2-1)/\sqrt{k}\}=0$ and
	\begin{align*}
	&\var\left\{\sum_{l=1}^k(n\check{z}_{i_l}^2-1)/\sqrt{k}\right\}\\
	=&\frac{1}{k}\sum_{l,t=1}^k\left[\frac{1}{n^2\sigma_{i_li_l}\sigma_{i_ti_t}}\sum_{s_1,s_2,t_1,t_2=1}^n\E\{(X_{s_1i_l}-\mu_{i_l})(X_{s_2i_l}-\mu_{i_l})(X_{t_1i_t}-\mu_{i_t})(X_{t_2i_t}-\mu_{i_t})\}-1\right]\\
	=&\frac{1}{k}\sum_{l,t=1}^k\left[\frac{\E\{(X_{1i_l}-\mu_{i_l})^2(X_{1i_t}-\mu_{i_t})^2\}}{n\sigma_{i_li_l}\sigma_{i_ti_t}}+\frac{n(n-1)}{n^2}+\frac{2n(n-1)}{n^2}\frac{\sigma_{i_li_t}^2}{\sigma_{i_li_l}\sigma_{i_ti_t}}-1 \right]\\
	=&\frac{2}{k}\sum_{l,t=1}^k\frac{\sigma_{i_li_t}^2}{\sigma_{i_li_l}\sigma_{i_ti_t}}+O(k/n)=\frac{2}{k}\tr(\R_{S,S}^2)+O(k/n)=O(1).
	\end{align*}
	Hence, $n\sum_{l=1}^k\check{z}_{i_l}^2=O_p(k)$. As $\check{z}_i$ is independent of the non-zero locations of $\bmu$, $n\sum_{l=1}^k\check{z}_{i_l^*}^2=O_p(k)$. Further, by the assumption that $\sum_{l=1}^k\mu_{i_l^*}^2/\sigma_{i_l^*i_l^*}\ge(2k+\epsilon)\log(p)/n$, we obtain that
	\begin{align}\label{A.18}
	\check{T}_k\ge O_p(k)+(2k+\epsilon)\log p-O_p\{\sqrt{(2k+\epsilon)\log p}\}.
	\end{align}
	
	{ 
 As will be shown in Theorem \ref{wb} (\romannumeral1) below, the bootstrap statistic $T_k^*$ imitates the sampling distribution of $\max_{1\le i_1<\ldots<i_k\le p}\sum_{l=1}^kn\check{z}_{i_l}^2$.} Further, due to the fact that
	\begin{align*}
	\max_{1\le i_1<\ldots<i_k\le p}\sum_{l=1}^kn\check{z}_{i_l}^2\le k\max_{1\le i\le p}n\check{z}_i^2=2k\log p-k\log(\log p)+O_p(1),
	\end{align*}
	we obtain
	\begin{align}\label{A.19}
	c_\alpha^*(k)\le 2k\log p-k\log(\log p)+O_p(1),
	\end{align}
	where $c_\alpha^*(k):=\inf\{x>0:\Pr(T_k^*\le x|\X_1^n)\ge 1-\alpha\}$. Combining \eqref{A.16} and \eqref{A.19}, we get
	\begin{align}\label{A.20}
	\Pr(\check{T}_k>c_\alpha^*(k))\to 1.
	\end{align}
	
	\noindent\emph{Step 2.} Show that the difference between $T_k$ and $\check{T}_k$ is asymptotically negligible. Define $\tilde{\mu}_i$ and $\tilde{z}_i$ by replacing $\sigma_{ii}$ with $\hat{\sigma}_{ii}$ in $\check{\mu}_i$ and $\check{z}_i$. Then,
	\begin{align*}
	|T_k-\check{T}_k|=&\left|\max_{1\le i_1<\ldots<i_k\le p}n\sum_{l=1}^k\bar{X}_{.i_l}^2(\hat{\sigma}_{i_li_l}^{-1}-\sigma_{i_li_l}^{-1})\right|\\
	\le&nk\max_{1\le i\le p}\bar{X}_{.i}^2|\hat{\sigma}_{ii}^{-1}-\sigma_{ii}^{-1}|\\
	=&nk\max_{1\le i\le p}|(\tilde{z}_i-\check{z}_i+\tilde{\mu}_i-\check{\mu}_i)(\tilde{z}_i+\check{z}_i+\tilde{\mu}_i+\check{\mu}_i)|.
	\end{align*}
	According to the proof of \eqref{A.2}, we derive $|\bar{\X}|_\infty=O_p\{\sqrt{\log(p)/n}\}$, which together with \eqref{A.6}, implies that
	\begin{align*}
	&\max_{1\le i\le p}|\tilde{z}_i-\check{z}_i|\le \max_{1\le i\le p}|\bar{X}_{.i}-\mu_i|\cdot\max_{1\le j\le p}|\sigma_{jj}^{-1/2}-\hat{\sigma}_{jj}^{-1/2}|=O_p\{\log(p)/n\},~~\text{and}\\
	&\max_{1\le i\le p}|\tilde{\mu}_i-\check{\mu}_i|\le O_p\{\sqrt{\log(p)/n}\}\cdot\max_{1\le i\le p}|\mu_i|.
	\end{align*}
	Define the event $A:=\{\max_{1\le i\le p}|\mu_i|<C_0\sqrt{\log(p)/n}\}$ for some larger enough constant $C_0>0$. Then, on $A$, we have $\max_{1\le i\le p}|\tilde{\mu}_i-\check{\mu}_i|=O_p\{\log(p)/n\}$. Further, on $A$, $|T_k-\check{T}_k|=O_p\{kn^{-1/2}\log^{3/2}(p)\}$. That is for any $\epsilon>0$, we can find some constant $C''>0$ such that
	\begin{align*}
	\Pr(|T_k-\check{T}_k|\le C''kn^{-1/2}\log^{3/2}(p)|A)\ge 1-\epsilon.
	\end{align*}
	Hence,
	\begin{align}\label{A.21}
	\Pr(T_k\ge c_\alpha^*(k)|A)\ge&\Pr(\check{T}_k\ge c_\alpha^*(k)+|T_k-\check{T}_k||A)\nonumber\\
	\ge&\Pr(\check{T}_k\ge c_\alpha^*(k)+C''kn^{-1/2}\log^{3/2}(p)|A)-\epsilon.
	\end{align}
	By Nazarov's inequality, under the assumption that $k^{7/2}n^{-1/2}\log^{7/2}(np/k)=o(1)$, we get
	\begin{align*}
	&\Pr(\check{T}_k\ge c_\alpha^*(k)|A)-\Pr(\check{T}_k\ge c_\alpha^*(k)+C''kn^{-1/2}\log^{3/2}(p)|A)\\
	=&\Pr(c_\alpha^*(k)\le \check{T}_k< c_\alpha^*(k)+C''kn^{-1/2}\log^{3/2}(p)|A)=o(1),
	\end{align*}
	which, combining with \eqref{A.20} and \eqref{A.21}, implies that
	\begin{align}\label{A.22}
	\Pr(T_k\ge c_\alpha^*(k)|A)\ge \Pr(\check{T}_k\ge c_\alpha^*(k)|A)-o(1)-\epsilon\to 1-\epsilon.
	\end{align}
	
	Suppose $\max_{1\le i\le p}|\mu_i|=|\mu_{i_1^*}|$. On $A^c$, we have for larger enough $C_0$, with probability tending to one,
	\begin{align*}
	\check{T}_k\ge n(\check{z}_{i_1^*}^2+\check{\mu}_{i_1^*}^2+2\check{z}_{i_1^*}\check{\mu}_{i_1^*})\ge C_0\log p>2k\log p-k\log(\log p).
	\end{align*}
	That is $\Pr(\check{T}_k\ge 2k\log p-k\log(\log p)|A^c)\to 1$, which combining with \eqref{A.19}, using similar argument to prove \eqref{A.22}, leads to
	\begin{align}\label{A.23}
	\Pr(T_k\ge c_\alpha^*(k)|A^c)\to 1.
	\end{align}
	By \eqref{A.22} and \eqref{A.23}, we obtain
	\begin{align*}
	\Pr(T_k\ge c_\alpha^*(k))=\Pr(A)\Pr(T_k\ge c_\alpha^*(k)|A)+\Pr(A^c)\Pr(T_k\ge c_\alpha^*(k)|A^c)\to 1-\epsilon \Pr(A),
	\end{align*}
	from which the desired result follows as $\epsilon$ is arbitrary.
\end{proof}

\subsection{Proof of Theorem 3}
\begin{proof}

{ 
\noindent\emph{Step 1.} Recalling the definitions of $T_k$ and $\check{T}_k$, we have
\begin{align*}
	\left|\frac{T_{\lceil\gamma p\rceil}-\check{T}_{\lceil\gamma p\rceil}}{\sqrt{p\sigma_{\gamma\gamma}}}\right|=&\frac{1}{\sqrt{p\sigma_{\gamma\gamma}}}\left|\max_{1\le i_1<\ldots<i_{\lceil\gamma p\rceil}\le p}n\sum_{l=1}^{\lceil \gamma p\rceil}\bar{X}_{.i_l}^2 \left(\frac{1}{\hat{\sigma}_{i_li_l}}-\frac{1}{\sigma_{i_li_l}}\right)\right| \\
	\le &\frac{1}{\sqrt{p\sigma_{\gamma\gamma}}}\left|\max_{1\le i_1<\ldots<i_{\lceil\gamma p\rceil}\le p}n\sum_{l=1}^{\lceil \gamma p\rceil}\bar{X}_{.i_l}^2 \sigma_{i_li_l}^{-2}({\sigma}_{i_li_l}-\hat\sigma_{i_li_l})\right|  \\
    &+\frac{1}{\sqrt{p\sigma_{\gamma\gamma}}}\left|\max_{1\le i_1<\ldots<i_{\lceil\gamma p\rceil}\le p}n\sum_{l=1}^{\lceil \gamma p\rceil}\bar{X}_{.i_l}^2 \sigma_{i_li_l}^{-2}({\sigma}_{i_li_l}-\hat\sigma_{i_li_l})(\hat{\sigma}_{i_li_l}^{-1}\sigma_{i_li_l}-1)\right| \\
    =&O_p(pn^{-2}+n^{-1/2}\log p)+O_p(p^{1/2}n^{-1}\log^{2}(p)) \to 0,
\end{align*}
 where the first term are followed by taking the same procedure as the proof of Theorem 1 in \cite{feng2015two} and the second term are followed by $|\bar{\X}|_\infty=O_p(\sqrt{\log p/n})$ and $\max_{1\le i\le p}|{\sigma}_{ii}-\hat\sigma_{ii}|=O_p(\sqrt{\log p/n})$.

 \noindent\emph{Step 2.} Under $H_0$, we will establish the asymptotic normality of $p^{-1/2}\check{T}_{\lceil \gamma p\rceil}$. By the definition, we can rewrite
\begin{align*}
(\lceil \gamma p\rceil)^{-1}\check{T}_{\lceil \gamma p\rceil}=(\lceil \gamma p\rceil)^{-1}\sum_{i=1}^p \check{t}_i^2I(\check{t}_i^2\ge \check{v}_\gamma)
\end{align*}
where $\check{v}_\gamma=\check{t}_{(p+1-\lceil \gamma p\rceil)}$ is the sample quantile estimator of $v_\gamma$  based on $\{\check{t}_i^2\}_{i=1}^p$. Let $v_\gamma$ be the ($1-\gamma$)-quantile of $\chi_1^2$. Thus, using the similar arguments in the proof of Theorem 2.1 in \cite{Wang2023NonparametricEO}, we have
\begin{align*}
&(\lceil \gamma p\rceil)^{-1}\sum_{i=1}^p
\check{t}_i^2I(\check{t}_i^2\ge \check{v}_\gamma)\\
&\qquad=(\gamma p)^{-1}\sum_{i=1}^p
\{(\check{t}_i^2-v_\gamma)I(\check{t}_i^2\ge v_\gamma)+\gamma v_\gamma\}
+o_p(p^{-3/4+\nu}),
\end{align*}
where $\nu>0$. Define
\begin{align*}
\mu_\gamma
&:=\sum_{i=1}^p\E_{H_0}\{(\check{t}_i^2-v_\gamma)
I(\check{t}_i^2\ge v_\gamma)+\gamma v_\gamma\},\\
\sigma_{\gamma\gamma}
&:=\var_{H_0}\left\{p^{-1/2}\sum_{i=1}^p
(\check{t}_i^2-v_\gamma)I(\check{t}_i^2\ge v_\gamma)\right\}.
\end{align*}
For each $i$, $\var(\check{t}_i^2)=O(1)$ and
$\{\check{t}_i^2\}_{i=1}^p$ is $\alpha$-mixing. Hence, Lemma
\ref{alpha mixing inequality} gives $\sigma_{\gamma\gamma}=O(1)$. Thus, we have
\begin{align}\label{A.24}
    \frac{\check{T}_{\lceil\gamma p\rceil}-\mu_\gamma}{\sqrt{p\sigma_{\gamma\gamma}}}=\frac{1}{\sqrt{p}}\sum_{i=1}^pT_{i,\gamma}+o_p(1),
\end{align}
where $T_{i,\gamma}:=\sigma_{\gamma\gamma}^{-1/2}[(\check{t}_i^2-v_\gamma)I(\check{t}_i^2\ge {v}_\gamma)-\E\{(\check{t}_i^2-v_\gamma)I(\check{t}_i^2\ge {v}_\gamma)\}]$. Next, we employ the Bernstein's block technique \citep{Ibragimov1971IndependentRV} to show that $p^{-1/2}\sum_{i=1}^pT_{i,\gamma}\cd N(0,1)$. Partition the set $\{1,\cdots, p\}$ into $r$ blocks, where each block contain $b$ elements such that $rb\le p<(r+1)b$. Further, for each $1\le l\le r$, we partition the $l$-th block into two sub-blocks with a larger one containing $b_1$ elements and a smaller one containing $b_2=b-b_1$ elements. We require $r\to\infty$, $b_1\to\infty$, $b_2\to\infty$, $rb_1/p\to 1$ and $rb_2/p\to 0$ as $p\to\infty$. Define for each $1\le l\le r$,
\begin{align*}
	&V_{\gamma,1}(l):=\sum_{(l-1)b<i\le(l-1)b+b_1}T_{i,\gamma},~~V_{\gamma,2}(l):=\sum_{(l-1)b+b_1<i\le lb}T_{i,\gamma},\\
    &\mathcal{L}_{1}:=\frac{1}{\sqrt{p}}\sum_{l=1}^rV_{\gamma,1}(l),~~\mathcal{L}_{2}:=\frac{1}{\sqrt{p}}\sum_{l=1}^rV_{\gamma,2}(l)~~\text{and}~~\mathcal{L}_3:=\frac{1}{\sqrt{p}}\sum_{rb<i\le p}T_{i,\gamma}.
\end{align*}
Now, we have the decomposition
\begin{align}\label{A.25}
	\frac{1}{\sqrt{p}}\sum_{i=1}^pT_{i,\gamma}=\mathcal{L}_{1}+\mathcal{L}_{2}+\mathcal{L}_{3}.
\end{align}
We note that $\E(\mathcal{L}_2)=\E(\mathcal{L}_3)=0$ and by the $\alpha$-mixing property, as $p\to \infty$,
\begin{align*}
    \var(\mathcal{L}_2)=&p^{-1}\sigma_{\gamma\gamma}^{-1}\var\left\{\sum_{l=1}^r\sum_{i=1}^{b_2}(\check{t}^2_{(l-1)b+b_1+i}-v_\gamma)\mathbb{I}(\check{t}^2_{(l-1)b+b_1+i}\ge v_\gamma)\right\}\\
    =&p^{-1}\sigma_{\gamma\gamma}^{-1}\sum_{l=1}^r\var\left\{\sum_{i=1}^{b_2}(\check{t}^2_{(l-1)b+b_1+i}-v_\gamma)\mathbb{I}(\check{t}^2_{(l-1)b+b_1+i}\ge v_\gamma)\right\}\{1+o(1)\}\\
    =&O(rb_2/p)\to 0,
\end{align*}
and 
\begin{align*}
    \var(\mathcal{L}_3)=O\{(p-rb)/p\}\to 0.
\end{align*}
Thus, $\mathcal{L}_2=o_p(1)$ and $\mathcal{L}_3=o_p(1)$.

For $\mathcal{L}_1$, by Theorem 17.2.1 and the arguments on page 338 in \cite{Ibragimov1971IndependentRV}, we have for properly chosen $r$ and $b_1$, for some positive constant $C$,
\begin{align}\label{A.26}
	\left|\E\{\exp(it\mathcal{L}_1)\}-\prod_{l=1}^r\E\left[\exp\{itp^{-1/2}V_{\gamma,1}(l)\}\right] \right|\le Cr\alpha_X(b_2),
\end{align}
which implies that there exist independent random variables $\{y_l\}_{l=1}^r$ such that $y_l$ and $V_{\gamma,1}(l)$ are identically distributed and $\mathcal{L}_1$ has the same asymptotic distribution as $p^{-1/2}\sum_{l=1}^ry_l$. From the moment bounds for mixing random variables in Theorem 1 in \cite{COX1995Momentbounds}, the $\alpha$-mixing assumption in (A3) implies for all $1\le l\le r$,
\begin{align}\label{A.27}
	\E(y_l^4)=&\E[\{V_{\gamma,1}(l)\}^4]\nonumber\\
	=&\sigma_{\gamma\gamma}^{-2}\E\left\{\left|\sum_{(l-1)b+1<i\le (l-1)b+b_1}\{(\check{t}^2_i-v_\gamma)\mathbb{I}(\check{t}^2_i\ge v_\gamma)-\E\{(\check{t}^2_i-v_\gamma)\mathbb{I}(\check{t}^2_i\ge v_\gamma)\}\}\right|^4\right\}\nonumber\\
	\le &\sigma_{\gamma\gamma}^{-2}b_1^2\left[C_1+C_2\sum_{i=1}^\infty i\{\alpha_X(i)\}^{\epsilon/(2+\epsilon)}\right]=O(b_1^2),
\end{align}
where $C_1$ and $C_2$ are constants. Similarly, $\var(y_l)=E(y_l^2)=O(b_1)$. Thus we have 
\begin{align*}
    \sum_{l=1}^r\E(y_l^4)/\left\{\sum_{l=1}^r\var(y_l)\right\}^2=O(r^{-1})=o(1),
\end{align*}
which, by the Liapounov central limit theorem, leads to $p^{-1/2}\sum_{l=1}^ry_l\cd N(0,1)$. In summary, we have \begin{align*}
    \frac{\check{T}_{\lceil\gamma p\rceil}-\mu_\gamma}{\sqrt{p\sigma_{\gamma\gamma}}}\cd N(0,1).
\end{align*}
The Cramer-Wold Theorem implies the asymptotic joint distribution of different $\check{T}_{\lceil\gamma p\rceil}$'s.

}
	
	\noindent\emph{Step 3.} Notice that with probability approaching one, we have
	\begin{align*}
	&\left|\check{T}_{\lceil\gamma p\rceil}-\max_{1\le i_1<\ldots<i_{\lceil\gamma p\rceil}}\sum_{l=1}^{\lceil\gamma p\rceil}\frac{n\mu_{i_l}^2}{\sigma_{i_li_l}}-\max_{1\le i_1<\ldots<i_{\lceil\gamma p\rceil}}\sum_{l=1}^{\lceil\gamma p\rceil}\frac{n(\bar{X}_{.i_l}-\mu_{i_l})^2}{\sigma_{i_li_l}}\right|\\
	=&\left|\max_{1\le i_1<\ldots<i_{\lceil\gamma p\rceil}}\sum_{l=1}^{\lceil\gamma p\rceil}\frac{2n\bar{X}_{.i_l}\mu_{i_l}}{\sigma_{i_li_l}}\right|\\
	\le&2\left(\max_{1\le i_1<\ldots<i_{\lceil\gamma p\rceil}}\sum_{l=1}^{\lceil\gamma p\rceil}\frac{n\bar{X}_{.i_l}^2}{\sigma_{i_li_l}} \right)^{1/2}\cdot\left(\max_{1\le i_1<\ldots<i_{\lceil\gamma p\rceil}}\sum_{l=1}^{\lceil\gamma p\rceil}\frac{n\mu_{i_l}^2}{\sigma_{i_li_l}} \right)^{1/2}\\
	\le&2\left(\lceil\gamma p\rceil\max_{1\le i\le p}\frac{n\bar{X}_{.i}}{\sigma_{ii}}\right)^{1/2}\left(\max_{1\le i_1<\ldots<i_{\lceil\gamma p\rceil}}\sum_{l=1}^{\lceil\gamma p\rceil}\frac{n\mu_{i_l}^2}{\sigma_{i_li_l}} \right)^{1/2}\\
	\le&C\sqrt{\gamma p\log p}\left(\max_{1\le i_1<\ldots<i_{\lceil\gamma p\rceil}}\sum_{l=1}^{\lceil\gamma p\rceil}\frac{n\mu_{i_l}^2}{\sigma_{i_li_l}} \right)^{1/2}=o(\sqrt{p}).
	\end{align*}
	due to the assumption that $\gamma \log(p) \max_{1\le i_1<\ldots<i_{\lceil\gamma p\rceil}}\sum_{l=1}^{\lceil\gamma p\rceil}\frac{n\mu_{i_l}^2}{\sigma_{i_li_l}}=o(1)$.
	
	In summary, note that
	\begin{align*}
	&\frac{T_{\lceil\gamma p\rceil}-\mu_\gamma-\max_{1\le i_1<\ldots<i_{\lceil\gamma p\rceil}}\sum_{l=1}^{\lceil\gamma p\rceil}\frac{n\mu_{i_l}^2}{\sigma_{i_li_l}}}{\sqrt{p\sigma_{\gamma\gamma}}}\\
	=&\frac{\max_{1\le i_1<\ldots<i_{\lceil\gamma p\rceil}}\sum_{l=1}^{\lceil\gamma p\rceil}\frac{n(\bar{X}_{.i_l}-\mu_{i_l})^2}{\sigma_{i_li_l}}-\mu_\gamma}{\sqrt{p\sigma_{\gamma\gamma}}}+\frac{T_{\lceil\gamma p\rceil}-\check{T}_{\lceil\gamma p\rceil}}{\sqrt{p\sigma_{\gamma\gamma}}}\\
	&+\frac{1}{\sqrt{p\sigma_{\gamma\gamma}}}\left\{ \check{T}_{\lceil\gamma p\rceil}-\max_{1\le i_1<\ldots<i_{\lceil\gamma p\rceil}}\sum_{l=1}^{\lceil\gamma p\rceil}\frac{n\mu_{i_l}^2}{\sigma_{i_li_l}}-\max_{1\le i_1<\ldots<i_{\lceil\gamma p\rceil}}\sum_{l=1}^{\lceil\gamma p\rceil}\frac{n(\bar{X}_{.i_l}-\mu_{i_l})^2}{\sigma_{i_li_l}}\right\},
	\end{align*}
	for which, the first term is asymptotically standard normal due to the result of \emph{Step 2}, the others are negligible due to the results of \emph{Step 1} and \emph{Step 3}. Then, the proof is complete.
\end{proof}

\subsection{Proof of (\romannumeral1) in Theorem 4}
\begin{proof}
{  
According to the proof of Theorem 3, we have
 \begin{align*}
    \frac{T_{\lceil\gamma p\rceil}-\mu_\gamma}{\sqrt{p\sigma_{\gamma\gamma}}}=\frac{1}{\sqrt{p}}\sum_{i=1}^pT_{i,\gamma}+o_p(1),
\end{align*}
where $T_{i,\gamma}$ is defined in \eqref{A.24}. Therefore, by \eqref{A.7} and Lemma \ref{independence of main term}, it suffices to show that $\check{t}_{(p+1-s)}^2$ is asymptotically independent with $p^{-1/2}\sum_{i=1}^pT_{i,\gamma}$.

Define $A_p=A_p(y)=\{p^{-1/2}\sum_{i=1}^pT_{i,\gamma}\le y\}$ and $B_i=B_i(x)=\{\check{t}_i^2>b_p+x\}$.
}
According to the proof of Theorems 1 and 3, we have $\Pr(A_p)\to\Phi(y)$ and
	\begin{align*}
	\Pr\left(\bigcup_{1\le i_1<i_2<\cdots<i_s\le p}B_{i_1}\cdots B_{i_s}\right)\to 1-\Lambda(x)\sum_{i=0}^{s-1}\frac{\{\log \Lambda^{-1}(x)\}^i}{i!}.
	\end{align*}
	Our goal is to prove that
	\begin{align*}
	\Pr\left(\bigcup_{1\le i_1<i_2<\cdots<i_s\le p}A_pB_{i_1}\cdots B_{i_s}\right)\to \Phi(y)\left[1-\Lambda(x)\sum_{i=0}^{s-1}\frac{\{\log \Lambda^{-1}(x)\}^i}{i!}\right].
	\end{align*}
	
	By the Bonferroni inequality given in Lemma \ref{Bonferroni inequality}, we observe that for any integer $1\le d<[(p-s)/2]$,
	\begin{align}\label{C.2}
	\sum_{j=s}^{2d+s+1}(-1)^{s+j}\binom{j-1}{s-1}\sum_{1\le i_1<\cdots<i_j\le p}&\Pr(A_pB_{i_1}\cdots B_{i_j})\le \Pr\left(\bigcup_{1\le i_1<i_2<\cdots<i_s\le p}A_pB_{i_1}\cdots B_{i_s}\right)\nonumber\\
	\le& \sum_{j=s}^{2d+s}(-1)^{s+j}\binom{j-1}{s-1}\sum_{1\le i_1<\cdots<i_j\le p}\Pr(A_pB_{i_1}\cdots B_{i_j}).
	\end{align}
	Further,
	\begin{align}\label{C.3}
	&\sum_{j=s}^{2d+s}(-1)^{s+j}\binom{j-1}{s-1}\sum_{1\le i_1<\cdots<i_j\le p}\Pr(A_pB_{i_1}\cdots B_{i_j})\nonumber\\
	=&\sum_{j=s}^{2d+s-1}(-1)^{s+j}\binom{j-1}{s-1}\sum_{1\le i_1<\cdots<i_j\le p}\Pr(A_p)\Pr(B_{i_1}\cdots B_{i_j})\nonumber\\
	&+\sum_{j=s}^{2d+s-1}(-1)^{s+j}\binom{j-1}{s-1}\sum_{1\le i_1<\cdots<i_j\le p}\{\Pr(A_pB_{i_1}\cdots B_{i_j})-\Pr(A_p)\Pr(B_{i_1}\cdots B_{i_j})\}\nonumber\\
	&+\binom{2d+s-1}{s-1}\sum_{1\le i_1<\cdots<i_{2d+s}\le p}\Pr(A_p)\Pr(B_{i_1}\cdots B_{i_{2d+s}})\nonumber\\
	\le &\Pr(A_p)\Pr\left(\bigcup_{1\le i_1<i_2<\cdots<i_s\le p}B_{i_1}\cdots B_{i_s}\right)+\sum_{j=s}^{2d+s-1}\zeta(p,j)+\binom{2d+s-1}{s-1}H(p,2d+s),
	\end{align}
	where
	\begin{align*}
	&\zeta(p,j):=\binom{j-1}{s-1}\sum_{1\le i_1<\cdots<i_j\le p}|\Pr(A_pB_{i_1}\cdots B_{i_j})-\Pr(A_p)\Pr(B_{i_1}\cdots B_{i_j})|~~\text{and}\\
	&H(p,j):=\sum_{1\le i_1<\cdots<i_j\le p}\Pr(B_{i_1}\cdots B_{i_j}).
	\end{align*}
	According to the proof of \eqref{A.10} and \eqref{A.11}, we have for each $j$,
	\begin{align}\label{C.4}
	\lim_{p\to\infty}H(p,j)=\pi^{-j/2}\frac{1}{j!}\exp\left(-\frac{jx}{2}\right).
	\end{align}
	We claim that for each $j$,
	\begin{align}\label{C.5}
	\lim_{p\to\infty}\zeta(p,j)\to 0.
	\end{align}    	
	Combining \eqref{C.2}-\eqref{C.5}, by letting $p\to \infty$, we have
	\begin{align*}
	\limsup_{p\to\infty}\Pr\left(\bigcup_{1\le i_1<i_2<\cdots<i_s\le p}A_pB_{i_1}\cdots B_{i_s}\right)\le \Phi(y)\left[1-\Lambda(x)\sum_{i=0}^{s-1}\frac{\{\log \Lambda^{-1}(x)\}^i}{i!}\right].
	\end{align*}
	Likewise, we can prove
	\begin{align*}
	\liminf_{p\to\infty}\Pr\left(\bigcup_{1\le i_1<i_2<\cdots<i_s\le p}A_pB_{i_1}\cdots B_{i_s}\right)\ge \Phi(y)\left[1-\Lambda(x)\sum_{i=0}^{s-1}\frac{\{\log \Lambda^{-1}(x)\}^i}{i!}\right].
	\end{align*}
	Hence, the desired result follows.
	
	It remains to prove that the claim \eqref{C.5} indeed holds. {  Recalling the decomposition in \eqref{A.25} that $p^{-1/2}\sum_{i=1}^pT_{i,\gamma}=\mathcal{L}_1+\mathcal{L}_2+\mathcal{L}_3$, we further divide $\mathcal{L}_1$ as $\mathcal{L}_1=\mathcal{L}_{11}+\mathcal{L}_{12}$, where  
    \begin{align*}
	    &V_{\gamma,11}(l):=\sum_{\substack{(l-1)b<i\le(l-1)b+b_1\\i\notin\{i_1,\cdots,i_j\}}}T_{i,\gamma},~~V_{\gamma,12}(l):=\sum_{\substack{(l-1)b<i\le(l-1)b+b_1\\i\in\{i_1,\cdots,i_j\}}}T_{i,\gamma},\\
        &\mathcal{L}_{11}:=\frac{1}{\sqrt{p}}\sum_{l=1}^rV_{\gamma,11}(l),~~\mathcal{L}_{12}:=\frac{1}{\sqrt{p}}\sum_{l=1}^rV_{\gamma,12}(l).
    \end{align*}
    Now, we have the decomposition
	\begin{align*}
	p^{-1/2}\sum_{i=1}^pT_{i,\gamma}=\mathcal{L}_{11}+(\mathcal{L}_{12}+\mathcal{L}_{2}+\mathcal{L}_{3})=:\mathcal{L}_{11}+\mathcal{L}_{small}.
	\end{align*}
	}
    Next, we show that for any $\epsilon>0$, there exists a sequence of constants $c:=c_p>0$ with $c_p\to \infty$ such that
	\begin{align}\label{C.6}
	\Pr(|\mathcal{L}_{small}|>\epsilon)\le p^{-c}.
	\end{align}
	For $\mathcal{L}_2$, similarly as \eqref{A.26}, we have for properly chosen $r$ and $b_1$, for some positive constant $C$,
	\begin{align*}
	\left|\E\{\exp(it\mathcal{L}_2)\}-\prod_{l=1}^r\E\left[\exp\{itp^{-1/2}V_{\gamma,2}(l)\}\right] \right|\le Cr\alpha_X(b_1),
	\end{align*}
	i.e. there exist independent random variables $\{z_l\}_{l=1}^r$ such that $z_l$ and $V_{\gamma,2}(l)$ are identically distributed and $\mathcal{L}_2$ has the same asymptotic distribution as $p^{-1/2}\sum_{l=1}^rz_l$. Similarly as \eqref{A.27}, $E(z_l^4)=O(b_2^2)$ and $\sigma_{z}^4:=\var\{(p^{-1/2}\sum_{l=1}^rz_l)^2\}=O(r^2p^{-2}b_2^2)$. By Markov inequality, we have for some constant $K>0$,
	\begin{align*}
	\Pr(|\mathcal{L}_2|>\epsilon)\le& \Pr\left(\left|p^{-1/2}\sum_{l=1}^rz_l\right|^2>\epsilon^2\right)+Cr\alpha_X(b_1)\\
	\le&\exp(-\epsilon^2\sigma_z^{-2})\cdot\E\left[\exp\left\{\sigma_z^{-2}\left(p^{-1/2}\sum_{l=1}^rz_l\right)^2\right\}\right]+Cr\alpha_X(b_1)\\
	\le&K[\exp\{-\epsilon^2p^2r^{-2}b_2^{-2}\}\cdot \log r+r\alpha_X(b_1)],
	\end{align*}
	where the last step follows from the fact that
	\begin{align*}
	\sigma_z^{-2}\left(\sum_{l=1}^rz_l\right)^2/\log(\log r)\to 0, a.s.
	\end{align*}
	by the law of the iterated logarithm of zero-mean square- integrable martingale (see Theorem 4.8 in \cite{Hall1980MartingaleLT}). Recalling that $r(b_1+b_2)\le p<(r+1)(b_1+b_2)$ and $r\to\infty$, $b_1\to\infty$, $b_2\to\infty$, $rb_1/p\to 1$, $rb_2/p\to 0$ as $p\to\infty$, we can set $b_1/b_2\gg \log p$ and $r\ll p$ so that $K[\exp\{-\epsilon^2p^2r^{-2}b_2^{-2}\}\cdot \log r+r\alpha_X(b_1)]\le p^{-c}$. Therefore,
	\begin{align*}
	\Pr(|\mathcal{L}_2|>\epsilon)\le p^{-c}.
	\end{align*}
	Similarly, we can prove $\Pr(|\mathcal{L}_3|>\epsilon)\le p^{-c}$ and $\Pr(|\mathcal{L}_{12}|>\epsilon)\le p^{-c}$. Thus, \eqref{C.6} is proved. Further, we have
    \begin{align*}
	    &\Pr(A_p(y)B_{i_1}\cdots B_{i_j})\\
	    \le& \Pr(A_p(y)B_{i_1}\cdots B_{i_j},|\mathcal{L}_{small}|\le\epsilon)+p^{-c}\\
	    \le& \Pr(\mathcal{L}_{11}\le y+\epsilon,B_{i_1}\cdots B_{i_j})+p^{-c}\\
	    =&\Pr(\mathcal{L}_{11}\le y+\epsilon)\Pr(B_{i_1}\cdots B_{i_j})+p^{-c}\\
	    \le&\left\{\Pr(\mathcal{L}_{11}\le y+\epsilon,|\mathcal{L}_{small}|\le\epsilon)+p^{-c}\right\}\Pr(B_{i_1}\cdots B_{i_j})+p^{-c}\\
	    \le &\Pr(A_p(y+2\epsilon))\Pr(B_{i_1}\cdots B_{i_j})+2p^{-c}.
	\end{align*}
	Likewise,
	\begin{align*}
	    \Pr(A_p(y)B_{i_1}\cdots B_{i_j})\ge \Pr(A_p(y-2\epsilon))\Pr(B_{i_1}\cdots B_{i_j})-2p^{-c}.
	\end{align*}
	Hence,
	\begin{align*}
	|\Pr(A_p(y)B_{i_1}\cdots B_{i_j})-\Pr(A_p(y))\Pr(B_{i_1}\cdots B_{i_j})|\le \Delta_{p,\epsilon}\cdot\Pr(B_{i_1}\cdots B_{i_j})+2p^{-c},
	\end{align*}
	where
	\begin{align*}
	\Delta_{p,\epsilon}
	&=\left|\Pr\left(A_p(y)\right)-\Pr\left(A_p(y+2\epsilon)\right) \right|+\left|\Pr\left(A_p(y)\right)-\Pr\left(A_p(y-2\epsilon)\right) \right|\\
	&=\Pr\left(A_p(y+2\epsilon)\right)-\Pr\left(A_p(y-2\epsilon)\right)
	\end{align*}
	since $\Pr(A_p(y))$ is increasing in $y$. By running over all possible combinations of $1\le i_1<...<i_j\le p$, we have
	\begin{equation*}
	\zeta(p,j)\le \binom{j-1}{s-1}\left\{\Delta_{p,\epsilon}\cdot H(p,j)+2\binom{p}{j}\cdot p^{-c}\right\}.
	\end{equation*}
	Since $\Pr(A_p(y))\to\Phi(y)$, we have $\lim_{\epsilon\downarrow 0}\limsup_{p\to\infty}\Delta_{p,\epsilon}=\lim_{\epsilon\downarrow 0}\{\Phi(y+2\epsilon)-\Phi(y-2\epsilon)\}=0$. Combining with \eqref{C.4}, first sending $p\to\infty$ and then letting $\epsilon\downarrow 0$, \eqref{C.5} is proved.

\end{proof}

\subsection{Proof of (\romannumeral2) in Theorem 4}
\begin{proof}
	It follows from (\romannumeral1) in Theorem 4 and (\romannumeral2) in Theorem 1, immediately.
\end{proof}

\subsection{Proof of Theorem 5}
\begin{proof}
    { By the proof of Theorem 3, if $\gamma \log(p) \max_{1\le i_1<\ldots<i_{\lceil\gamma p\rceil}}\sum_{l=1}^{\lceil\gamma p\rceil}\frac{n\mu_{i_l}^2}{\sigma_{i_li_l}}=o(1)$, then we have
	\begin{align*}
	\frac{T_{\lceil\gamma p\rceil}-\mu_\gamma-\max_{1\le i_1<\ldots<i_{\lceil\gamma p\rceil}}\sum_{l=1}^{\lceil\gamma p\rceil}\frac{n\mu_{i_l}^2}{\sigma_{i_li_l}}}{\sqrt{p\sigma_{\gamma\gamma}}}
	=&\frac{\max_{1\le i_1<\ldots<i_{\lceil\gamma p\rceil}}\sum_{l=1}^{\lceil\gamma p\rceil}\frac{n(\bar{X}_{.i_l}-\mu_{i_l})^2}{\sigma_{i_li_l}}-\mu_\gamma}{\sqrt{p\sigma_{\gamma\gamma}}}+o_p(1)\\
    =&\frac{1}{\sqrt{p}}\sum_{i=1}^pT_{i,\gamma}+o_p(1),
	\end{align*}
    which, according to the proof of Theorem 4, implies that the asymptotic independence still hold.
    }
\end{proof}

{  
\section{Accuracy of the Bootstrap}
In this section, we present the theoretical justification of the proposed bootstrap method.

\renewcommand{\thethm}{6}
\begin{thm}\label{wb}
	Suppose Assumptions (A1)-(A3) hold. Under $H_0$, we have
 
    (\romannumeral1) if $k^{7/2}n^{-1/2}\log^{7/2}(np)=o(1)$, 
	\begin{align*}
	\sup_{x\ge 0}\left|\Pr(T_k\le x)-\Pr(T_k^*\le x|\X_1,\ldots,\X_n)\right|=o_p(1).
	\end{align*}

    (\romannumeral1) if $p=o(n)$, 
	\begin{align*}
	\sup_{x\ge 0}\left|\Pr(T_{\lceil \gamma p\rceil}\le x)-\Pr(T_{\lceil \gamma p\rceil}^*\le x|\X_1,\ldots,\X_n)\right|=o_p(1).
	\end{align*}
\end{thm}

\begin{proof}
    \emph{Proof of (\romannumeral1). Case \uppercase\expandafter{\romannumeral1}}. Suppose $\{\X_i\}_{i=1}^n\stackrel{i.i.d.}{\sim}\mathcal{N}_p(\bm 0,\bms)$. The triangle inequality yields that 
    \begin{align*}
        &\sup_{x\ge 0}|\Pr(T_k\le x)-\Pr(T_k^*\le x|\X_1^n)|\\
        \le &\sup_{x\ge 0}|\Pr(\check{T}_k\le x)-\Pr(T_k\le x)|+\sup_{x\ge 0}|\Pr(\check{T}_k\le x)-\Pr(T_k^*\le x|\X_1^n)|,
    \end{align*}
    where $\X_1^n=\{\X_1,\ldots,\X_n\}$ and $\check{T}_k=\max_{1\le i_1<\ldots<i_k\le p}\sum_{l=1}^kn\bar{X}_{.i_l}^2/\sigma_{i_li_l}$.
    
    \emph{Step 1.} Analyze $|\Pr(\check{T}_k\le x)-\Pr(T_k\le x)|$. By \eqref{A.6} and the fact that $|\bar{\X}|_\infty=O_p\{\sqrt{\log(p)/n}\}$, we deduce
	\begin{align*}
	|T_k-\check{T}_k|\le &nk\max_{1\le i\le p}|\bar{X}_{.i}^2/\hat{\sigma}_{ii}-\bar{X}_{.i}^2/\sigma_{ii}|\\
	\le & nk|\bar{\X}|_\infty^2\cdot \max_{1\le i\le p}|\hat{\sigma}_{ii}^{-1}-\sigma_{ii}^{-1}|=O_p\{kn^{-1/2}\log^{3/2}(p)\},
	\end{align*}
    which, under the assumption that $n^{-1/2}\{k\log(np)\}^{7/2}=o(1)$, implies that we can pick $\zeta_1$ and $\zeta_2$ such that 
    \begin{align}\label{A.16}
        \Pr(|T_k-\check{T}_k|>\zeta_1)\le\zeta_2~~\text{with}~~\zeta_1k^{5/2}\log^2(np)=o(1)~~\text{and}~~\zeta_2=o(1).
    \end{align}
	Define the event $\mathcal{B}=\{|T_k-\check{T}_k|\le \zeta_1\}$, then we have
    \begin{align*}
        |\Pr(\check{T}_k\le x)-\Pr(T_k\le x)|
        \le&\Pr(\check{T}_k\le x, T_k>x)+ \Pr(\check{T}_k> x, T_k\le x)\\
        \le&\Pr(\check{T}_k\le x, T_k>x,\mathcal{B})+ \Pr(\check{T}_k> x, T_k\le x,\mathcal{B})+2\zeta_2\\
        \le &\Pr(x-\zeta_1<\check{T}_k\le x)+\Pr(x<\check{T}_k\le x+\zeta_1)+2\zeta_2.
    \end{align*}
    Define $\check{V}:=n^{-1/2}\diag^{-1/2}(\bms)\sum_{j=1}^n\X_j$, $\mathbb{S}^{p-1}:=\{b\in\mathbb{R}^p:|b|=1\}$ and $\mathcal{S}:=\{v\in\mathbb{S}^{p-1}:\Vert v\Vert_0\le k\}$. For any $1\le k\le p$, define $\mathcal{A}(x,k):=\bigcap_{v\in\mathcal{S}}\{w\in\mathbb{R}^p:w^\top v\le\sqrt{x}\}$. Notice that $\{\check{T}_k\le x\}=\{\check{V}\in\mathcal{A}(x,k)\}$. By Lemma \ref{lemma 6.1 in Z15}, we can find an $\epsilon$-net $\mathcal{S}_\epsilon$ of $\mathcal{S}$ such that $\Vert\mathcal{S}_\epsilon\Vert_0\le \{(2+\epsilon)ep/(\epsilon k)\}^k$ and
	\begin{align*}
	A_1(x):=\bigcap_{v\in\mathcal{S}_\epsilon}\{w\in\mathbb{R}^p:w^\top v\le(1-\epsilon)\sqrt{x}\}\subseteq\mathcal{A}(x,k)\subseteq\bigcap_{v\in\mathcal{S}_\epsilon}\{w\in\mathbb{R}^p:w^\top v\le\sqrt{x}\}=:A_2(x).
	\end{align*}
	We set $\epsilon=1/n$ throughout the following arguments. Then, we have
    \begin{align*}
        &\Pr(x-\zeta_1<\check{T}_k\le x)\\
        \le&\Pr(\check{V}\in A_2(x))-\Pr(\check{V}\in A_1(x-\zeta_1))\\
        \le&\Pr((1-\epsilon)(\sqrt{x}-\sqrt{\zeta_1})\le\max_{v\in\mathcal{S}_\epsilon}v^\top \check{V}\le\sqrt{x})\\
        \le&\Pr((1-\epsilon)(\sqrt{x}-\sqrt{\zeta_1})\le\max_{v\in\mathcal{S}_\epsilon}v^\top \check{V}\le(1-\epsilon)\sqrt{x})+\Pr((1-\epsilon)\sqrt{x}\le\max_{v\in\mathcal{S}_\epsilon}v^\top \check{V}\le\sqrt{x})\\
        :=&I_1+I_2.
    \end{align*}
    By Nazarov's inequality (see Lemma A.1 in \cite{Chernozhukov2017CLT}), we have 
    \begin{align*}
        I_1\le C\sqrt{\zeta_1}\sqrt{\log(\Vert\mathcal{S}_\epsilon\Vert_0)}=O\{\sqrt{\zeta_1k\log(np/k)}\}=o(1)
    \end{align*}
    To deal with $I_2$, we note that when $x\le k^3\log^2(np/k)$, $\epsilon\sqrt{x}\le k^{3/2}\log(np/k)/n$. Again by Nazarov's inequality, we have
    \begin{align*}
        I_2\le \Pr(\sqrt{x}-k^{3/2}\log(np/k)/n\le\max_{v\in\mathcal{S}_\epsilon}v^\top \check{V}\le\sqrt{x})=O\{k^2\log^{3/2}(np/k)/n\}=o(1).
    \end{align*}
    When $x> k^3\log^2(np/k)$, by Lemma 7.4 in \cite{fanQW15}, we have
    \begin{align*}
        I_2\le \Pr((1-\epsilon)\sqrt{x}\le\max_{v\in\mathcal{S}_\epsilon}v^\top \check{V})\le\frac{\E(\max_{v\in\mathcal{S}_\epsilon}v^\top \check{V})}{(1-\epsilon)k^{3/2}\log(np/k)}\le \frac{C\sqrt{k\log(np/k)}}{(1-\epsilon)k^{3/2}\log(np/k)}=o(1).
    \end{align*}
    Thus, we obtain $\Pr(x-\zeta_1<\check{T}_k\le x)=o(1)$. Similarly, $\Pr(x<\check{T}_k\le x+\zeta_1)=o(1)$. Therefore, we obtain
    \begin{align*}
        \sup_{x\ge 0}|\Pr(\check{T}_k\le x)-\Pr(T_k\le x)|=o(1).
    \end{align*}
 
    \emph{Step 2.} Analyze $|\Pr(\check{T}_k\le x)-\Pr(T_k^*\le x|\X_1^n)|$. Define $V^*:=n^{-1/2}\diag^{-1/2}(\hat{\bms})\sum_{j=1}^n(\X_j-\bar{\X})\xi_j$ with $\xi_1,\ldots,\xi_n$ being i.i.d. Rademacher variables and independent of $\X_1^n$. Define $\Delta_1:=\max\{|\Pr(\check{V}\in A_1(x))-\Pr(V^*\in A_1(x)|\X_1^n)|,|\Pr(\check{V}\in A_2(x))-\Pr(V^*\in A_2(x)|\X_1^n)|\}$. Using similar arguments in \emph{Step 1}, we have
	\begin{align*}
	\Pr(T_k^*\le x|\X_1^n)=&\Pr(V^*\in\mathcal{A}(x,k)|\X_1^n)\\
	\le&\Pr(V^*\in A_2(x)|\X_1^n)\\
	\le&\Pr(\check{V}\in A_2(x))+\Delta_1\\
	\le&\Pr(\check{V}\in\mathcal{A}(x,k))+\{\Pr(\check{V}\in A_2(x))-\Pr(\check{V}\in A_1(x))\}+\Delta_1\\
	=&\Pr(\check{T}_k\le x)+o(1)+\Delta_1.
	\end{align*}
	Similarly we have $\Pr(T_k^*\le x|\X_1^n)\ge \Pr(\check{T}_k\le x)-o(1)-\Delta_1$. Thus,
	\begin{align*}
	|\Pr(T_k^*\le x|\X_1^n)-\Pr(\check{T}_k\le x)|\le \Delta_1+o(1).
	\end{align*}
	To bound $\Delta_1$, according to the proof of (39) in \cite{Chernozhukov2017CLT}, we obtain
	\begin{align*}
	&\sup_{x\ge 0}|\Pr(\check{V}\in A_1(x))-\Pr(V^*\in A_1(x)|\X_1^n)|\\
	=&\sup_{x\ge 0}\left|\Pr\left(\max_{v\in\mathcal{S}_\epsilon}v^\top\check{V}\le(1-\epsilon)\sqrt{x}\right)-\Pr\left(\max_{v\in\mathcal{S}_\epsilon}v^\top V^*\le(1-\epsilon)\sqrt{x}|\X_1^n\right)\right|\\
	\le&C\Delta_2^{1/3}k^{2/3}\log^{2/3}(np/k),
	\end{align*}
	where $\Delta_2:=\max_{u,v\in\mathcal{S}}|u^\top(\hat{\R}-\R)v|$. Similarly,
	\begin{align*}
	\sup_{x\ge 0}|\Pr(\check{V}\in A_2(x))-\Pr(V^*\in A_2(x)|\X_1^n)|\le C\Delta_2^{1/3}k^{2/3}\log^{2/3}(np/k).
	\end{align*}
	For any $u,v\in\mathcal{S}$, by \eqref{A.6},
	\begin{align*}
	&|u^\top(\hat{\R}-\R)v|\\
	\le&|u^\top\diag^{-1/2}(\hat{\bms})(\hat{\bms}-\bms)\diag^{-1/2}(\hat{\bms})v|\\
	&+|u^\top\{\diag^{-1/2}(\hat{\bms})-\diag^{-1/2}(\bms)\}\bms\diag^{-1/2}(\hat{\bms})v|\\
	&+|u^\top\diag^{-1/2}(\bms)\bms\{\diag^{-1/2}(\hat{\bms})-\diag^{-1/2}(\bms)\}v|\\
	\le&\sqrt{k}\Vert\diag^{-1/2}(\hat{\bms})u\Vert_2\cdot\Vert\diag^{-1/2}(\hat{\bms})v\Vert_\infty\cdot\max_{1\le i\le p}\Vert(\hat{\bms}_i-\bms_i)\Vert_1\\
	&+\Vert\bms\Vert_2\cdot\Vert\diag^{1/2}(\hat{\bms})\Vert_2\cdot\max_{1\le i\le p}|\sigma_{ii}^{-1/2}-\hat{\sigma}_{ii}^{-1/2}|\\
	&+\Vert\bms\Vert_2\cdot\Vert\diag^{1/2}(\bms)\Vert_2\cdot\max_{1\le i\le p}|\sigma_{ii}^{-1/2}-\hat{\sigma}_{ii}^{-1/2}|\\
	=&O_p(\sqrt{k\log(p)/n})+O_p(\sqrt{\log(p)/n}).
	\end{align*}
	Therefore, under the assumption that $k^{7/2}n^{-1/2}\log^{7/2}(np)=o(1)$, we have $\Delta_1=o_p(1)$. Thus, 
    \begin{align}\label{A.17}
        \sup_{x\ge 0}|\Pr(\check{T}_k\le x)-\Pr(T_k^*\le x|\X_1^n)|=o_p(1).
    \end{align}
    The proof is thus completed by combining \emph{Step 1} and \emph{Step 2}.

    \emph{Case \uppercase\expandafter{\romannumeral2}}. When $\X$ has sub-Gaussian-type or sub-exponential-type or sub-polynomial-type tails, \eqref{A.6} and the fact that $|\bar{\X}|_\infty=O_p\{\sqrt{\log(p)/n}\}$ still hold. Therefore using the same arguments in \emph{Step 1} for \emph{Case \uppercase\expandafter{\romannumeral1}}, we can pick $\zeta_1$ and $\zeta_2$ satisfying \eqref{A.16}. Further, we still have
    \begin{align*}
        |\Pr(\check{T}_k\le x)-\Pr(T_k\le x)|
        \le\Pr(x-\zeta_1<\check{T}_k\le x)+\Pr(x<\check{T}_k\le x+\zeta_1)+2\zeta_2.
    \end{align*}
	By Corollary 2.1 in \cite{Chernozhukov2017CLT}, we have
    \begin{align*}
&\Pr(x-\zeta_1<\check{T}_k\le x)\\
\le{}&\Pr\bigl((1-\epsilon)(\sqrt{x}-\sqrt{\zeta_1})
\le\max_{v\in\mathcal{S}_\epsilon}v^\top\check{V}
\le(1-\epsilon)\sqrt{x}\bigr)\\
&+\Pr\bigl((1-\epsilon)\sqrt{x}
\le\max_{v\in\mathcal{S}_\epsilon}v^\top\check{V}\le\sqrt{x}\bigr)\\
\le{}&\Pr\bigl((1-\epsilon)(\sqrt{x}-\sqrt{\zeta_1})
\le\max_{v\in\mathcal{S}_\epsilon}v^\top W
\le(1-\epsilon)\sqrt{x}\bigr)\\
&+\Pr\bigl((1-\epsilon)\sqrt{x}
\le\max_{v\in\mathcal{S}_\epsilon}v^\top W\le\sqrt{x}\bigr)+c_{n,p,k},
\end{align*}
    where $W\sim N(\bm 0,\R)$ and $c_{n,p,k}=C\{k\log(np/k)\}^{7/6}/n^{1/6}=o(1)$. Thus following the arguments in \emph{Step 1} for \emph{Case \uppercase\expandafter{\romannumeral1}}, we obtain
    \begin{align*}
        \sup_{x\ge 0}|\Pr(\check{T}_k\le x)-\Pr(T_k\le x)|=o(1).
    \end{align*}

    On the other hand, again using Corollary 2.1 in \cite{Chernozhukov2017CLT}, we have
    \begin{align*}
        \Pr(\check{V}\in A(x,k))-\Pr(W\in A(x,k))\le &\Pr(\max_{v\in\mathcal{S}_\epsilon}v^\top \check{V}\le\sqrt{x})-\Pr(\max_{v\in\mathcal{S}_\epsilon}v^\top W\le(1-\epsilon)\sqrt{x})\\
        \le&\Pr((1-\epsilon)\sqrt{x}<\max_{v\in\mathcal{S}_\epsilon}v^\top W\le\sqrt{x})+c_{n,p,k}=o(1),
    \end{align*}
    and 
    \begin{align*}
        \Pr(W \in A(x,k))-\Pr(\check{V}\in A(x,k))\le &\Pr(\max_{v\in\mathcal{S}_\epsilon}v^\top W\le\sqrt{x})-\Pr(\max_{v\in\mathcal{S}_\epsilon}v^\top \check{V}\le(1-\epsilon)\sqrt{x})\\
        \le&\Pr((1-\epsilon)\sqrt{x}<\max_{v\in\mathcal{S}_\epsilon}v^\top W\le\sqrt{x})+c_{n,p,k}=o(1),
    \end{align*}
    which, using the same arguments for the proof of \eqref{A.17}, imply that \eqref{A.17} still hold. In summary, we get the desired result (\romannumeral1).

    \emph{Proof of (\romannumeral2).} Recalling that $\check{t}_i=\frac{1}{\sqrt{n\sigma_{ii}}}\sum_{j=1}^nX_{ji}$ and $t_i^*=\frac{1}{\sqrt{n\hat{\sigma}_{ii}}}\sum_{j=1}^n(X_{ji}-\bar{X}_{.i})\xi_j$. Note that $(\check{t}_1,\ldots,\check{t}_p)^\top$ has zero mean and covariance matrix $\R$. And given $\X_i^n$, $(t^*_1,\ldots,t^*_p)^\top$ has zero mean and covariance matrix $\hat{\R}$.

    According to the proof of Theorem 3, we have 
    \begin{align*}
        \frac{T_{\lceil\gamma p\rceil}-\mu_\gamma}{\sqrt{p\sigma_{\gamma\gamma}}}=\frac{\check{T}_{\lceil\gamma p\rceil}-\mu_\gamma}{\sqrt{p\sigma_{\gamma\gamma}}}+o_p(1)=\frac{1}{\sqrt{p}}\sum_{i=1}^pT_{i,\gamma}+o_p(1)\cd N(0,1).
    \end{align*}
    The derivation is distribution-free with respect to $(\check{t}_1,\ldots,\check{t}_p)^\top$, needing only mild moment condition and $\alpha$-mixing assumption. Hence, conditioning on $\X_1^n$, $T_{\lceil\gamma p\rceil}^*$ admits analogous result that
    \begin{align*}
        \frac{T_{\lceil\gamma p\rceil}^*-\mu^*_\gamma}{\sqrt{p\sigma^*_{\gamma\gamma}}}=\frac{1}{\sqrt{p}}\sum_{i=1}^pT^*_{i,\gamma}+o_p(1)\cd N(0,1),
    \end{align*}
    where $T^*_{i,\gamma}:=[(t_i^{*2}-v_\gamma)I(t_i^{*2}\ge {v}_\gamma)-\E\{(t_i^{*2}-v_\gamma)I(t_i^{*2}\ge {v}_\gamma)\}]/\sqrt{\sigma^*_{\gamma\gamma}}$, $\mu^*_\gamma:=\sum_{i=1}^p\E\{(t_i^{*2}-v_\gamma)I(t_i^{*2}\ge {v}_\gamma)+\gamma v_\gamma\}$ and $\sigma^*_{\gamma\gamma}:=\var\{p^{-1/2}\sum_{i=1}^p(t_i^{*2}-v_\gamma)I(t_i^{*2}\ge {v}_\gamma)\}$. Here, for brevity of notation, we make the conditioning implicit and write $\E(\cdot)$ and $\var(\cdot)$ instead of $\E(\cdot|\X_1^n)$ and $\var(\cdot|\X_1^n)$, respectively.

    Next, we analyze the difference between $\mu_\gamma$ and $\mu^*_\gamma$, and $\sigma_{\gamma\gamma}$ and $\sigma^*_{\gamma\gamma}$. The Berry-Esseen bound implies
    \begin{align*}
        &\E\{(\check{t}_i^2-v_\gamma)I(\check{t}_i^2\ge {v}_\gamma)\}-\E\{(t_i^{*2}-v_\gamma)I(t_i^{*2}\ge {v}_\gamma)\}\\
        =&\E(\check{t}_i^2)-\E(t_i^{*2})-[\E\{(\check{t}_i^2-v_\gamma)I(\check{t}_i^2< {v}_\gamma)\}-\E\{(t_i^{*2}-v_\gamma)I(t_i^{*2}< {v}_\gamma)\}]\\
        =&-\int_{-\sqrt{v_\gamma}}^{\sqrt{v_\gamma}}(t^2-v_\gamma)\{f_{n,i}(t)-f_{n,i}^*(t)\}\dif t=O(n^{-1/2}), \text{and}\\
        &\E\{(\check{t}_i^2-v_\gamma)^2I(\check{t}_i^2\ge {v}_\gamma)\}-\E\{(t_i^{*2}-v_\gamma)^2I(t_i^{*2}\ge {v}_\gamma)\}\\
        =&\E(\check{t}_i^4)-\E(t_i^{*4})-\int_{-\sqrt{v_\gamma}}^{\sqrt{v_\gamma}}(t^2-v_\gamma)^2\{f_{n,i}(t)-f_{n,i}^*(t)\}\dif t=O(n^{-1}+n^{-1/2})
    \end{align*}
    where $f_{n,i}(\cdot)$ and $f_{n,i}^*(\cdot)$ are the probability density functions of $\check{t}_i$ and $t_i^*$, respectively. Thus, 
    \begin{align*}
        (\mu_\gamma-\mu^*_\gamma)/\sqrt{p\sigma_{\gamma\gamma}}=O(\sqrt{p/n})=o(1),
    \end{align*}
    under the assumption that $p=o(n)$, and 
    \begin{align*}
        p^{-1}\sum_{i=1}^p[\var\{(\check{t}_i^2-v_\gamma)I(\check{t}_i^2\ge {v}_\gamma)\}-\var\{(t_i^{*2}-v_\gamma)I(t_i^{*2}\ge {v}_\gamma)\}]=O(n^{-1/2}).
    \end{align*}
    Similarly, the multivariate Berry-Esseen bound implies
    \begin{align*}
        \frac{2}{p}\sum_{1\le i<j\le p}&|\cov\{(\check{t}^2_i-v_{\gamma})\mathbb{I}(\check{t}^2_i\ge v_{\gamma}), (\check{t}^2_j-v_{\gamma})\mathbb{I}(\check{t}^2_j\ge v_{\gamma}) \}\\
        &- \cov\{(t_i^{*2}-v_{\gamma})\mathbb{I}(t_i^{*2}\ge v_{\gamma}), (t_j^{*2}-v_{\gamma})\mathbb{I}(t_j^{*2}\ge v_{\gamma}) \}|\\
        \le\frac{2}{p}\sum_{1\le i<j\le p}&|\cov\{(\check{t}^2_i-v_{\gamma})\mathbb{I}(\check{t}^2_i\ge v_{\gamma}), (\check{t}^2_j-v_{\gamma})\mathbb{I}(\check{t}^2_j\ge v_{\gamma}) \}|\\
        &\times\max_{1\le i<j\le p}\left|1-\frac{\cov\{(t_i^{*2}-v_{\gamma})\mathbb{I}(t_i^{*2}\ge v_{\gamma}), (t_j^{*2}-v_{\gamma})\mathbb{I}(t_j^{*2}\ge v_{\gamma}) \}}{\cov\{(\check{t}^2_i-v_{\gamma})\mathbb{I}(\check{t}^2_i\ge v_{\gamma}), (\check{t}^2_j-v_{\gamma})\mathbb{I}(\check{t}^2_j\ge v_{\gamma}) \}}\right|=o(1).
\end{align*}
Thus, $\sigma_{\gamma\gamma}/\sigma^*_{\gamma\gamma}=1+o(1)$. In summary, conditioning on $\X_1^n$, $T_{\lceil\gamma p\rceil}^*$ has the same asymptotic distribution as $T_{\lceil\gamma p\rceil}$. 
\end{proof}

}

\clearpage
\bibliographystyle{agsm}
\bibliography{refs}

\end{document}